\documentclass{article}

\usepackage{amsthm,amsmath,stmaryrd,bbm,hyperref,geometry,color}
\usepackage[utf8]{inputenc}
\usepackage[english]{babel}
\usepackage{graphicx}
\usepackage{amsfonts,amssymb}
\usepackage{verbatim}
\usepackage{enumitem}
\usepackage{algorithm}
\usepackage{authblk}
\usepackage{algpseudocode}
\usepackage{multirow}
\usepackage{xcolor}
\usepackage[autostyle]{csquotes}
\usepackage{caption}
\usepackage{subcaption}
\usepackage{siunitx}
\usepackage[export]{adjustbox}
\usepackage{ulem}
\usepackage{braket}
\usepackage{cite}
\usepackage{booktabs}
\usepackage{tabularx}
\usepackage{ragged2e}
\usepackage{array}

\DeclareSIUnit\angstrom{\text {Å}}

\newcommand{\veps}{\varepsilon}

\hypersetup{
    unicode=false,          
    pdftoolbar=true,        
    pdfmenubar=true,      
    pdffitwindow=false,    
    pdfstartview={FitH},
    citecolor=blue,
    colorlinks=true,       
    linkcolor=blue,
    filecolor=blue,      
    urlcolor=blue          
}

\title{The Convergence Frontier: Integrating Machine Learning and High Performance Quantum Computing for Next-Generation Drug Discovery}
\date{\today}

\author[1]{Narjes Ansari}
\author[1]{César Feniou}
\author[1]{Nicolaï Gouraud}
\author[1]{Daniele Loco}
\author[1]{Siwar Badreddine}
\author[1]{Baptiste Claudon}
\author[1]{Félix Aviat}
\author[1]{Marharyta Blazhynska}
\author[2]{Kevin Gasperich}
\author[1]{Guillaume Michel}
\author[1]{Diata Traore}
\author[1]{Corentin Villot}
\author[3]{Thomas Plé}
\author[3]{Olivier Adjoua}
\author[1,3]{Louis Lagardère}
\author[1,3,*]{Jean-Philip Piquemal}

\affil[1]{Qubit Pharmaceuticals, Advanced Research Department, 75014 Paris, France}
\affil[2]{Qubit Pharmaceuticals Inc, Advanced Research Department, Chicago, IL, USA}
\affil[3]{Sorbonne Université, Laboratoire de Chimie Théorique, UMR 7616 CNRS, 75005 Paris, France}
\affil[*]{Contact author: jean-philip.piquemal@sorbonne-universite.fr}
\begin{document}
\maketitle

\begin{abstract}
Integrating quantum mechanics into drug discovery marks a decisive shift from empirical trial-and-error toward quantitative precision. However, the prohibitive cost of \textit{ab initio} molecular dynamics has historically forced a compromise between chemical accuracy and computational scalability. This paper identifies the convergence of High-Performance Computing (HPC), Machine Learning (ML), and Quantum Computing (QC) as the definitive solution to this bottleneck. While ML foundation models, such as FeNNix-Bio1, enable quantum-accurate simulations, they remain tethered to the inherent limits of classical data generation. We detail how High-Performance Quantum Computing (HPQC), utilizing hybrid QPU-GPU architectures, will serve as the ultimate accelerator for quantum chemistry data. By leveraging Hilbert space mapping, these systems can achieve true chemical accuracy while bypassing the heuristics of classical approximations. We show how this tripartite convergence optimizes the drug discovery pipeline, spanning from initial system preparation to ML-driven, high-fidelity simulations. Finally, we position quantum-enhanced sampling as the \textquote{beyond GPU} frontier for modeling reactive cellular systems and pioneering next-generation materials.

\end{abstract}

\section{Introduction}
In modern pharmacology, quantum mechanics and drug discovery have become an essential duo. What was once a metaphorical search for a key in a dark room has been illuminated by the predictive precision of quantum physics. However, because these calculations are incredibly resource-heavy, they are rarely used in isolation. To make them practical, researchers now rely on a powerful convergence of acceleration strategies: High-Performance Computing (HPC), Machine Learning (ML), and the emerging capabilities of Quantum Computing (QC). Indeed, while artificial intelligence has significantly advanced the prediction of static protein structures through models like AlphaFold\cite{AlphaFold3}, characterizing the dynamic behavior and interactions of these systems in a chemically reactive realistic cellular environment remains a major computational challenge. Molecular Dynamics (MD) \cite{alder1959studies,mccammon1977dynamics} is the primary method for unraveling these motions, yet researchers face a difficult trade-off: full quantum physics/chemistry simulation, i.e. ab initio MD (AIMD) \cite{PhysRevLett.55.2471,kuhne2020cp2k} is too computationally intensive for large-scale biological systems, while classical approximations, i.e. the traditional empirical force fields,  often lack accuracy and require tedious manual parameterization. In this evolving landscape, the first major breakthrough involves coupling Machine Learning with GPU-accelerated HPC (GPU=Graphic Processing Unit). This synergy unlocked the use of Neural Network Potentials (NNPs)\cite{behler2007generalized,Behler-review}, which act as a powerful alternative to traditional force fields\cite{MACKERELL200591} methods by (1) achieving the high-precision accuracy of quantum mechanics at a fraction of the cost and (2) providing automated parameterization. Initially, existing NNPs often struggled with long-range interactions and charged species\cite{smith2017ani}; besides, despite GPU-acceleration, the condensed-phase biological simulations required for drug discovery predictions were hindered by the low scalability. 

To address these limitations, some of us recently presented FeNNix-Bio1\cite{Ple2025}, a foundation machine learning model engineered for high-speed, reactive atomistic molecular dynamics simulations of biological systems with unprecedented scalability and quantum precision.
In practice, the underlying Quantum Chemistry methodologies dictate NNP accuracy by providing the high-quality training data these models require.

Because exact solutions to the Schrödinger equation remain computationally intractable, researchers must rely on approximations whose cost creates a bottleneck, limiting the volume of high-fidelity data available for training.

This is where further convergence with quantum computing becomes a game changer. Quantum Computing is increasingly viewed as the ultimate computational frontier, the "beyond GPUs" accelerator, poised to revolutionize specific, high-complexity applications. In our case, it holds significant potential for electronic-structure problems, allowing to drastically enhance the quality of the reference NNPs databases. The primary advantage of QC lies in its ability to map an exponentially large Hilbert space, representing the possible states of electrons, onto a linear number of qubits. Quantum algorithms can theoretically perform ground-state wave-function calculations that are far more accurate than those possible on classical computers, and could reach the so-called \textit{chemical accuracy}, a level of precision  comparable to experimental predictions.

As a final step, by accelerating the core statistical mechanics algorithms designed to sample and measure observables of interest on biomolecular systems, quantum computing is also set to transform molecular simulations, pushing the boundaries of what is possible in MD and Monte Carlo sampling\cite{claudon2026quantumcircuitsmetropolishastingsalgorithm,claudon2026quantumcircuitsmetropolishastingsalgorithm, qp_cqt_quantinuum}. 
\begin{figure}[h!]
    \centering
\includegraphics[width=0.65\linewidth]{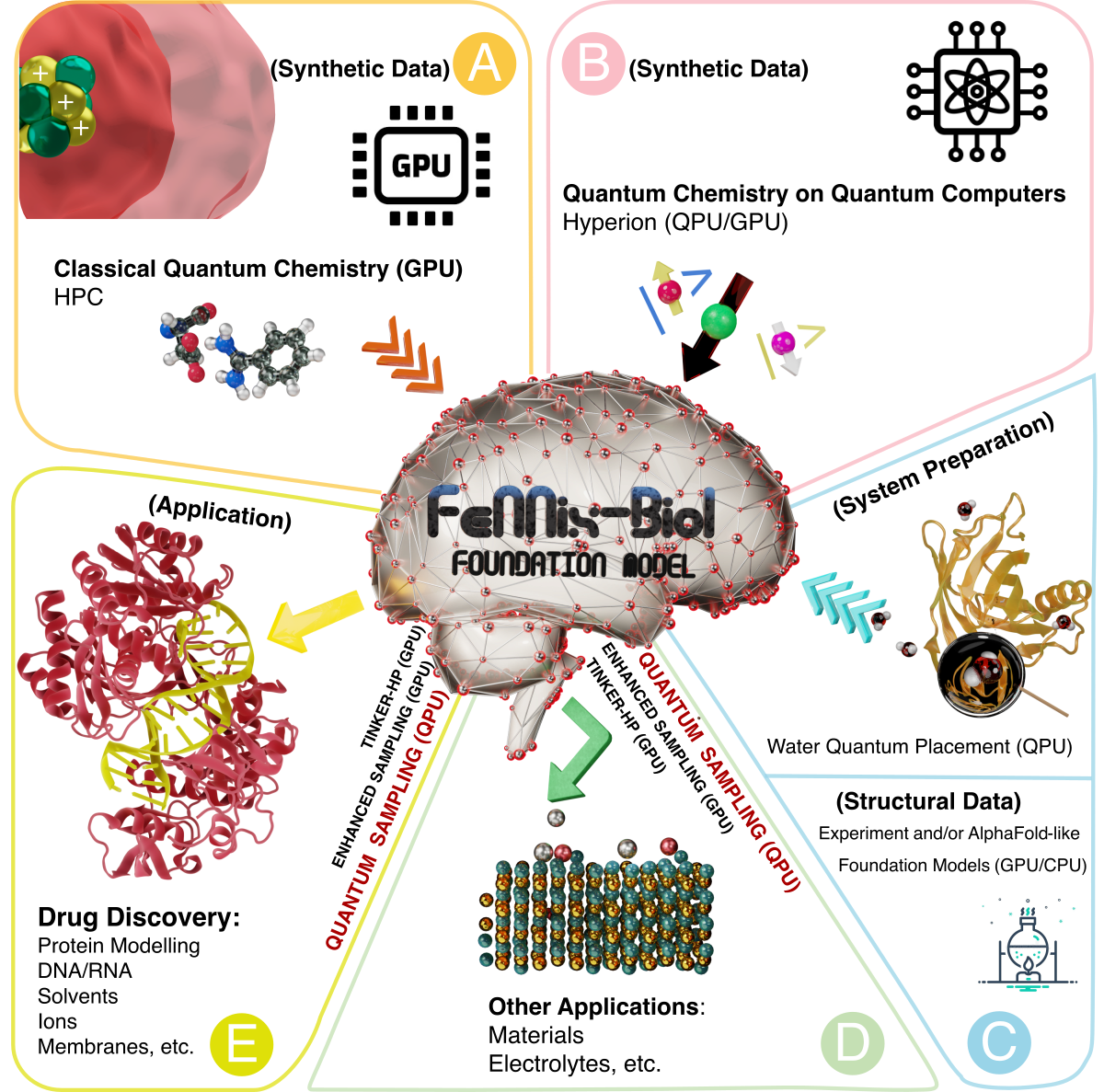}
\caption{ ML/HPC/QC convergence leveraging all type of computing plateforms: from classical Graphics Card Units (GPUs) and Central Processing Units (CPUs) to Quantum Processing Units (QPUs) for simulations in drug design and beyond, from synthetic data (A/B) to system preparation (C) and applications (D/E).}
    \label{fig:summary}
\end{figure}

In this work, we present an integrated drug-discovery pipeline that leverages the synergy between Machine Learning, HPC, and Quantum Computing (Figure \ref{fig:summary}). We demonstrate that the benefits of quantum integration are immediate; rather than waiting for fault-tolerant systems, we show that current NISQ-era hardware is already capable of driving meaningful scientific progress. We also present Hyperion\cite{Hyperion}, a quantum emulator designed to bridge the gap between the NISQ era and FTQC. By leveraging exact and quasi-exact emulation of noiseless logical qubits, Hyperion enables rigorous algorithm development in a controlled environment, enhancing current workflows while preparing for future fault-tolerant hardware.

This here paper is organized to follow a streamlined, end-to-end drug design workflow utilizing both classical and quantum computing platforms. 
We begin by discussing the preparation of molecular systems, including the water placement problem in quantum computing. 
We then analyze quantum algorithms for quantum chemistry as providers of high-accuracy training data for quantum-based Neural Network Potentials. A focus on the Hyperion quantum emulator is provided alongside the discussion on quantum hardware.
We then focus on the development of these specific Machine Learning models for Molecular Dynamics simulations, introducing the FeNNix-Bio1 foundation model and its underlying Ignis database, and showing the transition from classical to exact quantum algorithms for training data generation.

Finally, we provide an overview of modern enhanced sampling techniques and their projected quantum speedups.
Throughout each section, we contrast classical and quantum implementations while providing detailed performance and resource estimations.

\section{Quantum Algorithm for the Preparation of Systems in Drug Discovery Simulations}
\subsection{A Short Overview of Quantum Computing Architectures}
The fundamental divide in quantum hardware lies between the NISQ (Noisy Intermediate-Scale Quantum) era \cite{preskill2018quantum} and the future of FTQC (Fault-Tolerant Quantum Computing \cite{campbell2017roads,reviewhard1} a transition defined by how systems manage operational errors. Current NISQ devices rely on \textquote{physical} qubits that are highly susceptible to environmental noise and decoherence. In contrast, FTQC uses many physical qubits to encode a single \textquote{logical} qubit. This redundancy enables the system to detect and correct errors in real time using Quantum Error Correction (QEC) codes~\cite{google2025quantum}, and unlocks the potential for reliable, error-free computations in deep quantum circuits. Multiple hardware technologies are being pursued toward the common goal of fault tolerance, each with its own trade-offs. \textbf{Superconducting qubits}~\cite{kjaergaard2020superconducting} (e.g.\ Transmons~\cite{kim2023evidence}) offer fast gate speeds and well-developed fabrication techniques, but require cooling down to $\sim$10,mK and have limited coherence times. \textbf{Cat qubits} store information in specially designed quantum states that naturally suppress bit-flip error\cite{reglade2024quantum}s, a promising route toward hardware-efficient fault tolerance. \textbf{Trapped ions}~\cite{bruzewicz2019trapped,moses2023race} achieve very high gate fidelities and long coherence times, although gate operations are slower and scaling to large systems remains challenging. \textbf{Neutral atoms}~\cite{Henriet2020quantumcomputing,reviewhard1} allow large numbers of qubits using arrays of laser traps that can be rearranged dynamically. \textbf{Photonic systems}~\cite{bartolucci2023fusion,krasnok2026emerging} operate at room temperature and connect naturally to quantum communication networks, though reliable two-qubit gates and photon losses remain important challenges. Finally, \textbf{spin qubits} in silicon offer long coherence times\cite{burkard2023semiconductor,schmidt202513} and can be fabricated using techniques similar to those of conventional semiconductor technology, making them a promising candidate for large-scale integration. Although FTQC is often seen as the field’s ultimate goal, it is not a strict prerequisite for progress. Many quantum algorithms can already run on today’s NISQ hardware and provide useful capabilities in the near term. In the following, we discuss how these systems may progressively integrate into drug discovery workflows as both hardware and algorithms continue to improve.
\subsection{Preparation of Systems for Drug Discovery Simulations in the NISQ Era}\label{sec:prep}
Preparing a biological system for molecular dynamics (MD)\cite{Prep-Brooks} or docking\cite{corbeil2012variability} is a meticulous process that transforms a static structural snapshot -- whether experimentally derived via X-ray Crystallography, NMR, or Cryo-EM from the Protein Data Bank\cite{wwpdb2019protein}, or predicted through machine learning foundation models like AlphaFold -- into a dynamic, chemically realistic model. It begins with structure cleaning, where researchers resolve missing loops, add hydrogen atoms, usually absent in X-ray crystallography, and assign appropriate protonation states to amino acids based on the local pH. The system is then solvated in a box of explicit water molecules and neutralized by adding counter-ions (like $Na^+$ or $Cl^-$) to mimic physiological ionic strength. Once the environment is set, the system undergoes energy minimization to clear any steric clashes or "bad contacts" inherited from the initial setup. Finally, the system is subjected to equilibration phases, where temperature and pressure are gradually stabilized under the chosen force field parameters, ensuring the system reaches a steady state before the actual production run begins. Among the myriad steps in computer-aided drug design (CADD)\cite{CADD_RV_1,CADD_RV_2,CADD_RV_3}, the \textbf{"water problem"} (or \textbf{hydration-site prediction problem}), i.e. the challenge of accurately and efficiently placing water molecules around proteins, remains a formidable hurdle, as water serves as a dynamic participant rather than a mere background solvent. Because water molecules frequently act as structural glue or interfacial mediators between a protein and a ligand, their behavior dictates the thermodynamic success of a drug candidate. Both the initial conditions needed to perform MD simulations and the biomolecule's structure used in docking are therefore impacted by the presence of water molecules, which is hard to detect in buried cavities.

\subsubsection{Revisiting the Water Placement Problem using Quantum Computing}

Accurately capturing the configurational sampling of water molecules is computationally expensive due to their fluctuating networks of hydrogen bonds and the dual nature of water as bulk fluid and site-stabilized entity~\cite{blazhynska2025water}. While classical molecular dynamics and Monte Carlo simulations provide the groundwork for exploring this vast conformational space, high-precision free-energy estimations require GPU-accelerated enhanced sampling~\cite{ansari2025targeting,ansari2025lambda}. CADD pipelines demand novel methodologies to reduce this computational overhead without sacrificing accuracy. Although machine learning approaches promise to streamline such task~\cite{bigi2025flashmdlongstrideuniversalprediction,plainer2025consistentsamplingsimulationmolecular}, they not always surpass the accuracy of high-performance physics-based simulations~\cite{annurevnoe,QAB_1}.

Some of the authors have recently proposed a new method based on \textbf{quantum optimisation} to address the problem with current NISQ hardware~\cite{QP}. To leverage a \textbf{neutral atom analog device}, the hydration-site prediction problem is formulated as a Gaussian Mixture Model (GMM) optimisation and mapped onto the Rydberg Hamiltonian. A classically pre-computed, continuous 3D Reference Interaction Site Model (RISM) density \cite{R2_3DRISM,R2_2_3DRISM2} is used as target density to mimic, by minimizing the $L^2$ norm difference between such continuous density and the discrete GMM. The optimisation problem is defined by a quadratic objective function $C(\mathbf{x})$, cast into a standard QUBO (Quadratic Unconstrained Binary Optimisation) equation:

\begin{equation}
    \min_{\mathbf{x} \in \{0, 1\}^N} C(\mathbf{x}) = \sum_{i=1}^N Q_{ii} x_i + 2\sum_{i<j}^N Q_{ij} x_i x_j.
    \label{eq:cost_function}
\end{equation}

The symmetric QUBO matrix $\mathbf{Q}$ maps the Gaussian Mixture Model (GMM) for localized water positions to a discrete optimisation problem. The $Q_{ii}$ term represents the energetic cost or benefit of placing a single water molecule based on its protein environment, while $Q_{ij}$ encodes pairwise interaction energy between sites to constrain the water network's configuration. 
We proposed a quantum adiabatic evolution based and a variational algorithm to find the minimum of this cost function \cite{QP} using \textbf{Pasqal's Rydberg atoms analog technology} \cite{Henriet2020quantumcomputing}.
This choice allowed us to take advantage of the inherent Rydberg blockade mechanism, to enforce physical distance constraints between water molecules, preventing un-physical overlapping.

\begin{figure}[h!]
    \centering
\includegraphics[width=0.65\linewidth]{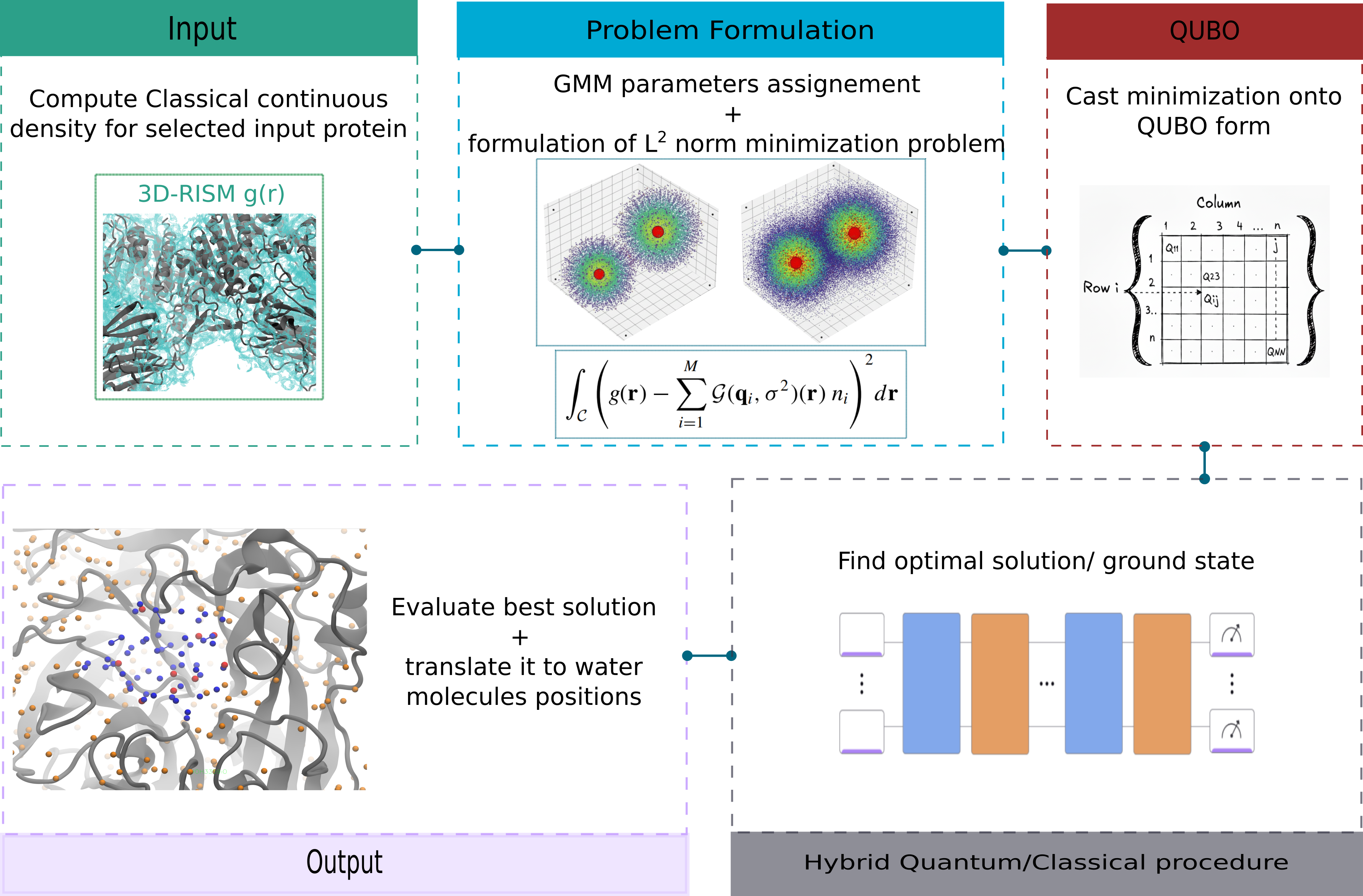}
\caption{Detailed, step-by-step workflow followed to implement our protein's hydration pocket hybrid NISQ/classical approach; each step is briefly described in a simple flow diagram, whose components are grouped and attached to one of the five main phases in which we schematically partition the procedure: Input, Problem Formulation, QUBO, Hybrid NISQ/Classical procedure and Output. Pictures and formulas are used to illustrate the most relevant steps. The notation used for the formulas can be found in the original paper presenting the method~\cite{QP}.}
    \label{fig:opt_scheme}
\end{figure}
In a recent work \cite{Qp2}, we have extended the methodology to the \textbf{digital, gate-based quantum computing paradigm}. We integrated all required steps to map the hydration-site prediction problem as a QUBO into a numerical pipeline to take advantage of advanced quantum optimisation tools~\cite{FireOpal1,FireOpal2}, demonstrating the performances of our method by running hybrid quantum-classical optimisation up to 123 qubits on \textbf{IBM's 156-qubit Heron processors (Pittsburgh and Kingston}). We tested our digital implementation on a set of FDA-approved drug targets, to reproduce experimental crystal water positions in the binding sites. The overall pipeline is illustrated on Figure \ref{fig:opt_scheme}. 
\subsubsection{Performances of the Quantum Water Placement using NISQ Hardware}
\begin{figure}[h!]
    \centering
\includegraphics[width=0.5\textwidth]{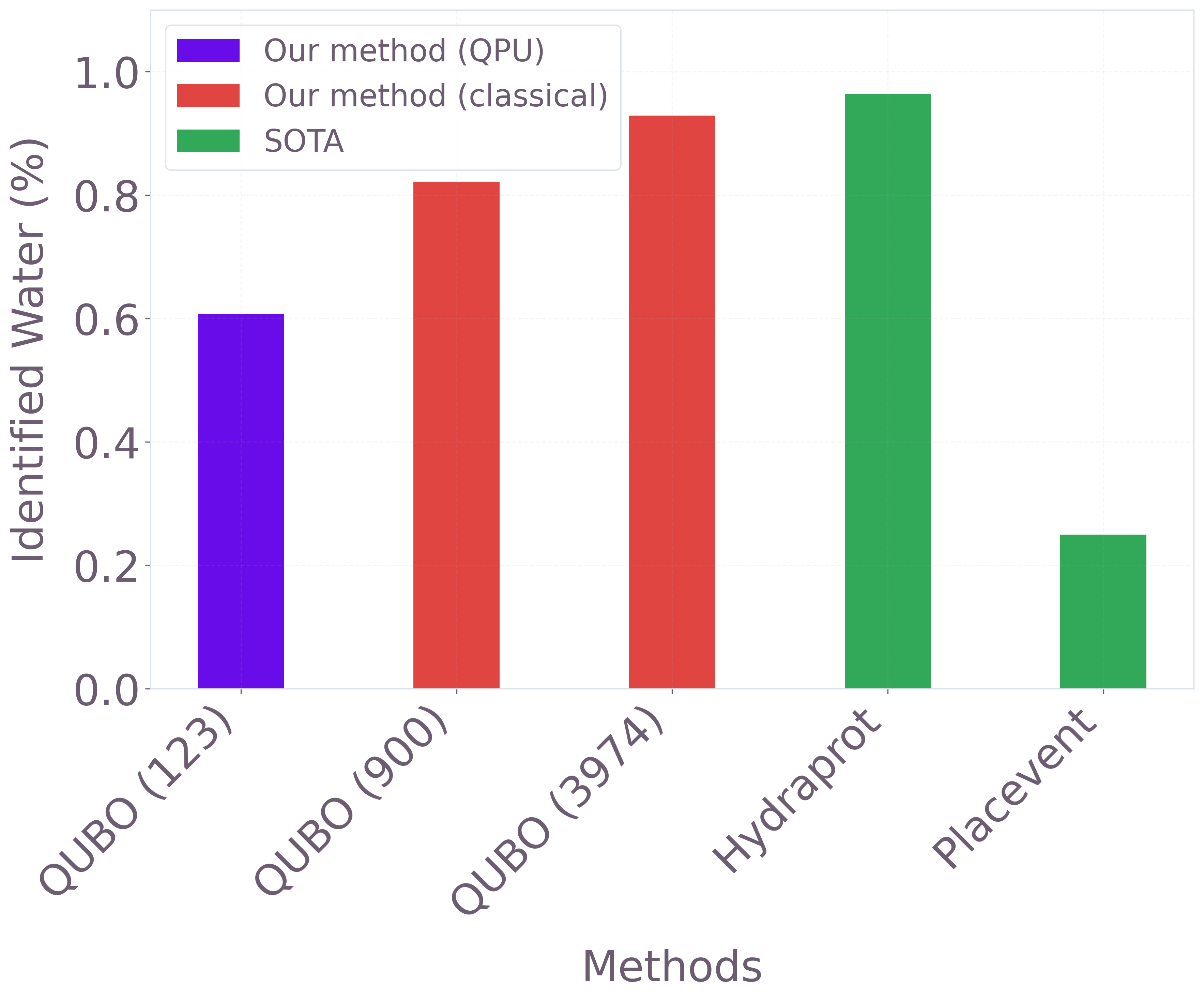}
\caption{Example of accuracy comparison between: our method performing a quantum optimisation on IBM Heron R3 Pittsburgh (violet) on a simplified protein instance encompassing 123 binary variables (qubits), to meat the current hardware limitations; our method performing the optimisation classically (red), to scale-up (900 and 3974 binary variables, respectively) to reach the required complexity in the Drug Discovery application; two of the existing alternative solutions (green), namely Hydraprot~\cite{hydraprot} and Placevent~\cite{Placevent}); the bar plot represents the percentage of crystal water molecules identified in the binding pocket of the 3be7 drug-protein complex structure (PDB). The results demonstrate how, increasing the QUBO complexity to reach the order of 1000 variables, our approach perform comparably to the neural-network Hydraprot.}
    \label{fig:utility}
\end{figure}

\textbf{As first outcome of our study}, we validated our method’s accuracy at scales relevant to drug discovery, identifying crystal water molecules using simplified $\sim$100-variable (qubit) QUBO instances on the currently available Heron QPU, see Fig.~\ref{fig:utility}. While our approach on the Heron yielded satisfactory results—outperforming state-of-the-art classical methods like Placevent, it was itself surpassed by Hydraprot. This discrepancy is primarily attributed to current qubit limitations. To explore full-scale complexity, we conducted classical simulations mimicking future quantum performance with 900 and 3,974 qubits (Fig. \ref{fig:utility}). These simulations correctly identified 80–90\% of crystallographic water molecules using up to $\sim$4000 variables, matching or exceeding the precision of current top-tier tools. Consequently, we demonstrate that \textbf{accuracy scales systematically with qubit count}, suggesting \textbf{a clear path toward quantum utility}.

\begin{figure}[h!]
    \centering
\includegraphics[width=0.4\textwidth]{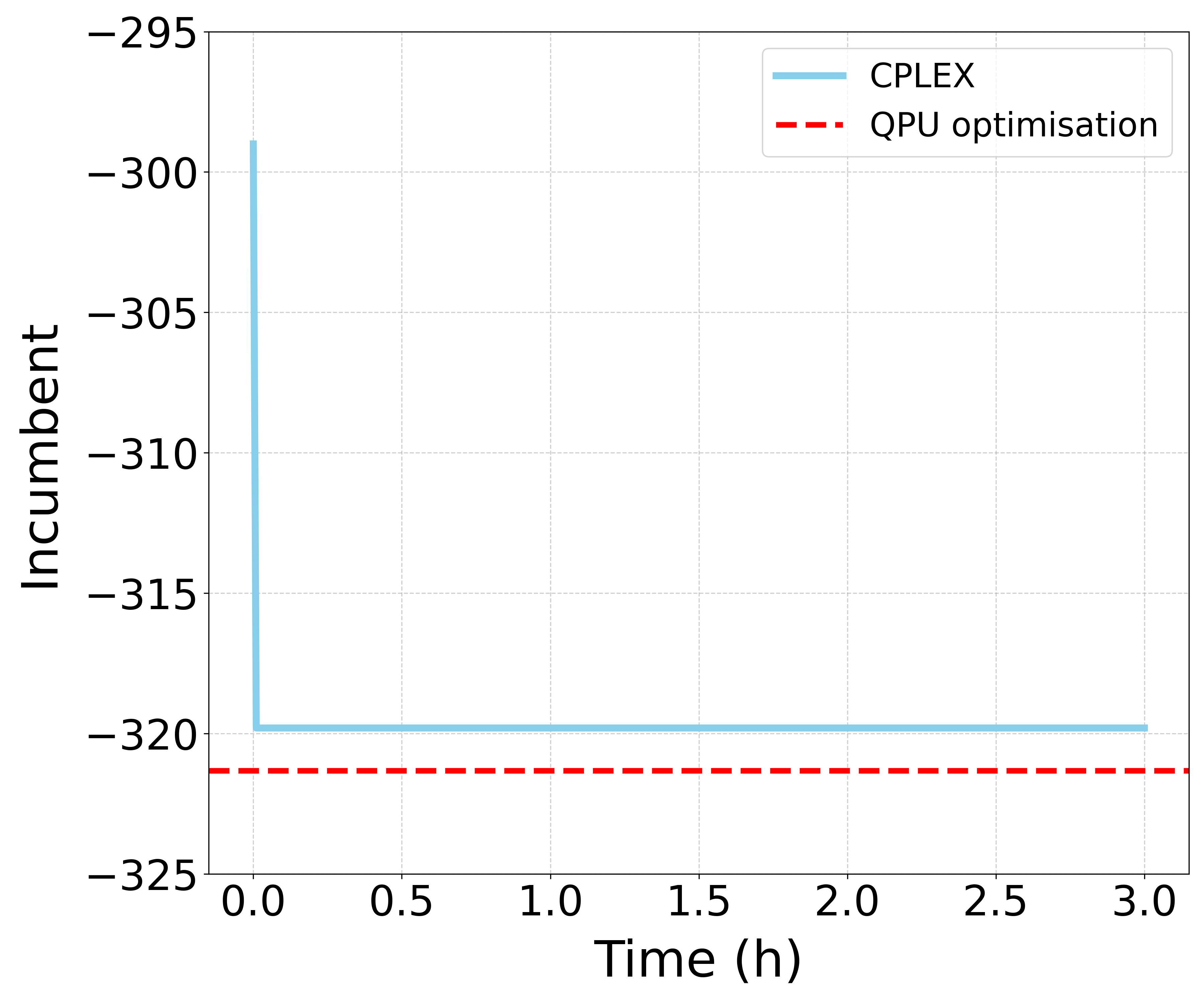}
\includegraphics[width=0.4\textwidth]{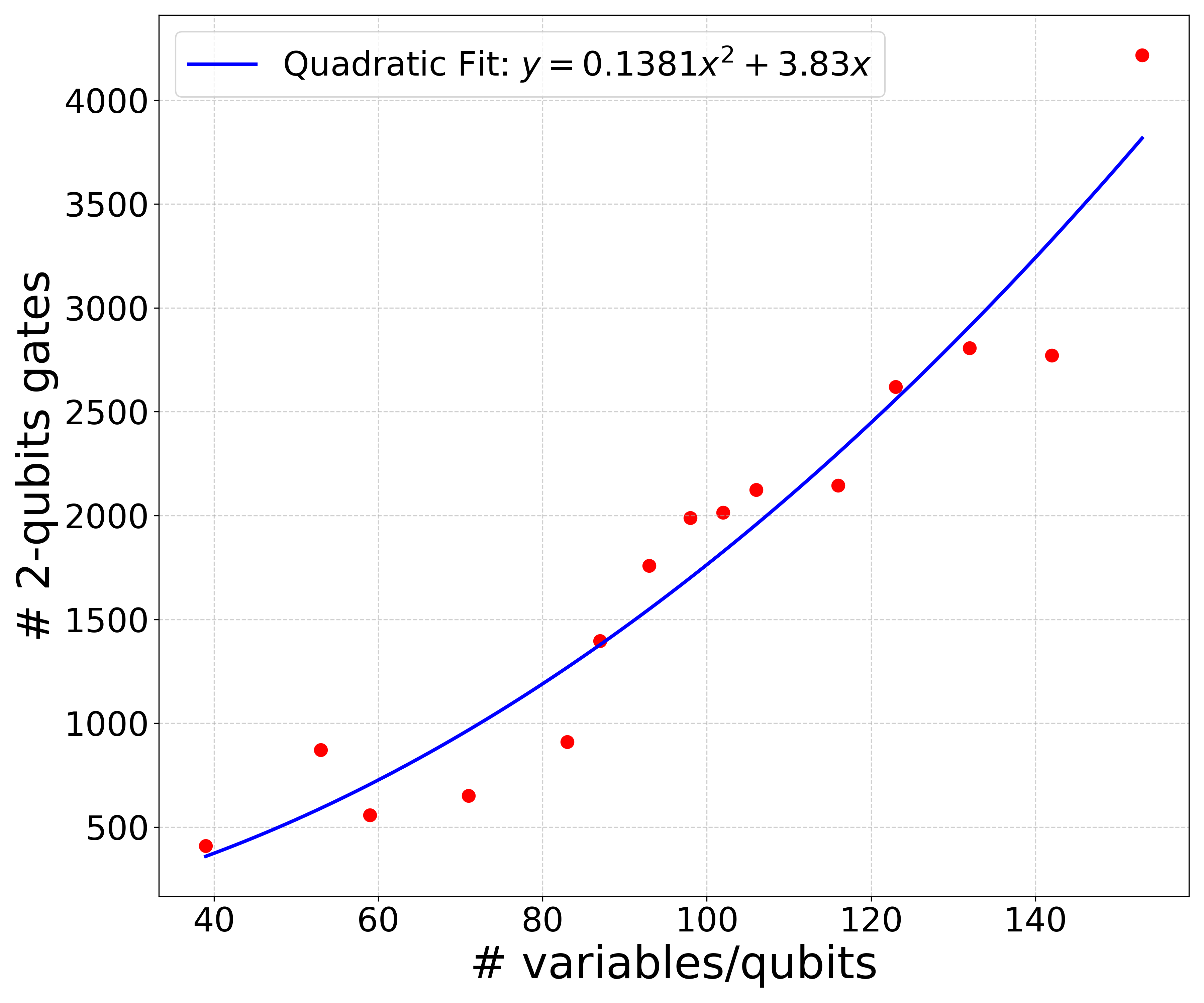}
\caption{Left: Exact classical (CPLEX) vs QPU optimisation results on a 123 variables/qubit hydration site prediction problem formulated as a QUBO. Top panel shows the incumbent solution (light-blue line) over time compared to the best solution identified using the QPU optimisation on IBM R3 Pittsburgh (dashed red line). Bottom panel shows the optimality gap over time, during the 3h optimisation run.~\cite{Qp2}. Right: scaling and resources estimation performed on a protein complex, varying QUBO complexity; the number of 2-qubits gates is plot against the number of QUBO variables/qubits.~\cite{Qp2}}
    \label{fig:scaling}
\end{figure}

\textbf{As second outcome}, we achieved a relevant milestone towards quantum utility, by outperforming in one problem instance the exact classical solver CPLEX. As shown in Fig.~\ref{fig:scaling} (left panel), the solution obtained on the Heron R3 (Pittsburgh) QPU reached a more favorable cost for a 123-variable (qubit) instance. This shows the potential of quantum optimization to outperform exact classical solvers as problem complexity increases beyond this threshold.\\
\textbf{As third outcome}, \textbf{we estimated the quantum resources needed to reach the utility threshold for drug discovery}, identified at around 1000 variables. Extrapolating from benchmarks up to 156 qubits (Fig.~\ref{fig:scaling}, right panel), we found that roughly 100,000 gates will be required for our 900-variable instance. Based on IBM’s roadmap and projected improvements in fidelity and gate reduction, we predict utility-scale implementation is achievable by 2028.

\subsubsection{Perspectives: QUBO, Quantum optimization and Drug Design Pipeline}
Accurately localizing water molecules within binding pockets is a prerequisite for high-fidelity Molecular Dynamics and can significantly enhance binding mode predictions in molecular docking \cite{Dock_rev_Nat}. By adapting the initial input density to encode small-molecule interactions with target proteins, our method could be extended into a quantum-native docking algorithm. While current literature explores both quantum optimization \cite{PhysRevApplied.21.034036} and quantum-inspired approaches \cite{QID_2024}, our results reinforce the transformative potential of quantum computing (QC) in drug discovery. As demonstrated by our algorithmic benchmarks, quantum optimization is poised to disrupt existing pipelines \cite{QC_DD_1,QC_DD_2,QC_DD_3}, initially augmenting classical hardware before systematically replacing traditional algorithms as quantum advantage is realized.

\section{Quantum Algorithm for Quantum Chemistry: Overcoming the Exponential Wall to generate Quantum-accurate Training Data}\label{sec:electronic}

\paragraph{The case for quantum chemistry data} In typical computer-assisted Drug Design pipelines, sampling and characterizing dynamic behavior of biological systems of interest is usually performed through Molecular Dynamics (MD), with traditional MD simulations relying on classical force fields. Such models describe interatomic interactions using simple, fixed functional forms, which are computationally efficient but unable to capture the complexity of electronic effects. 
A promising alternative is to train neural network potentials on accurate quantum chemistry data, allowing the model to learn the underlying potential energy surface (PES) of a molecular system.
Once trained, such models can reproduce near-quantum accuracy while retaining the efficiency needed for large-scale simulations.

Achieving this requires highly accurate reference calculations : in quantum chemistry, the standard benchmark is chemical accuracy, corresponding to an energy error of about 1 kcal/mol relative to the exact solution of the electronic Schrödinger equation. The latter corresponds to solving the many-electron problem 

\begin{equation}
    \hat{H} \ket{\Psi(\textbf{x}_1, \textbf{x}_2, ..., \textbf{x}_{N_{\text{elec}}})} = E \ket{\Psi(\textbf{x}_1, \textbf{x}_2, ..., \textbf{x}_{N_{\text{elec}}})}
\end{equation}
where $\Psi(\textbf{x}_1, \textbf{x}_2, ..., \textbf{x}_{N_{\text{elec}}})$ ($\Psi$ in the following) is the $N_{\text{elec}}$-electrons wave function with $\textbf{x} = \{\textbf{r}, \sigma\}$, $\hat{T}$ is the kinetic energy operator, $\hat{V}_{\text{ne}}$ is the electron-nucleus interaction, $\hat{W}_{\text{ee}}$ is the electron-electron interaction and $E$ is the energy of the state $\Psi$.
In order to solve the electronic structure problem, one needs to face several challenges: (i) chemical accuracy requires a complete treatment of electron correlation (at the Full Configuration Interaction, or Full-CI, level) and (ii) also requires an infinite basis set to represent the electronic wavefunction, known as the Complete Basis Set (CBS) limit, see 6.1.4 and ref.~\cite{traore2024shortcut} for more details. To understand the challenges of reaching this level, we must examine the computational bottlenecks inherent in modeling the many-body problem.

\subsection{Classical Computing State of the Art in Quantum Chemistry}
\label{Classical}
\subsubsection{Quantum Chemistry and the Quest for Accuracy}
\begin{figure}[h!]
    \adjustimage{width=0.55\textwidth,center}{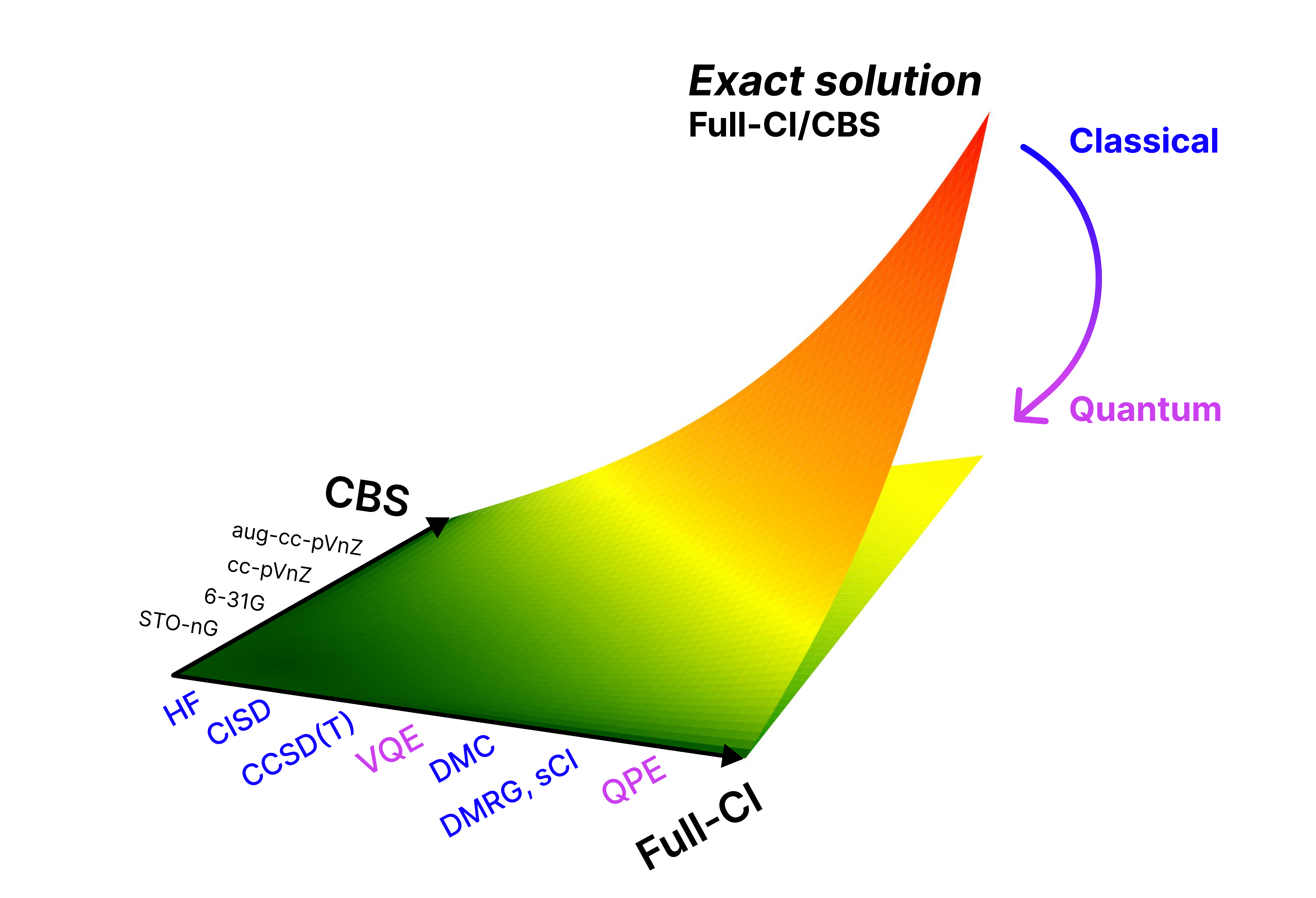}
    \caption{Pople-style diagram illustrating the exponential wall encountered in classical electronic-structure methods, and the potential mitigation offered by quantum computing approaches. The surface color and elevation qualitatively represent the computational complexity.}
    \label{fig:pople}
\end{figure}

In quantum chemistry, most computational methods can be viewed as successive approximations to the Full-CI/CBS limit, trading reduced treatments of electron correlation and finite basis sets for computational savings. This trade-off is commonly illustrated by the so-called Pople diagram, which represents the computational complexity as a function of these two approximation axes (see Fig.~\ref{fig:pople}). As one moves toward more accurate descriptions, the computational complexity increases rapidly, leading to what is commonly referred to as the exponential wall. The simultaneous need for increasingly accurate treatments of electron correlation and larger basis sets thus creates a severe computational bottleneck: while chemical accuracy can be reached for small gas-phase molecules, extending such calculations to large biomolecular systems or condensed-phase materials remains extremely challenging.
At the lowest level, Hartree–Fock (HF) theory treats electrons within a mean-field approximation, neglecting their instantaneous correlated motion. Density Functional Theory (DFT)~\cite{koch2015chemist} partially accounts for correlation effects through approximate functionals of the electron density. Achieving higher accuracy requires post–Hartree–Fock approaches that explicitly treat electron correlation. Methods such as Coupled Cluster—particularly CCSD(T)~\cite{ccsdt}, often considered the “gold standard”—provide high accuracy but exhibit steep computational scaling (typically \(O(N^7)\)) and can fail for strongly correlated, multireference systems. Alternative high-accuracy approaches include the Density Matrix Renormalization Group (DMRG)~\cite{White1992,DMRG,Schollwck2011}, which efficiently captures strong correlations within carefully chosen active spaces, and Diffusion Monte Carlo (DMC)\cite{DMC}, a stochastic method that naturally avoids basis-set incompleteness errors by working in real-space.
Selected Configuration Interaction (sCI) methods \cite{huron1973iterative,garniron2019quantum} construct systematically improvable multideterminant wavefunctions by selecting the most important determinants from the Full-CI expansion. While this approach provides a controllable path toward the exact solution, it also clearly exposes the exponential wall: achieving progressively smaller improvements in energy requires an exponentially growing number of determinants as the Full-CI limit is approached. Additional details about the methods and the basis sets can be found in the technical Appendix \ref{sec:appendix} at the end of the paper.

\subsubsection{Classically Intractable Systems in Quantum Chemistry}
As we have seen, the boundary of classical intractability in quantum chemistry is defined by the exponential growth of the configuration space required to describe "strong correlation", a threshold where classical approximations of electron behavior fundamentally collapse. In "weakly correlated" systems, such as most organic molecules at their equilibrium geometry, electrons move relatively independently within a mean field, allowing DFT, the workhorse of the field, to model thousands of atoms by mapping the complex many-body problem onto a simpler electron density. While DFT and high-accuracy methods like CCSD(T) function well for systems up to roughly 20–50 atoms before CCSD(T)'s $O(N^7)$ scaling or functional inaccuracies become a bottleneck, they evaporate in "strongly correlated" scenarios, such as transition metal clusters, lanthanides, or molecules during bond dissociation. In these cases, multiple electronic states are nearly equal in energy, and electrons become so inextricably linked that the single-determinant foundation of DFT fails to capture the physics. As explained, advanced classical techniques like DMRG and sCI attempt to truncate the Hilbert space by following only the most "entangled" paths, while DMC uses stochastic sampling to avoid explicit wavefunction storage. Yet, these methods hit a final wall: DMRG is limited by the "area law" of entanglement in 3D systems, SCI suffers from a combinatorial explosion of necessary determinants, and DMC is plagued by the Fermion Sign Problem, which leads to exponential noise. This definitive breakdown typically occurs when the "active space" exceeds approximately 50 electrons in 50 orbitals, at which point the possible configurations surpass the memory and stochastic capacity of any classical supercomputer, shifting the problem from a matter of hardware speed to a fundamental mathematical impossibility that necessitates quantum computing simulation.
\subsection{Quantum Computing: Overview of Recent Quantum Algorithms for Electronic Structure}
\label{Overview}
Quantum computers offer a promising way forward: the apparent exponential wall can be built with only a linear number of qubits, allowing molecular wavefunctions in the Full-CI space to be encoded efficiently (see Figure \ref{fig:pople}). Quantum algorithm development for the ground-state quantum chemistry problem consists in converting this representation capability into computational speedup over the classical methods introduced in the former section. 

For the ground-state quantum chemistry problem, the field splits into two broad classes, revealing a clear tradeoff between theoretical guarantees and near-term implementability: (i) NISQ-oriented algorithms such as the Variational Quantum Eigensolver (VQE)\cite{peruzzo2014variational}, and related hybrid quantum-classical schemes (QSCI, QSD) that pair ansatz preparation and measurement on the QPU with classical CPU/GPU processing. (ii) Algorithms such as the Quantum Phase Estimation (QPE)~\cite{lee2021even}, on the other hand, assume access to large and fault-tolerant quantum computers, and extract the ground-state energy with deterministic success, at a computational cost that depends on the system size, the desired accuracy, and the quality of the initial state.
Below, we review contributions in both paradigms, aiming to reduce resource demands and bring quantum advantage closer to practice.

\subsubsection{Near-Term Hybrid Quantum Algorithms: the Search for Optimal Ansätze}
The VQE approximates the molecular ground state by minimizing the energy over a truncated subspace of the Full-CI Hilbert space, in close analogy with classical CI methods. A key difference is that VQE constructs this variational space using an ansatz composed of unitary, exponentially parameterized operators, whereas CI relies on linear expansions in excitation operators. In the VQE workflow, a classical optimizer iteratively updates the parameters to minimize the Hamiltonian expectation value, seeking convergence to the lowest-energy state within the variational subspace defined by the ansatz, ideally approximating the true ground state of the system. Quantum advantage with VQE would then mean that, for some practically relevant problems, the variational space spanned by the quantum circuit provides a better heuristic than classical ansätze~\cite{leimkuhler2025exponential}, assuming efficient parameter optimisation. Then, the central challenge in VQE is ansatz design: it must balance expressivity (large variational space), trainability (enabling efficient parameter optimization), and circuit depth, as NISQ devices allow only limited numbers of qubits and gates due to decoherence. So far, practical implementations continue to face numerous challenges, several of which are outlined below.
First, commonly used ansätze can be insufficiently expressive for strongly correlated systems, particularly when higher-order excitations such as triples are required to achieve chemical accuracy. Second, the operator pools from which ansätze are constructed are often far larger than algebraically necessary, resulting in substantial computational overhead. Last but not least, VQE relies on a global, non-convex optimization landscape that is often rough and largely flat, due both to barren plateaus arising from high dimensionality and to excited-state contamination intrinsic to energy minimization, leading to significant convergence difficulties in multi-parameter optimization.

To address these limitations, we developed a set of complementary strategies, all developed/tested at the emulation level using our Hyperion quantum emulator (see 3.3), that directly target each of the aforementioned challenges. 
\paragraph{Fixed ansatz approaches}First, we consider \textit{fixed-ansatz} approaches, which employ a \underline{predefined circuit structure} encompassing tunable single-qubit rotation gates (the rotation angles being the parameters to optimise). A conventional example is the tUCCSD ansatz~\cite{UCCSD}, that takes the form \[U_{\mathrm{tUCCSD}}(\boldsymbol{\theta}) 
    = \prod_{\mu \in \mathcal{S}} e^{\theta_\mu (\tau_\mu - \tau_\mu^\dagger)}
      \prod_{\nu \in \mathcal{D}} e^{\theta_\nu (\tau_\nu - \tau_\nu^\dagger)},\] 
where $\tau_\mu$ and $\tau_\nu$ are single and double excitation cluster operators, and $\tau - \tau^\dagger$ ensures unitarity.

To improve expressivity in strongly correlated regimes, we extended this standard tUCCSD by incorporating triple excitations, leading to a \textbf{tUCCSDT} formulation~\cite{UCCSDT} capable of capturing higher-order correlation effects required for chemical accuracy. Second, to leverage hardware-efficient operators, i.e., operators that admit shallower circuit decompositions than the fermionic excitations in tUCC ansätze, the \textbf{non-iterative disentangled UCC} (NI-DUCC) ansatz was introduced~\cite{Haidar2025}, built from a highly expressive network ensuring algebraic completeness. Third, an efficient and systematic method was proposed to generate \textit{complete operator pools}, sets of operators associated with a given Lie algebra from which any unitary can be constructed~\cite{fastmcp}, a formerly technically demanding task as the number of qubit grows. This advance enables the practical deployment of \textbf{minimal complete pools} (MCPs) within NI-DUCC, referred to as NI-DUCC(MCPs). 

\begin{table*}[t]
\centering
\caption{Comparison of representative Variational Quantum Eigensolver (VQE) methods for quantum chemistry. 
Criteria: (1) Operator Pool, (2) Ansatz Structure, (3) Cost Function, (4) Operator Selection Criterion, (5) Optimisation, (6) Near-term Hardware Compatibility, (7) Target Chemical Regime, (8) Comments.}
\label{tab:compact_variational_methods}
\renewcommand{\arraystretch}{1.08}
\setlength{\tabcolsep}{5pt}
\small
\begin{tabularx}{\textwidth}{
>{\RaggedRight\arraybackslash}p{3.2cm}
>{\RaggedRight\arraybackslash}X
>{\RaggedRight\arraybackslash}X
}
\toprule
\textbf{Method} &  &  \\
\midrule

\textbf{tUCCSDT}\cite{UCCSDT}
&
\textbf{1.} Fermionic singles/doubles, $\mathcal{O}(N^4)$. \newline
\textbf{3.} Energy, $\mathcal{O}(N^4)$ Pauli strings. \newline
\textbf{5.} Global, non-convex. \newline
\textbf{7.} Small systems, moderate correlation.
&
\textbf{2.} Fixed. \newline
\textbf{4.} Pre-determined. \newline
\textbf{6.} Weak. \newline
\textbf{8.} Chemically inspired, lacks flexibility.
\\

\addlinespace
\textbf{ADAPT}\cite{grimsley2019adaptive}
&
\textbf{1.} Any, up to $\mathcal{O}(N^4)$. \newline
\textbf{3.} Energy, $\mathcal{O}(N^4)$ Pauli strings. \newline
\textbf{5.} Global, non-convex. \newline
\textbf{7.} Small systems, moderate/high correlation.
&
\textbf{2.} Adaptive, operator-wise. \newline
\textbf{4.} \textit{On-the-fly} energy gradient. \newline
\textbf{6.} Weak. \newline
\textbf{8.} Barren-plateaus resilience.
\\

\addlinespace
\textbf{GGA}\cite{feniou2025greedy}
&
\textbf{1.} Any. \newline
\textbf{3.} Energy, $\mathcal{O}(N^4)$ Pauli strings. \newline
\textbf{5.} Sequential, deterministic update. \newline
\textbf{7.} Small systems, weak correlation
&
\textbf{2.} Adaptive, operator-wise. \newline
\textbf{4.} \textit{On-the-fly} energy line-search.\newline
\textbf{6.} Moderate/strong. \newline
\textbf{8.} High trainability, low expressivity.
\\

\addlinespace
\textbf{Overlap-ADAPT}\cite{feniou2023overlap}
&
\textbf{1.} Any. \newline
\textbf{3.} Overlap, $\mathcal{O}(1)$ observable. \newline
\textbf{5.} Global, convex. \newline
\textbf{7.} Strong correlation levels.
&
\textbf{2.} Adaptive, operator-wise. \newline
\textbf{4.} \textit{On-the-fly} overlap gradient. \newline
\textbf{6.} Strong. \newline
\textbf{8.} Needs accurate target state.
\\

\addlinespace
\textbf{MB-ADAPT}\cite{fastmcp}
&
\textbf{1.} Minimal complete pool, $\mathcal{O}(N)$. \newline
\textbf{3.} Energy, $\mathcal{O}(N^4)$ Pauli strings. \newline
\textbf{5.} Global, non-convex. \newline
\textbf{7.} Large systems, all correlation levels.
&
\textbf{2.} Adaptive, batch of operators. \newline
\textbf{4.} \textit{On-the-fly} energy gradients. \newline
\textbf{6.} Intermediate. \newline
\textbf{8.} Reduced gradient costs w.r.t ADAPT.
\\

\addlinespace
\textbf{NI-DUCC(MCPs)}\cite{Haidar2025,fastmcp}
&
\textbf{1.} Minimal complete pool, $\mathcal{O}(N)$. \newline
\textbf{3.} Energy, $\mathcal{O}(N^4)$ Pauli strings. \newline
\textbf{5.} Global, non-convex. \newline
\textbf{7.} Large systems, all correlation levels.
&
\textbf{2.} Fixed, layered. \newline
\textbf{4.} Pre-determined. \newline
\textbf{6.} Intermediate. \newline
\textbf{8.} Very expressive ansatz.
\\

\bottomrule
\end{tabularx}
\end{table*}

\paragraph{Adaptive ansatz approaches}
In contrast to fixed-structure ansätze, adaptive approaches such as ADAPT-VQE~\cite{grimsley2019adaptive, tang2021qubit} iteratively construct the circuit by selecting and adding operators from a predefined pool based on a given criterion. This enables the generation of compact, problem-tailored circuits. In this context, MCPs provide a possible choice of operator sets. Building on this idea, we showed that selecting and adding operators in batches~\cite{sapova2022variational} significantly reduces the computational cost associated with gradient evaluations, leading to the \textbf{MB-ADAPT-VQE} framework~\cite{fastmcp}.

\textit{Taken together, NI-DUCC (with MCPs) and MB-ADAPT-VQE yield performance that is arguably superior to prior state-of-the-art across a range of tested molecular systems in relevant regimes.}

To address excited-state contamination and the resulting energy plateaus in ADAPT-VQE, we introduced \textbf{Overlap-ADAPT-VQE}~\cite{feniou2023overlap}, which grows the ansatz via variational projection onto an intermediate-quality reference state rather than direct energy minimization. This projection-based strategy yields significantly more compact circuits and accelerates subsequent convergence. More broadly, Overlap-ADAPT-VQE establishes a unified framework for \underline{learning and compressing quantum states} as ansätze. We exploited this capability for classical data loading by compressing sparse selected-CI wavefunctions into compact quantum representations that serve as high-quality initial states for both VQE and Quantum Phase Estimation~\cite{feniou2023overlap, feniou2024sparse}, two algorithms whose performance is critically dependent on initialization quality. 

\paragraph{VQE on present NISQ hardware: the example of Greedy-ADAPT-VQE}
We conclude by addressing the practical challenges of VQE on current quantum hardware, in particular the optimization difficulties caused by non-convex landscapes. We introduced \textbf{Greedy Gradient-Free Adaptive}-VQE (GGA-VQE)~\cite{feniou2025greedy}, that exploits analytical structure to convert the global variational optimization into an iterative greedy procedure in which each parameter can be determined locally without gradient estimation. This substantially reduces the classical optimization overhead while producing compact, hardware-aligned ansätze tailored to the target Hamiltonian. 

A proof-of-principle implementation was carried out on a \textbf{25-qubit IonQ trapped-ion processor} for transverse-field Ising simulations, constituting, to our knowledge, the first hardware realization of an adaptive VQE algorithm. 

\begin{figure}[h!]
\begin{center}
    \includegraphics[width=0.4\textwidth]{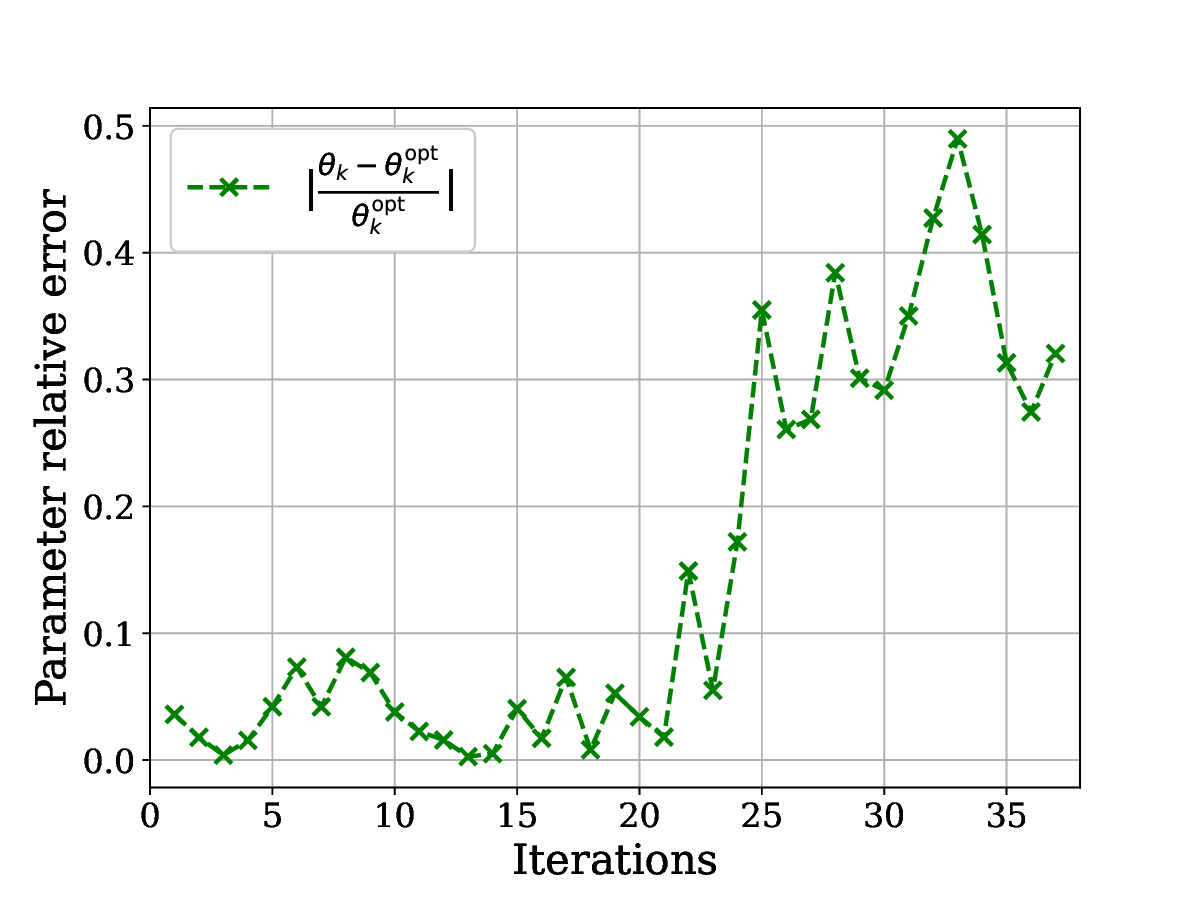}
\end{center}
    \caption{Relative error in optimizing the angle at iteration $k$ of the GGA-VQE algorithm using only the QPU. Data from ref. \cite{feniou2025greedy}.}
    \label{fig:revise_1}
\end{figure}

Despite significant fluctuations in the QPU-measured energies, the parameters extracted from hardware measurements remain close to the optimal noiseless values for the first iterations of the algorithm (Fig.~\ref{fig:revise_1}), indicating that the greedy analytical updates are robust to current processor noise levels. At the same time, recent analyses by Dalton \textit{et al.}~\cite{dalton2024quantifying} show that achieving a genuine quantum advantage with adaptive VQE would require several orders of magnitude reduction in current gate error rates. GGA-VQE can be viewed as a "minimal" adaptive strategy for hardware implementation because it removes the costly global non-convex optimization step, which is especially sensitive to noise. More generally, current hardware demonstrations of adaptive VQE remain confined, at best, to very small problems and minimal-basis descriptions, still far from the accuracy required for integration into our foundation model. In this context, emulation plays a central role in the development and validation of advanced methods, enabling algorithmic progress beyond current hardware limitations. We now turn to the longer-term perspective of fault-tolerant quantum computing, where Quantum Phase Estimation offers a route to near-exact quantum chemistry calculations.

\subsubsection{Scalable Algorithms for Fault-Tolerant Architectures: Towards Quantum Phase Estimation}
\begin{figure}[h!]
  \centering
  \includegraphics[width=0.5\linewidth]{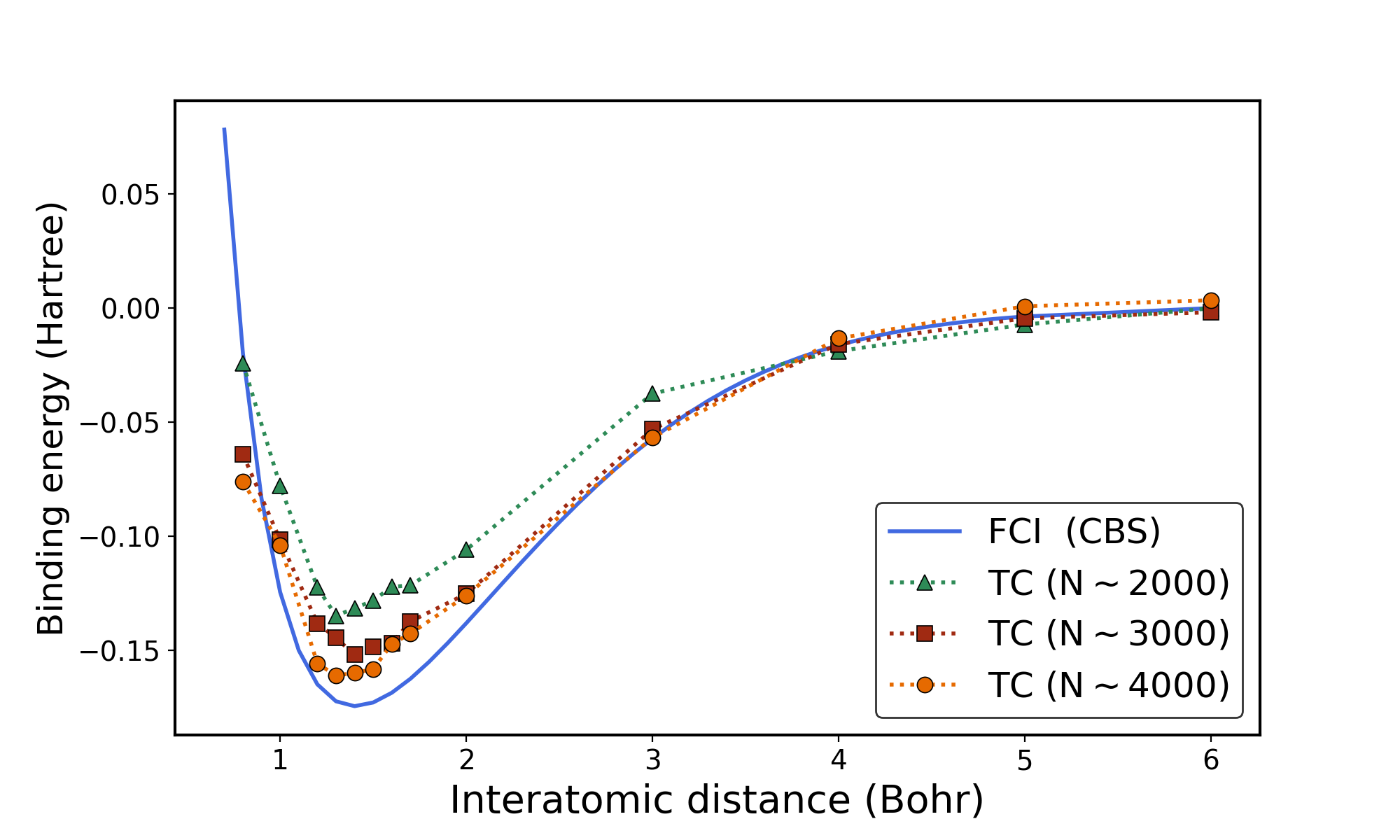}
\caption{Dissociation curve of H$_2$: binding energy versus internuclear distance for our transcorrelated (TC) method on nonuniform grids, compared to FCI at the CBS limit. Grids studied contain 2000, 3000, and 4000 total points (20, 30, and 40 radial points combined with 50 angular quadrature points per atom), corresponding to Hamiltonian matrices of dimensions $4$, $9$, and $16$ millions, respectively. Data from ref. \cite{feniou2025real}.}
  \label{fig:diss}
\end{figure}
\paragraph{Quantum Phase Estimation} At the opposite end of the spectrum, the Quantum Phase Estimation algorithm provides ground-state energies at Full-CI accuracy with controllable error assuming fault-tolerant hardware. Within a fixed basis, it yields the exact Full-CI solution using a circuit whose size scales with system size, target accuracy, and initial-state quality. Current resource estimates for chemically relevant problems indicate requirements on the order of billions of gates and thousands of logical qubits~\cite{lee2021even}, placing QPE well beyond near-term hardware capabilities, despite its favorable asymptotic scaling and the prospect of \textit{at least} polynomial quantum advantage. Ongoing efforts therefore focus on reducing these resource demands, a challenge that is twofold: improving the high-level mapping of chemical problems onto quantum operations, and improving the low-level gate decomposition of the fundamental building blocks of QPE. We identify that leading asymptotic formulations of QPE for chemistry rely on first-quantized plane-wave bases~\cite{babbush2019quantum}, which are natural for periodic systems but poorly suited for strongly correlated or highly localized molecular states. To address this, we introduce the use of adaptive molecular grids as a first-quantized basis for QPE, together with an operator transformation yielding a transcorrelated Hamiltonian whose ground states are smoother in real space~\cite{feniou2025real}. Together, these efforts aim to incorporate more chemical structure and physical insight into fault-tolerant quantum algorithms, helping bridge the gap between formal asymptotic advantage and practical chemical applications.

\begin{table}[h!]
    \centering
    \caption{Comparison between real-space first-quantized and molecular-orbital second-quantized frameworks}
    \label{tab:realspace_mo_comparison}
    \begin{tabularx}{\textwidth}{l >{\raggedright\arraybackslash}X >{\raggedright\arraybackslash}X}
        \toprule
        & \textbf{Real-space basis} & \textbf{Molecular-Orbital basis} \\
        \midrule
        \textbf{Quantization} & First quantization & Second quantization \\
        \addlinespace
        \textbf{Qubit count} & $\mathcal{O}(\eta \log N)$; logarithmic in basis & $\mathcal{O}(N)$; no explicit particle dependence \\
        \addlinespace
        \textbf{Hamiltonian terms} & Typically $\mathcal{O}(N^2)$; diagonal Coulomb term & Typically $\mathcal{O}(N^4)$; \\
        \addlinespace
        \textbf{Quantum algorithms} & QPE, QEVE (FTQC) & VQE (NISQ), QPE (FTQC) \\
        \addlinespace
        \textbf{Spatial resolution tricks} & Adaptive grids, Transcorrelated Hamiltonian & Adaptive basis sets (SABS), orbital rotations \\
        \bottomrule
    \end{tabularx}
\end{table}

A second aspect of this challenge concerns the optimization of low-level gate decompositions for key QPE building blocks. In this direction, we developed low-depth implementations of essential primitives, including a \textbf{multi-controlled NOT gate} with polylogarithmic circuit depth in the number of controls~\cite{Claudon_2024}, representing an exponential improvement over previous constructions. Similarly, we introduced a logarithmic-depth method for \textbf{quantum state preparation of polynomials}~\cite{Claudon26b}.

Although these constructions are primarily designed for fault-tolerant quantum computers, current NISQ devices can already serve as valuable testbeds for validating such algorithmic primitives at small scale. We illustrate this approach below with the case of polynomial state preparation.
\subsubsection{Assessing FTQC Building Blocks on NISQ Hardware}

As an example, we consider a method we introduced for preparing quantum states with amplitudes following a polynomial of degree $d$. The construction achieves logarithmic circuit depth in the number of qubits $n$, improving over previous approaches with linear depth. It relies on three FTQC-compatible components: a block-encoding of an affine diagonal operator relying on a modified linear-combination-of-unitaries (LCU) scheme enabling logarithmic-depth execution, and a generalized quantum eigenvalue transformation (GQET) to promote the affine operator to arbitrary degree polynomials. As polynomial approximations are ubiquitous in scientific computing, this primitive provides a scalable route to quantum state preparation across a wide range of quantum algorithms.

A proof-of-principle implementation~\cite{Claudon26b} was performed on the \textbf{56-qubit Quantinuum H2 trapped-ion processor}~\cite{Liu2025CertifiedRandomness}, using 14 qubits and over 500 primitive gates. The measured amplitudes closely match the expected polynomial profile, confirming the validity of the construction. This experiment demonstrates that NISQ hardware can already serve as a testbed for validating FTQC-oriented primitives at small scale.

\begin{figure}[h!]
\centering
\includegraphics[width=0.5\linewidth]{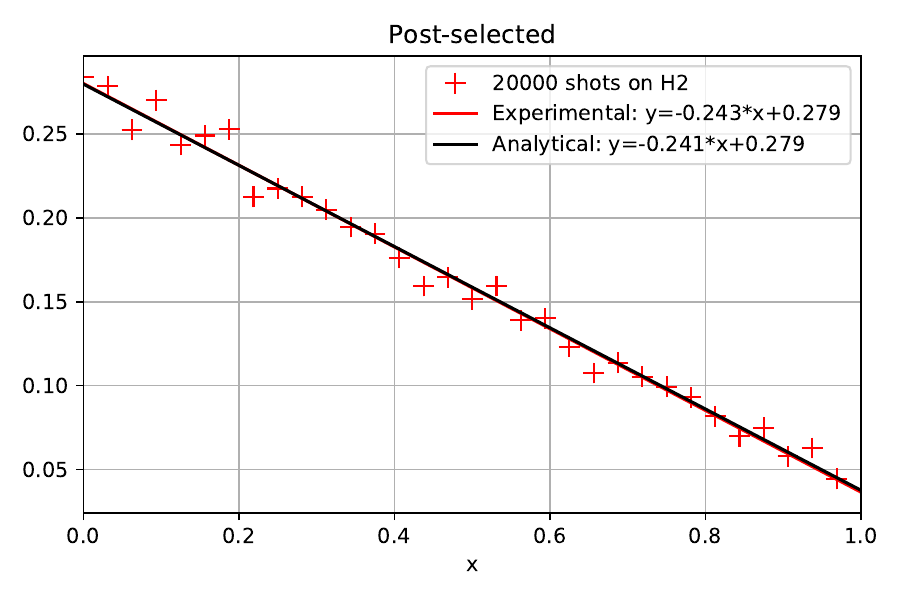}
\caption{Hardware demonstration on the Quantinuum H2 trapped-ion processor of a degree-one polynomial-amplitude state preparation. The plotted values correspond to the square roots of the postselected measurement frequencies of each computational-basis outcome, showing good agreement with the target polynomial profile. Data from ref. \cite{Claudon26b}}
\label{fig:quantinuum_poly}
\end{figure}

These results highlight both the necessity and the opportunity for tight co-design between chemistry-informed algorithm development and efficient quantum primitive operations.

\subsection{The GPU-accelerated Hyperion Quantum Platform: from Emulation to Orchestration}

Exploitation and development of the quantum and hybrid algorithms presented earlier depends on direct access to quantum computing hardware, a resource increasingly critical and strategic. Availability is, however, scarce, and still mainly distributed on NISQ-era machines. This supply gap has driven the demand for high-performance software packages capable of simulating quantum systems using classical computing resources.

These are known as \textbf{quantum emulators} and serve multiple purposes: validating novel algorithms, guiding algorithmic design and refinement, verifying outputs from physical quantum hardware, and providing a controlled environment for rigorous benchmarking. In response to this need, various emulation frameworks have emerged over the past five years \cite{Suzuki2021,McClean2020,pennylane,Qiskit,myqlm2024,Guerreschi2020,10313722}, with underlying strategies typically dictated by specific target applications.
\begin{figure}[h!]
    \centering
    \includegraphics[scale=0.6]{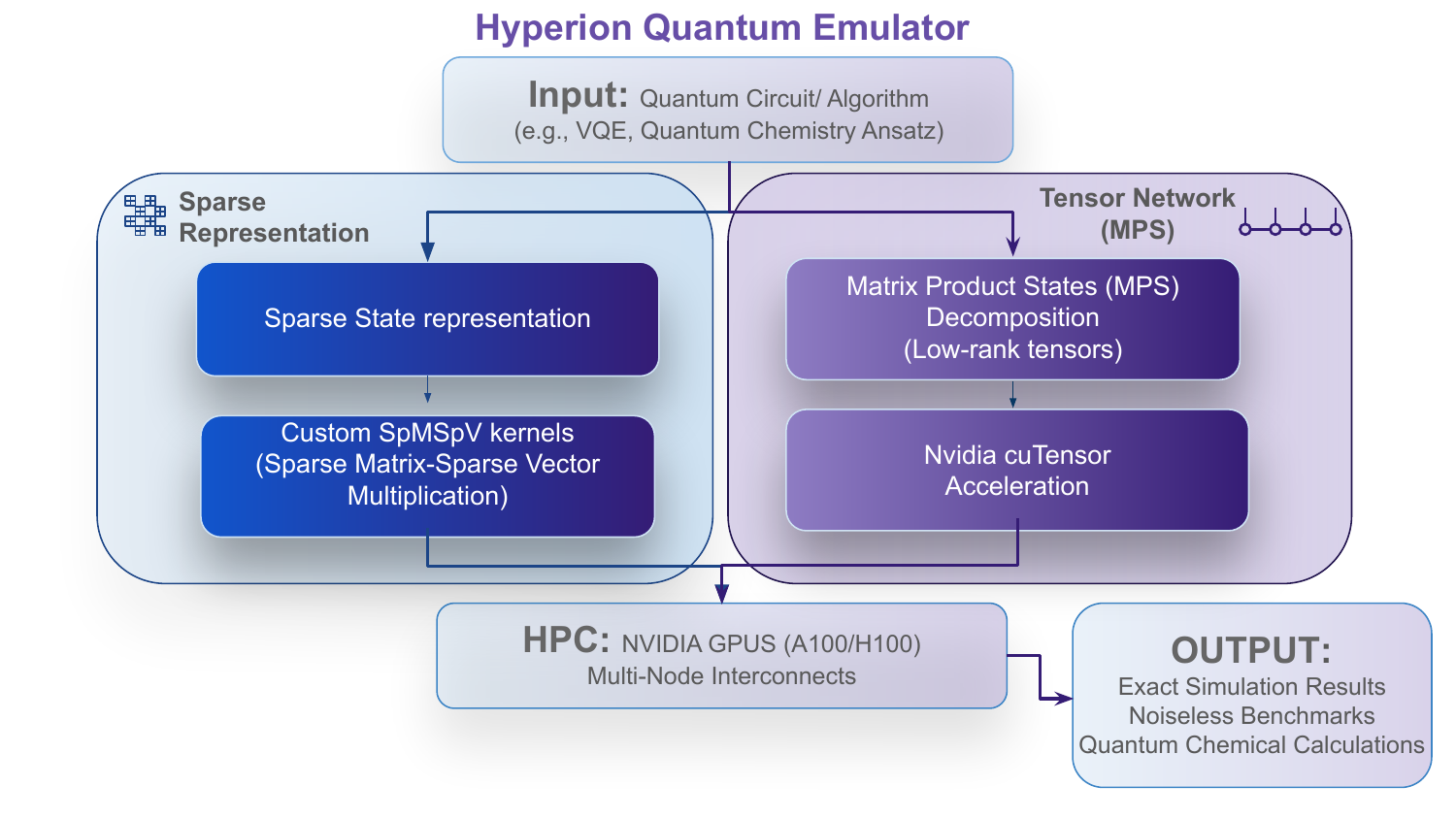}
    \caption{Overview of the Hyperion quantum emulator}
    \label{fig:overview}
\end{figure}
\subsubsection{Hyperion: High Performance GPU-accelerated Quantum Emulation}
The strategies employed by quantum emulators are often shaped by the specific use cases envisioned by their
developers. Among these,\textbf{ Hyperion} is our in-house quantum emulator, a \textbf{GPU-accelerated}, \textbf{high-performance software library} jointly developed by Qubit Pharmaceuticals and Sorbonne University\cite{Hyperion}. 

\paragraph{Hyperion's philosophy.}Hyperion is dedicated to simulating quantum chemistry problems using exact or quasi-exact strategies that emulate noiseless logical qubits. Our primary objective is to improve present NISQ methodologies and to prepare for the advent of FTQC by enabling algorithm design within a controlled environment. We have specifically focused on emulation strategies targeting quantum algorithms capable of addressing currently intractable quantum chemistry problems on classical supercomputers, specifically aiming for Full-CI/CBS accuracy (see Sections \ref{Classical} and \ref{Overview}). To contextualize Hyperion within the broader landscape, modern quantum emulators generally employ two primary approaches: state-vector-based emulation (using either dense or sparse representations for exact or approximate computations) and tensor network-based methods (where performance is governed by the entanglement structure of the simulated state). To maintain our focus on high-fidelity simulations that bridge the gap toward physical quantum hardware, Hyperion strictly uses these native quantum computing strategies. We deliberately avoid "quantum-inspired" classical approaches that rely heavily on traditional classical chemistry integral codes to revisit quantum algorithms. Hyperion therefore provides a \textbf{rigorous framework for assessing and validating novel quantum algorithms}. At its core, the emulator relies on highly optimized linear algebra kernels to accelerate operations between state-vectors and Hamiltonian matrices. An overview of Hyperion is given in~Figure~\ref{fig:overview}. Hyperion includes several emulators that we will detail below.

\subsubsection{The Hyperion-1 Emulator: State-vector and MPS for Chemistry}

Within the \textbf{state-vector paradigm}, the full state-vector (or brute-force) approach represents an $n$-qubit quantum state as a dense vector of size $2^n$. This approach provides an exact representation of quantum states and enables efficient gate
operations by using straightforward linear algebra, which is well-suited for GPU optimization. Therefore it provides \textbf{exact - noiseless- logical qubits} that can be compared to incoming FTQC results. While this method provides an exact representation and enables efficient, GPU-optimized  operations, it becomes memory-bound and unpracticable with regular resources for systems exceeding approximately 32-34 qubits due to the necessity of storing all $2^n$ complex amplitudes. Alternatively, when the specific algorithm or underlying physical problem exhibits structural sparsity, quantum states can be represented as sparse vectors to significantly reduce memory overhead, an approach commonly used in sparse quantum
emulators. 
In this representation, the state-vector is updated via indexing operations, thereby avoiding the need to explicitly store the full operator matrices during computation and  providing a substantial computational speedup over dense arithmetics.

However, it is worth noting that the fundamental limitation of state-vector based methods
persists: the memory overhead scales exponentially with the system size. Even when fully
leveraging sparsity, these methods consistently hit hard memory limits. 
A more promising avenue
to alleviate this problem involves methods such as tensor networks \cite{Markov2008,Vidal2003,Schollwck2011,Cirac2021}. Tensor network approaches have emerged as a compelling alternative for quantum emulation, offering a flexible and memory efficient framework. These methods, widely used in physics and mathematics, can help overcome the so-called \textit{curse of dimensionality} in certain scenarios and have become very important in the quantum computing field. 
In this paradigm, a quantum circuit is cast as a network of tensors where the desired output is obtained through a sequence of pairwise contractions (the tensor equivalent of matrix multiplication). The primary advantage of these data structures lies in their adaptability, contraction orders can be heuristically optimized to prioritize either computational speed or memory constraints. By representing a quantum state as a tensor network with fixed input and open output indices, one can efficiently compute specific amplitudes and expectation values through targeted contractions.
Beyond general tensor contractions, specific decompositions like \textbf{Matrix Product States (MPS)} enable significant data compression, provided that the system exhibits limited entanglement. By using a 1D chain of low rank tensors, MPS can represent a quantum state using a memory footprint that scales polynomially rather than exponentially. Using the MPS backend enables the substitution of exact state-vectors with memory-controlled approximations. 

In this context, our Hyperion emulator supports primarily sparse state-vector emulation while still offering approximate representations of state-vectors using MPS. Consequently, the performance of Hyperion is directly dependent on the efficiency of its underlying linear and multilinear algebra kernels.
Although adopting sparse representations can significantly enhance simulation speed, it introduces considerable complexity into the underlying linear algebra. For example, sparse matrix–dense vector multiplication (SpMV) is extensively documented and supported by highly optimized libraries such as Intel MKL.
In contrast, sparse matrix–sparse vector multiplication (SpMSpV), a central operation in our applications, remains computationally demanding. Because SpMSpV performance is heavily dictated by the sparsity patterns of both operands and has limited applicability in traditional scientific computing, it lacks optimized library support tailored for sparse quantum simulations. Hyperion  addresses this critical gap directly by implementing \textbf{custom kernels specifically engineered for efficient SpMSpV operations}. Furthermore, the library integrates tensor network strategies with a focus on MPS. To maximize computational throughput, these tensor operations are \textbf{GPU-accelerated  using the Nvidia} \textbf{cuTensor library} \cite{nvidia_cutensor}. 

To provide context, Table \ref{tab:hyperion-perfs} outlines the capabilities of Hyperion-1 in terms of logical qubit capacity. As a robust benchmarking tool for novel quantum and hybrid algorithms, Hyperion-1 was notably used to evaluate the Greedy Gradient-free Adaptive VQE (GGA-VQE) algorithm \cite{feniou2025greedy}. By computing the exact ground state of a 25-qubit Ising model, it provided a critical noiseless baseline for assessing ansatz quality on physical hardware. Furthermore, the emulator enabled large-scale simulations of up to 32 qubits in ref. \cite{traore2024shortcut}, which used a novel approach embedding a quantum computing ansatz into density-functional theory via density-based basis-set corrections. This method yielded quantitative quantum chemistry results for molecular systems that would otherwise necessitate brute-force calculations involving hundreds of logical qubits (see Section \ref{sec:database} for further details). Hyperion-1 also demonstrated its capacity for intensive workloads in \cite{feniou2024sparse}, performing a rigorous 28-qubit Overlap-ADAPT-VQE simulation \cite{feniou2023overlap} for state preparation in strongly correlated systems. Finally, our emulator has been used in validating new variational algorithms, such as MB-ADAPT-VQE \cite{fastmcp} and NI-DUCC-VQE \cite{Haidar2025}, pushing scalable molecular simulations to 26 qubits for accurate modeling of systems like the water molecule. Note that Hyperion-1 already enables to simulate ten of thousands of CNOT gates thanks to its GPU-acceleration (see Table \ref {tab:cnot}, section 4 for examples). Several applications of Hyperion-1 can be found in section 4.3.2.

In summary, "Hyperion-1 facilitates the rapid emulation of complex algorithms, such as ADAPT-VQE, at scales reaching 32 qubits. Its GPU acceleration allows for thousands of iterations within quantum chemistry workflows, ensuring true convergence and high-fidelity results. However, to overcome the exponential memory scaling and resource requirements of standard state-vector emulators,\textbf{ a novel hybrid simulation strategy} was introduced and led to the Hyperion-2 emulator.

\subsubsection{Pushing the Limits of State-Vector Emulation with Hyperion-2: a New Partitioned State-Vector/MPS Approach}
 Hyperion-2 uses a new State-Vector/MPS Approach partitioned approach\cite{Hyperion}. There, the molecular Hamiltonian is hierarchically partitioned: \textbf{non-interactive local blocks are evaluated exactly using a sparse state-vector representation}, while \textbf{complex  interactive terms are approximated using compressed MPS}. 

The approach begins by constructing a compact representation of the $n$-qubit Hamiltonian matrix, $\mathbf{H} \in \mathbb{R}^{2^n \times 2^n}$. This is achieved through a hierarchical left-right decomposition. Given this  partitioning of the Hamiltonian operator, our approach evaluates the most computationally expensive components of ADAPT-VQE using a combination of exact and approximate methods. These steps, specifically, the gradient measurements and the frequent energy evaluations required by the classical optimizer, are computed as follows.

First, the gradient measurements are evaluated as:
The ADAPT-VQE algorithm constructs a compact wavefunction ansatz by iteratively selecting operators from a predefined pool. At each step, it chooses the operator that yields the maximum energy gradient magnitude. This gradient is determined by the expectation value of the commutator between the Hamiltonian and the candidate operator, expressed as $\left| \frac{\partial E}{\partial \theta_i} \right| = \left| \langle \psi | [\mathbf{H}, \mathbf{P}_i] | \psi \rangle \right|$, where $\mathbf{P}_i$ denotes an admissible Hermitian generator. Our emulator can evaluate this quantity using either sparse state-vector techniques or a hybrid approach. 

 The primary goal of this partitioning is \textbf{to overcome the severe memory limitations inherent in the sparse state-vector method while maintaining accuracy}. Even highly sparse Hamiltonians, when stored in a distributed fashion, still necessitate multiple GPU resources for larger systems.

With this partitioned methodology, we overcome the need to store the entire Hamiltonian. We only need to store the non-interactive blocks, which scale efficiently as $\mathcal{O}(2^{\frac{n}{2^\eta}})$, where $\eta$ is the chosen partitioning level. The storage requirements for these blocks are significantly reduced. Conversely, the interactive Hamiltonian blocks are represented in a low-rank approximated Matrix Product Operator (MPO) form, and all operations involving these interactive components are performed using tensor network arithmetics.

The key advantage of this method lies in twofold: first, it reduces the memory overhead typically associated with state-vector approaches when storing the Hamiltonian. Second, it enables to minimize numerical errors by performing exact evaluations for certain blocks  using state-vector methods, rather than relying only on approximate MPS operations for all evaluations. 
We recall that while an MPS-based representation scales linearly with the system size, the algebraic operations required for a single ADAPT-VQE iteration possess varying complexities. Let $n$ be the number of qubits, and let $r$ and $R$ denote the maximum tensor ranks (commonly referred to as TT-ranks or bond dimensions) of the ansatz MPS and the Hamiltonian MPO, respectively. Table \ref{tab:hyperion-perfs} benchmarks the maximum logical qubit capacities of Hyperion-1 and Hyperion-2, based on recent ADAPT-VQE simulations executed on NVIDIA H100 (80GB) GPUs. Clearly, Hyperion-2 delivers considerable memory savings, enabling quasi-exact 28-qubit simulations on a single 80GB GPU. Furthermore, it  reduces the hardware footprint for 32-qubit simulations, scaling down from 128 GPUs in Hyperion-1 to just 16 GPUs. This  is achieved while preserving the  accuracy of exact state-vector methods throughout the ADAPT-VQE iterations, making large-scale quantum emulation highly accessible on standard HPC clusters. Current a priori limits for ADAPT-VQE with Hyperion-2 (SV/MPS) is 40 qubits (single ADAPT-VQE iteration) while iterated simulations were performed until 36 qubits.

\begin{table}[h!]
    \centering
    \caption{Simulation capacity limit in logical qubits on NVIDIA H100 (80\,GB) GPUs. The Table compares the initial Hyperion-1 state-vector approach to the new Partitioned SV-MPS architecture of Hyperion-2 (from Ref \cite{Hyperion}. \textbf{Both implementations can reach CNOT counts beyond 13000.}}
    \label{tab:hyperion-perfs}
    \begin{tabular}{@{} l r c c c @{}}
        \toprule
        & & \multicolumn{2}{c }{\textbf{Maximum Logical Qubits}} \\
        \cmidrule(l){3-5}
        \textbf{H100 GPUs} & \textbf{Total Memory} & \textbf{Hyperion-1} & \textbf{Hyperion-2 (MPS)} & \textbf{Hyperion-2 (Partitioned)}\\ 
        \midrule
        1   & \SI{80}{\giga B}    & 24 & 30& 28 \\
        4   & \SI{320}{\giga B}   & - & 32& 30 \\
        8   & \SI{640}{\giga B}   & 28 & -& 31 \\
        16  & \SI{1280}{\giga B}  & - & 36*& 32 \\
        64  & \SI{5120}{\giga B}  & - & -& 36 \\
        128 & \SI{10240}{\giga B} & 32 & -& - \\
        256 & \SI{20480}{\giga B} & - & 40& 40** \\
        \bottomrule
  \end{tabular}  
    \footnotesize
    \vspace{0.5cm}

\footnotesize
{$^*$ While the implementation can handle 40 qubits (and more), 36 qubits is the stability limit for the present pure MPS implementation in the context of ADAPT-VQE computations.\\
$^{**}$ 40 qubits is the implementation limit for the partitioned SV/MPS approach (single ADAPT-VQE iteration)}

\end{table}

\

\paragraph{Perspectives.} Scaling toward 50 qubits represents the theoretical frontier for state-vector approaches on exascale supercomputers \cite{de2025universal}. Currently, Hyperion-2 utilizes the CuQuantum SDK \cite{bayraktar2023cuquantum} to push beyond our present 36-qubit partitioned SV-MPS mark; however, this approach's reliance on massive computational power makes it unsustainable for broader industrial use. To achieve chemical accuracy for systems between 40 and 100 qubits without prohibitive costs, our emulator must evolve beyond its current MPS-based architecture. Indeed, although the Partitioned SV-MPS, and MPS can  solve, in certain cases, the memory bottleneck for intermediate-scale ADAPT-VQE simulations, its one-dimensional structure limits its ability to model the complex entanglement of highly correlated molecules.  Indeed, while MPS are a powerful classical tool for simulating quantum systems, their primary limitation is the exponential growth of the bond dimension $D$ required to track entanglement. In a quantum computer, a circuit can easily generate "volume law" entanglement where every qubit is highly correlated with every other qubit, for an MPS to accurately mimic this, the size of its internal tensors must scale exponentially with the number of qubits, eventually exhausting classical memory. Furthermore, MPS is inherently a one-dimensional architecture, making it highly inefficient at simulating the complex 2D or 3D connectivity found in modern quantum processors. Consequently, as a quantum circuit increases in depth and connectivity, the compression required to keep the MPS simulation running leads to a rapid loss of fidelity, meaning it can only truly mimic "shallow" or weakly entangled quantum processes before the classical overhead becomes insurmountable. Therefore,  one can  include more flexible structures, such as Tree Tensor Networks (TTNs) and 2D tensor graphs. \textbf{Such an evolution is mandatory to bridge the gap between theoretical accuracy and industrial utility, ensuring that complex simulations remain viable on sustainable GPU resources}. Because processing these complex networks are computationally demanding, current focuses on developing a flexible  software engine. This upgraded \textbf{Hyperion-3 platform} will rely on advanced approximate algorithms, to efficiently handle the massive entanglement of complex quantum chemistry problems. 
\paragraph{Qiskit Interface}Finally, we are currently finalizing a Qiskit interface for quantum circuit emulation. This will allow users to try our Hyperion emulator using sparse state-vector emulation for specifically structured circuits, MPS for circuits with low entanglement, and dense state-vector emulator for others.

\subsubsection{Hyperion's Competitive Landscape for Exact Quantum Emulation}
\paragraph{The 50 Qubits Limit for State-Vector Emulation.}The JUQCS-50 project recently defined the current frontier of State-Vector (SV) simulation, while simultaneously highlighting its practical constraints for domain-specific applications. This landmark 50-qubit simulation utilized the Booster module of the exascale JUPITER supercomputer, harnessing a massive fleet of 16,384 NVIDIA GH200 Grace Hopper Superchips across 4,096 nodes. While JUQCS-50 leverages the NVIDIA cuQuantum SDK—specifically the Host State Vector Migration features for explicit data movement—its primary demonstration involved Random Quantum Circuits (RQC). Although RQC is the standard benchmark for "quantum supremacy," it remains a synthetic task that does not address the specific topological and algebraic complexities of quantum chemistry. Consequently, while logical qubit count serves as a high-profile metric for emulator performance, it is largely meaningless in isolation. For quantum chemistry, performance must instead be measured by the ability to support a complete operational framework. This includes high-precision FP64 arithmetic, complex Hamiltonian mappings, and the iterative demands of the Variational Quantum Eigensolver (VQE)—all of which are necessary to produce chemically meaningful predictions. A rigorous metric must therefore account for circuit depth and the total number of CNOT gates simulated. Indeed, many HPC studies report high qubit counts by oversimplifying the underlying chemical problem, often failing to simulate circuits of practical depth. In the State-Vector (SV) regime, it is technically possible to reach 80 qubits or more \cite{steiger2024sparse}; however, such results rely on specialized Hamiltonian-based methods like Iterative Qubit Coupled Cluster (iQCC). These methods employ aggressive Hamiltonian cutoffs and "greedy" operator selection to keep the wavefunction artificially sparse by discarding essential interaction terms. While effective for weakly correlated systems, this strategy inevitably collapses in the presence of strong electronic correlation, where entanglement spreads across the entire state vector and eliminates any sparsity advantage. Ultimately, these molecule-dependent approximations do not offer a reliable or general path toward numerically exact solutions. Consequently, for general-purpose, high-fidelity SV emulation, the practical hard limit remains 50 qubits, as established by the JUQCS-50 project.
\paragraph{Accuracy limit for Matrix Product States}Regarding Matrix Product States (MPS), exact tensor network simulations struggle with deep circuits and high CNOT counts due to the inherent complexity of contraction path optimization. Consequently, these methods are often validated using shallow, small-scale circuits, which allow for the artificial inflation of qubit counts. For instance, Shang et al. \cite{shang2022large} developed a high-performance VQE simulator on the Sunway supercomputer by integrating MPS with Density Matrix Embedding Theory (DMET), nominally reaching 1,000 qubits. However, DMET is not a strictly accurate many-body method; its primary weakness lies in its fragment-based nature, which relies on an "embarrassingly parallel" scheme for quantum chemistry solvers. By partitioning the system into independent local clusters associated with isolated, small circuits, the methodology achieves high qubit counts at the expense of capturing the long-range correlation effects that define a true quantum advantage.

Similarly, high emulated MPS qubit counts can also be attained by strictly constraining circuit complexity. A prominent example is the seniority-zero unitary coupled cluster (upCCD) architecture \cite{lee2018generalized}, which, in the vein of the Doubly-Occupied Configuration Interaction (DOCI) approach \cite{weinhold1967reduced}, restricts the wave function expansion to configurations where electrons remain paired. While these methods yield simplified circuits with low gate counts and shallow CNOT depths—making them executable on standard workstations—they face severe operational limitations. These constraints arise primarily from the non-exact bosonic mapping which, by design, results in an artificial doubling of the reported qubit count. Furthermore, by confining simulations to the seniority-zero subspace, such methods rely on molecule-specific approximations that neglect critical correlation effects, rendering the FCI limit fundamentally inaccessible. If these approaches have been particularly useful and well-suited for early demonstration of quantum chemistry on current, resources constrained, quantum hardware\cite{zhao2023orbital}, the ultimate objective of quantum computing emulation in chemistry should remain the realization and test of predictive algorithms that aim at approaching the FCI/CBS limit. Indeed, for now, emulators offer more possibility to explore complex circuits than current hardware, e.g., Hyperion can simulate in excess of 10,000 CNOTs using ADAPT-VQE, see Table 10. Therefore achieving genuine chemical accuracy is paramount over attaining inflated qubit counts that yield only qualitative approximations.

Current research supports these limitations: a recent study by Google and the Chan group \cite{provazza2024fast} estimated the \textbf{stability limit for MPS at 36 qubits using ADAPT-VQE}. \textbf{Our experiments confirm these findings\cite{Hyperion}} (see below). Ultimately, there is "no free lunch" in quantum chemistry: simplification rarely equates to accuracy. While MPS emulation can reach higher qubit counts than State Vector (SV) methods by restricting the bond dimension, such simulations often lack the precision required for chemical applications. Furthermore, many large-scale simulations rely on molecule-dependent approximations or arbitrary cutoffs, which introduce uncontrolled errors and lack generality. Because MPS and other Tensor Network approaches are inherently non-exact, their computational costs explode when attempting to capture the high-degree entanglement characteristic of strongly correlated molecular systems.
\paragraph{Quantum Emulation for Quantum Chemistry applications.}Table \ref{tab:hyperion-compet} illustrates where Hyperion stands within the current state of the art of quantum emulation for chemistry applications.
 
Hyperion-1 and 2 are engineered as prioritizing full-fidelity representations over lossy compression\cite{Hyperion}. To overcome the massive overhead inherent in exact calculations, the platform integrates GPU acceleration with high-performance sparsity, a combination that is mandatory for the iterative nature of variational algorithms. Traditional CPU-based state-vector approaches are prohibitively slow, often failing to reach convergence within the thousands of iterations required by protocols like ADAPT-VQE. By offloading these intensive workloads to specialized GPU architectures, Hyperion-1 achieves the throughput necessary for complex molecular simulations that would otherwise remain computationally intractable on conventional CPU frameworks. Concerning brute force circuit simulations, \textbf{Hyperion already offers the possibilities to handle ten of thousands of CNOT gates} (see Table \ref {tab:cnot}, section 4 for examples).
Note that Hyperion embodies \textbf{qubits extrapolation techniques} allowing to go far beyond these qubit counts to provide a shortcut to chemical accurate. Offering up to \textbf{10-fold reduction in qubits}, these techniques allowed us in recent years to \textbf{explore the electronic structure of systems ranging beyond 300 qubits}. These approaches are described in section~\ref{subsec:DBBSC}.
\begin{table}[h!]
\centering
\caption{List of Quantum Emulator Packages with Demonstrated Applications to Quantum Chemistry (standard ADAPT-VQE), and Corresponding Features. SV=State-Vector, MPS= Matrix Product States}
\label{tab:quantum_emulators}
\begin{tabular}{|l|c|c|c|c|c||c|}
\hline
\textbf{Software} &  \textbf{Max Qubits} & \textbf{Accuracy} & \textbf{SV} & \textbf{MPS}& \textbf{GPUs}  \\ \hline
\textbf{Hyperion-1 (SV)}$^1$ & \textbf{32} & Exact & \checkmark  & $\times$ & \checkmark\\ \hline
\textbf{Hyperion-2 (Partitioned SV/MPS)}$^1$ & \textbf{36 (40$^1$)} & Locally Exact+Locally bounded & \checkmark  & \checkmark & \checkmark\\ \hline
\textbf{Hyperion-2 MPS}$^1$ & \textbf{36} &Globally bounded & $\times$  & \checkmark & \checkmark\\ \hline
quantumlib qsim SV (Google)$^2$ & 12 & Exact & \checkmark & $\times$ & $\times$ \\ \hline
cuStateVec SV (NVIDIA)$^3$ & 32 & Exact & \checkmark  & $\times$ & \checkmark \\ \hline
cuTensorNet MPS (NVIDIA)$^4$& 36 & Globally bounded & $\times$  & \checkmark & \checkmark \\ \hline
FQE MPS (Google)$^5$ & 36 & Globally bounded & $\times$ & \checkmark & $\times$\\ \hline
Microsoft sparse SV $^6$ & 8 & Exact & \checkmark  & $\times$ & $\times$\\ \hline
mpiQulacs SV (Fujitsu) $^7$ & 36 & Non-exact (cutoffs) & \checkmark  & $\times$ & $\times$\\ \hline
Captiva/myQLM-fermion SV (Bull/Eviden)$^{8}$ & 24 & Exact & \checkmark & $\times$ & $\times$\\ \hline
Pennylane Lightning SV $^{9}$ & 24 & Exact & \checkmark  & \checkmark & $\times$\\ \hline 
Qiskit Aer SV (IBM) $^{10}$ & 22 & Exact & \checkmark & \checkmark & $\times$ \\ \hline

\end{tabular}\label{tab:hyperion-compet}

\vspace{0.5cm}
\begin{flushleft}
\footnotesize
$^1$ 36 qubits "iterated" ADAPT-VQE simulations were performed using Hyperion-2 (Partitioned SV/MPS) on 64 H100 GPUs\cite{Hyperion}. Iterated 32 qubits simulations reported in \cite{traore2024shortcut,Hyperion} using Hyperion-1 on 128 H100 GPUs. Larger qubits counts can be obtained with Hyperion-2 on larger resources, either with the SV/MPS approach (single iterations up to 40 qubits) or through leveraging the cuQuantum SDK dense matrix operations\cite{bayraktar2023cuquantum}. For pure MPS, due to accuracy, the limit is the same as discussed in \cite{provazza2024fast}\\
$^2$ quantumlib \cite{qsimGoogle,2020} is an open-source ecosystem of quantum computing tools maintained by Google Quantum AI. Using this framework, a 12 qubit VQE simulation is achieved.\\
$^3,^4$ cuStateVec and cuTensorNet are parts of the cuQuantum SDK\cite{bayraktar2023cuquantum}. For MPS, due to accuracy,the limit is 36 \cite{Hyperion}. For SV, the limit is theoretically the one discussed for state-vector emulation: between 32 and 50, see the JUQCS-50\cite{de2025universal} discussion in 3.3.4. To our knowledge, reported applications of cuQuantum to quantum chemistry using SV are limited to 32 qubits\cite{nvidia2024cudaq}.\\
$^5$ Fermionic Quantum Emulator (FQE)\cite{provazza2024fast}. MPS simulations pose accuracy limitations for pure MPS. FQE uses the open-source pyblock3 and block2 librairies. \\
$^6$ 26 qubits max but 8 demonstrated for chemistry \cite{jaques2022leveraging}\\
$^7$ 36 qubits single-iteration ADAPT-VQE simulation and Hamiltonian terms cut-off method was used to slim down terms\cite{morita2024simulator,imamura2022mpiqulacs}.\\
$^8$ The distributed SV emulator from the Qaptiva plateform can perform quantum chemistry using the OpenVQE package \cite{haidar2023open} coupled to myQLM-fermion\cite{myqlm_fermion_github}. myQLM-fermion can emulate up to 30 qubits with CLinalg and a recent Captiva update lists some MPS/GPU capabilities. In both cases, quantum chemistry applications have not been reported. \\
$^9$ \cite{asadi2024hybrid}\\
$^{10}$  \cite{Qiskit}
\end{flushleft}
\end{table}

\subsubsection{Beyond Quantum Emulation: Unified Orchestration of Hybrid QPU and GPU Architectures and noise modeling.}
\par
\textbf{Orchestration} is another key layer of Hyperion. Its role is to coordinate how computations are distributed between quantum processors (QPUs) and classical computers (CPUs and GPUs). Because most near-term quantum algorithms combine quantum and classical steps~\cite{robledo2025chemistry}, orchestration breaks a calculation into smaller tasks, sends each one to the appropriate hardware, and manages the exchange of data between them.

\begin{figure}[h!]
    \centering
    \includegraphics[scale=0.45]{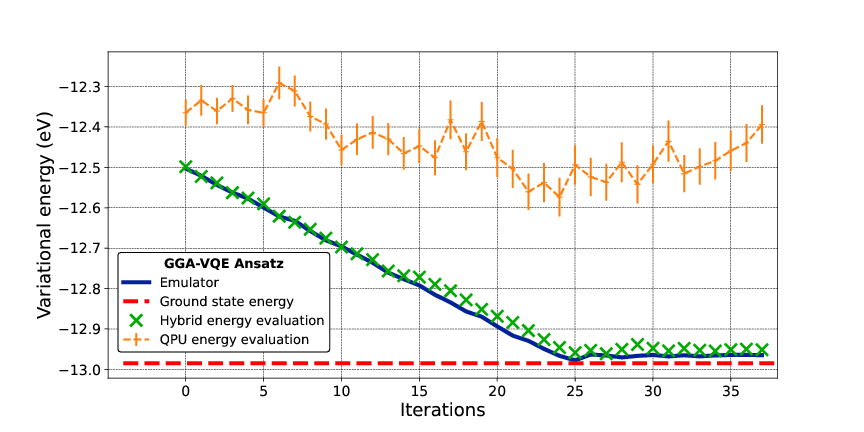}
    \caption{Convergence of the GGA-VQE algorithm's energy as a function of iteration count. The blue curve provides the classical ansatz reference. In contrast, the green and orange curves represent the GGA-VQE ansatz energy as evaluated via a hybrid approach and a 25-qubit QPU, respectively. The hybrid results were obtained by extracting the QPU-generated wavefunction and re-simulating it on the Hyperion-1 HPC emulator to calculate the variational energy. Data from ref. \cite{feniou2025greedy}.}
    \label{fig:isinghybrid}
\end{figure}

Orchestration should be reflected directly in algorithm workflows. As an example, we implemented such a workflow for the GGA-VQE algorithm, combining a trapped-ion IonQ quantum processor with classical GPU resources to perform hybrid energy evaluations. In this experiment~\cite{feniou2025greedy}, the GGA-VQE circuit was first executed on the IonQ Aria QPU to generate the variational ansatz state. The resulting circuit parameters were then retrieved and reproduced on the Hyperion emulator, where the variational energy was evaluated using HPC resources (see Sec.~3.2.1 and Fig.~\ref{fig:isinghybrid}). This \textbf{hybrid execution} illustrates how orchestration can seamlessly coordinate quantum hardware and classical computing resources within a single workflow.
\paragraph{Noise modeling}The integration of QPU orchestration and emulation modules enables precise modeling of statistical sampling noise - often termed shot noise - which remains a primary bottleneck for NISQ-era algorithms\cite{feniou2025greedy}. Since Hamiltonian expectation values are derived from finite measurements, the resulting cost function is inherently stochastic and non-linear. In algorithms such as ADAPT-VQE, this noise can cause energy convergence to plateau prematurely, often failing to reach chemical accuracy. Developing noise-resilient strategies - including local optimization and analytical extrapolation—is therefore critical to maintaining the tractability of high-dimensional optimization under sampling uncertainty. Furthermore, recognizing that device-specific hardware noise presents additional challenges, we are collaborating with our hardware partners to integrate these modeling capabilities into our framework.

On the classical side, Hyperion is hardware-agnostic and can run on GPUs from NVIDIA, Intel, and AMD, although current optimizations target NVIDIA architectures. On the quantum side, it follows a similar approach by interfacing with multiple types of QPUs and providing a unified environment for the quantum algorithms discussed in this work.

\section{Quantum Foundation Machine Learning Models for Molecular Dynamics}\label{sec:FM}
In recent decades, molecular simulations have relied on heavily parametrized empirical interatomic potentials. There has been a steady improvement of the quality of these models allowing an overall increase in the quality of their prediction. This comes from both better parametrization procedures~\cite{poltype2,chen2024advancing}) and more accurate functional forms that manage to better approximate the intrinsic quantum mechanical nature of the phenomena at stake ~\cite{AMOEBA03,shi2013polarizable,AMOEBA+1,AMOEBA+2,nawrocki2022protein,amezcua2021sampl7}. Among these, polarizable force fields have proven to provide a systematic quality improvement while still being associated to a manageable computational cost thanks to algorithmi\cite{Lagardere2015,TCG,ddCOSMO} and high performance computing implementation efforts~\cite{lagardere2018tinker,adjoua2021tinker}. They have also proven to be able to produce large scale predictive high-resolution simulations of complex biological systems~\cite{D1SC00145K,Dinawater,D4SC04364B,jing2021thermodynamics,schahl2025histidine,D1SC05892D}. Still, these models suffer from a lack of flexibility of their mathematical formulations as well as expert knowledge for their calibration. 
The advent of neural network potentials directly addresses both these shortcomings. 

Early scalable architectures such as the seminal work of Behler and Parrinello~\cite{behler2007generalized} laid the groundwork for modern models, which provide high-accuracy gas-phase simulations for small molecules at a significantly lower cost than traditional ab initio methods. However, these models often struggle when transitioning to condensed-phase biological environments where efficient sampling is critical. The MACE foundation models~\cite{Mace} can also be highlighted as a pioneering general-purpose machine learning potential for material science, though its derivative that focuses on organic small molecules, MACE-OFF, remains computationally expensive and limited for drug design due to its initial lack of support for charged species~\cite{Kovacs2025}. 

\subsection{FeNNix-Bio1, a Quantum Foundation Model for Drug Design}
In that context, some of us introduced FeNNix-Bio1~\cite{Ple2025}, a family of foundation machine learning models (S and M versions) designed to enable accurate and reactive atomistic simulations of biological systems with unprecedented speed and scalability. While traditional drug discovery has been hampered by the computational bottleneck of characterizing dynamic interactions between proteins and drug candidates, FeNNix-Bio1 bridges this gap by providing quantum-accurate molecular dynamics without the need for manual parameterization. Trained exclusively on the Ignis inhouse proprietary dataset~\cite{Ple2025} (see sec.~\ref{sec:ignis}) — a massive initial collection of over 2.2 million synthetic quantum chemistry configurations — the model employs a novel RaSTER (Range-Separated Transformer with Equivariant Representations) architecture~\cite{Ple2025}. Going beyond the initial 2023 FENNIX-OP1 local equivariant neural network\cite{ple2023fennix}, FeNNIx-Bio1's architecture utilizes a multi-resolution descriptor that resolves short-range covalent structures with an equivariant Transformer while addressing long-range interactions (up to 11 \AA) via a physics-inspired radial basis. 

To handle the complex charge states ubiquitous in biology, a multi-Qeq embedding was introduced. By non-locally distributing the total charge, it allows the model to differentiate between various charge states in condensed-phase environments.
The model's versatility was demonstrated across a wide spectrum of applications, including the precise prediction of hydration free energies (HFEs) for the Freesolv~\cite{mobley2014freesolv} dataset, where it significantly outperformed traditional GAFF force fields (see ref.~\cite{chen2024advancing} and references therein). 
Furthermore, FeNNix-Bio1 successfully modeled complex phenomena such as the reversible folding of the Chignolin protein—a first for a foundation ML model—and preserved stable binding modes for protein-ligand complexes like Benzamidine-Trypsin. The model also natively handles chemical reactivity, accurately sampling reactive pathways for the conversion of butadiene to cyclobutene through active-learning-driven finetuning. From a performance standpoint, FeNNix-Bio1 is optimized for massively parallel GPU-accelerated systems, showing inference speeds over an order of magnitude faster than existing models like MACE-OFF~\cite{Kovacs2025} while maintaining comparable or superior accuracy. By accounting for nuclear quantum effects (NQEs) through an adaptive Quantum Thermal Bath~\cite{mangaud2019fluctuation,mauger2021nuclear} and providing robust uncertainty quantification via the DPOSE method~\cite{kellner2024uncertainty}, FeNNix-Bio1 establishes a new standard for predictive pharmacological sciences, capable of simulating efficiently systems as large as the SARS-CoV-2 Spike protein with over 1.6 million atoms while providing quantum accurate properties.

\subsubsection{FeNNix-Bio1: Fast Neural Network Architectures for Quantum-accurate  Simulations}
In the context of molecular dynamics, beside its efficient neural architecture, a key innovation of FeNNix-Bio1 is to leverage acceleration methods that couple model distillation, multi-time-stepping and non-conservative forces offering unprecedented simulation capabilities at quantum accuracy.

\paragraph{Efficient Neural Architectures and Fast Efficient Multi-Time-Stepping.}

Let us summarize the general architecture of FeNNix-Bio1~\cite{Ple2025}. In a first step, each atom is embedded in a $N_f$-dimensional vector space, describing electronic structure and charge information. This initial geometry-independent embedding is then updated to include local geometric information via two message-passing layers of an equivariant transformer~\cite{vaswani2017attention,fuchs2020se,tholke2022torchmd,le2022equivariant,liao2022equiformer,liao2023equiformerv2,frank2024euclidean}.
The first layer considers a high-resolution description of very short range (within a $R_c^{(sr)}=\SI{3.5}{\angstrom}$ cutoff) geometry, while the second layer incorporates both short-range and medium-range messages up to a cutoff $R_c^{(lr)}=\SI{7.5}{\angstrom}$, resulting in a total receptive field of $\SI{11}{\angstrom}$. Then, atomic energies for each atom are obtained from mixture-of-experts~\cite{cai2024survey} multi-layer perceptrons $MLP_E$, taking the final embeddings $x_i$ as an input, with routing depending on the chemical group $G(Z_i)$ of each atom. Each perceptron $MLP_E^{G(Z_i)}$ outputs $N_{\text{ens}}$ scalars interpreted as an energy distribution, which are then averaged in the final output. Finally, an explicit screened nuclear repulsion term $E_{\text{rep}}$ is added, as well as fixed reference atomic energies $E^0_{Z_i}$. The expression of total energy is thus given by
\[ E_{\text{tot}} = \sum_{i=1}^{N_{\text{at}}} E_{Z_i}^0 + \sum_{i,j=1}^{N_{\text{at}}} E_{\text{rep}}(r_{ij}) + \frac{1}{N_{\text{ens}}}\sum_{n=1}^{N_{\text{ens}}}\sum_{i=1}^{N_\text{at}}[MLP_E^{G(Z_i)}(x_i)]_n\,,\]
where $N_{\text{at}}$ is the number of atoms in the system and $r_{ij}$ is the distance between atoms $i$ and $j$. The repulsion term $E_{\text{rep}}$ has an analytic expression, defined using the element-pair-specific NLH parametrization of~\cite{Nordlund2025}:
\[E_{\text{rep}}(r_{ij}) = \frac{Z_iZ_j}{4\pi\veps_0r_{ij}}\sum_{n=1}^3 a_{nij}e^{-b_{nij}r_{ij}}\,,\]
with $Z_i$ the atomic number of atom $i$ and $a_{nij},b_{nij}$ parameters that depend on the unordered pair of species $Z_i,Z_j$ that were obtained via a fit to \textit{ab initio} references. Compared to other NNPs, using different resolutions at different scales, namely highly precise short-range spatial descriptors and minimal physics-inspired longer-range features (i.e. radial basis that uses Coulomb and dispersion kernels, similar to the ones used in~\cite{huguenin2023physics}), allow for a drastic improvement in efficiency. We refer to~\cite{Ple2025} for more details.

While simulating an atomistic system with FeNNix-Bio1(M) is less expensive than using \textit{ab initio} methods by several orders of magnitude, evaluating neural networks remains slower than any classical empirical force field, which motivates the development of algorithmic methods to accelerate NNP-based simulations. In molecular dynamics, the number of evaluations the force field per unit of physical time is proportional to the time discretization step size. A well-known method used to increase the maximum usable time step of classical integrators (which is usually limited by the highest-frequency motions of the system such as chemical bond vibrations) is called multi time-stepping (MTS)~\cite{Tuckerman1992,Lagardere2019}. When applied to an empirical force field model, it consists in treating the fast-varying short-range forces within a small step size, while only applying the long-range, more regular forces at larger time step. This effectively increases the speed of simulation by reducing the rate at which expensive long-range forces are computed. In a NNP however, there is no natural decomposition of the forces into \textquote{cheap} and \textquote{expensive} parts. Instead, we proposed in~\cite{Cattin2026} a so-called distilled multi-time-step (DMTS) strategy for NNPs, based on distillation~\cite{Hinton2015,Gou2021} of the FeNNix-Bio1(M) foundation model~\cite{Ple2025} which was performed via the FeNNol~\cite{Ple2024} library. In this strategy, a small, less precise but fast-to-evaluate model is trained on data labeled with FeNNix-Bio1(M) instead of DFT. Then, following the MTS procedure, this small model, which uses a single 3.5 \AA-cutoff message passing layer that is sufficient to capture the fast-varying forces, is applied in several iterations of an inner loop, before being corrected by an application of the difference between the large and small force models in an external loop. Therefore, by reducing the number of computations of the large expensive model, simulation speed is increased by a factor 3 to 4 with respect to standard single-time-step, enabling longer stable simulations.

Another way of accelerating simulation speed is the use of non-conservative forces~\cite{hu2021forcenet,Gasteiger2021,liao2023equiformerv2,neumann2024orb,rhodes2025orb,eissler2025simple}, that are not constrained to deriving from a potential energy. By directly predicting force vector fields, one is able to bypass the costly backpropagation step required to compute energy derivatives, therefore improving computational efficiency at fixed model architecture. In~\cite{Gouraud2026}, we extend the DMTS procedure to the use of such forces, which we denote DMTS-NC.
Now, the large NNP model is distilled into a small non-conservative one, which still enforces several physical priors such as equivariance under rotation and cancellation of atomic force components. This architecture allows the distilled model to reach an excellent agreement (the final training mean absolute error being close to 1kcal/mol) with the reference one, on top of being faster to evaluate than the small conservative model used in DMTS. By lowering evaluation cost while improving stability (in particular, the occasional disagreements between the small and large models which required system-specific fine-tuning are greatly reduced), the DMTS-NC scheme is found to be more stable and efficient than its conservative counterpart with additional speedups reaching 15-30\% over DMTS. Note that similarly to DMTS, DMTC-NC is a model-agnostic strategy  and could be applied to distill any NNP model such as MACE~\cite{Mace}. \\
Overall, this acceleration strategy achieves speedups of up to 431\% over standard single-time-step methods while maintaining the near-quantum mechanical accuracy required for complex systems like solvated proteins and water boxes while ensuring FeNNix-Bio1 to be, by far, the fastest NNP approach.
\begin{table}[h!]
    \centering
    \caption{Summary of FeNNix-Bio1 Foundation Model Methodology}
    \label{tab:fennix_methodology}
    \begin{tabularx}{\textwidth}{l >{\raggedright\arraybackslash}X}
        \toprule
        \textbf{Component} & \textbf{Description} \\
        \midrule
        \textbf{Model Nature} & A foundation machine learning model designed for accurate, reactive atomistic simulations of biological systems. \\
        \addlinespace
        \textbf{Architectural Variants} & Consists of \textbf{FeNNix-Bio1(S)} (light-weight, high-throughput) and \textbf{FeNNix-Bio1(M)} (heavier, higher accuracy). \\
        \addlinespace
        \textbf{Training Data} & Trained on the \textbf{Ignis dataset} (~2.23M conformations) derived exclusively from synthetic DFT quantum chemistry data \\
        \addlinespace
        \textbf{Energy Decomposition} & Total energy is decomposed into atomic contributions using a \textbf{Mixture-of-Experts (MoE)} multilayer perceptron. \\
        \addlinespace
        \textbf{Charge Modeling} & Employs \textbf{multi-Qeq embedding} to globally distribute total charge, enabling non-local differentiation of charge states. \\
        \addlinespace
        \textbf{Receptive Field} & Uses two-layer equivariant message-passing with a total receptive field of \textbf{11\AA}, comparable to traditional force fields. \\
        \addlinespace
        \textbf{Range Resolution} & Features the \textbf{RaSTER} descriptor: equivariant Transformers for short-range and physics-inspired kernels for long-range effects. \\
        \addlinespace
        \textbf{Nuclear Quantum Effects} & Explicitly includes \textbf{Nuclear Quantum Effects (NQEs)} via the adaptive Quantum Thermal Bath (adQTB) method or Path Integrals. \\
        \addlinespace
        \textbf{Uncertainty Tracking} & Uses the \textbf{DPOSE} method to provide well-calibrated, inexpensive uncertainty estimates for predictions. \\
        \addlinespace
        \textbf{Hardware Target} & \textbf{Massively parallel}, \textbf{GPU-accelerated} implementation functional on NVIDIA, AMD, and Intel platforms. \\
        \addlinespace
        \textbf{Distilled Multi Time-Stepping (DMTS-NC)} & \textbf{4-time faster} than in the initial 1fs timestep published implementation. \\
        \bottomrule
    \end{tabularx}
\end{table}
\subsubsection{Applications of the FeNNix-Biol Foundation Model}
Potential applications of the FeNNix-Biol Foundation Model are numerous and are listed in Table \ref{tab:applicationsfennix}. FeNNix-Biol serves as a versatile foundation model for drug design, enabling accurate simulations of liquid water properties and ion solvation. It excels in predicting hydration free energies, absolute binding free energies, and chemical reactivity out-of-the-box. Furthermore, it facilitates the reversible folding of proteins and the refinement of static structures from AI models like Boltz or AlphaFold. Its scalability even allows for the simulation of massive complexes, such as the full SARS-CoV-2 Spike protein.
\begin{table}[h!]
\centering
\caption{Applications of the FeNNix-Biol Foundation Model}
\begin{tabular}{|p{5.25cm}|p{6cm}|p{5.5cm}|}
\hline
\textbf{Application Domain} & \textbf{Specific Description} & \textbf{Notable Performance / Result} \\ \hline
\textbf{Liquid Water Properties} & Calculation of radial distribution functions, density, and enthalpy of vaporization. & Reaches the quality of accurate polarizable models while outperforming other foundation models at lower cost. \\ \hline
\textbf{Ions in Solution} & Study of the behavior of $Na^{+}$ and $Cl^{-}$ ions in a water box. & Achieves agreement with polarizable force fields. \\ \hline
\textbf{Hydration Free Energies (HFEs)} & Prediction of HFEs for over 600 molecules in the Freesolv dataset. & Provides sub-kcal/mol accuracy and converges at least 60 times faster than other foundation models. \\ \hline
\textbf{Protein Folding} & Study of the reversible folding of the Chignolin protein (CLN025). & Recovers folded, misfolded, and unfolded states; first time achieved with a foundation ML model. \\ \hline
\textbf{Protein-Ligand Complexes} & MD simulation of complexes such as the Benzamidine-Trypsin complex. & Maintains a well-preserved binding mode over several nanoseconds of trajectory. \\ \hline
\textbf{Absolute Binding Free Energy} & Calculation of standard binding free energy for Phenol and Benzene to Lysozyme. & Obtains values close to experimental results, such as $-4.4$ kcal/mol for benzene-lysozyme. \\ \hline
\textbf{Chemical Reactivity} & Exploration of the gas-phase reaction converting butadiene to cyclobutene. & Models reaction pathways and bond order inversion out-of-the-box with high data efficiency. \\ \hline
\textbf{Large-Scale Simulations} & Simulation of the SARS-CoV-2 Spike protein with membrane and glycans ($>$ 1M-atom system). & Demonstrates scalability to systems exceeding 7 million atoms. \\ \hline
\textbf{Synergy with AI Models} & Refinement of structures predicted by AlphaFold or Boltz models. & Enables full relaxation of predicted structures to model side-chain dynamics in solution. \\ \hline
\end{tabular}
\label{tab:applicationsfennix}

\end{table}

\subsubsection{Software Base: the High Performance Tinker-HP Package.}
\begin{figure}[h!]
    \centering
    \includegraphics[scale=0.65]{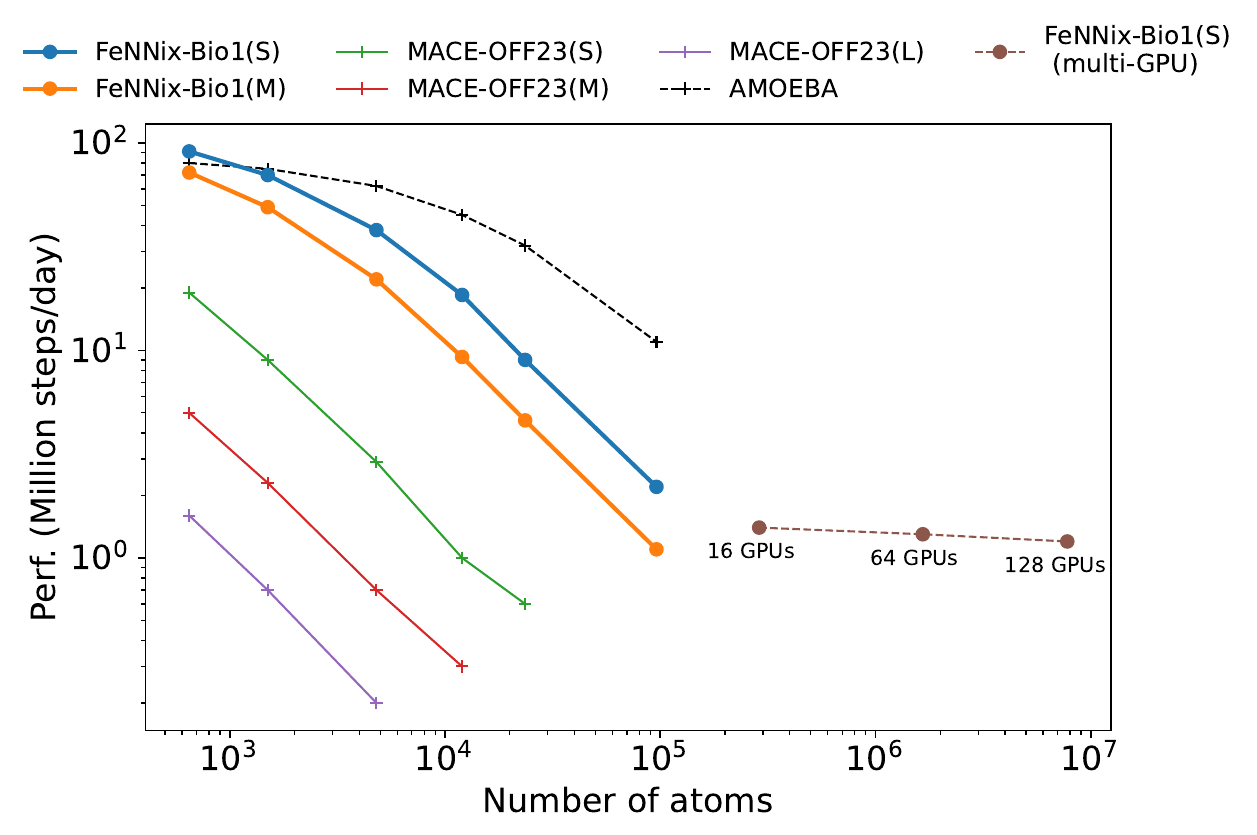}
    \caption{Performance in millions of steps per day of various Machine Learning Potentials, compared to the AMOEBA force field, with varying system size for 1fs simulations (i.e. no multi-timestep). Data from ref \cite{Ple2025}}
    \label{fig:perf_Deep-HP}
\end{figure}

FeNNix-Bio1 integrates the Tinker-HP software\cite{lagardere2018tinker,adjoua2021tinker} suite—specifically via the Deep-HP\cite{inizan2022scalable} scalable interface—to manage the extreme computational requirements of large-scale biological systems. This architecture bridges the gap between static structural prediction and high-fidelity dynamic simulations through two primary pillars:
\paragraph{High-Performance Scalability.} By utilizing the Deep-HP interface to link the FeNNol library\cite{Ple2024}, which embodies FeNNix-Bio1, with Tinker-HP, FeNNix-Bio1 leverages advanced 3D spatial domain decomposition and full GPU offloading. This synergy allows for:
\begin{itemize}
    \item \textbf{Massive System Sizes}: Efficient simulation of systems containing millions of atoms.
    \item \textbf{Optimized Workloads}: Offloading of intensive neighbor list computations and parallel communication to the Tinker-HP engine.
\item \textbf{Hardware Agnosticism}: Seamless execution across NVIDIA, Intel, and AMD GPU architectures.
\end{itemize}
\paragraph{Quantum Accuracy in Dynamics at Scale.} FeNNix-Bio1 foundation models incorporate Nuclear Quantum Effects (NQEs) through non-classical molecular dynamics, ensuring high-fidelity modeling of condensed-phase phenomena (e.g., the anomalous properties of liquid water). Key methods include:
\begin{itemize}
    \item a fast \textbf{adQTB} (Adaptive Quantum Thermal Bath)implementation\cite{mauger2021nuclear}: Provides \textbf{quantum dynamics at a computational cost comparable to classical simulations}.
    \item \textbf{Path Integral} Approaches: Integrated via the \textbf{Quantum-HP} module\cite{ple2023routine} for rigorous, high-resolution, treatment of quantum nature of dynamics in complex environments.
    \item Extended Functionality: Support for QM/MM multiscale modeling \cite{loco2019towards,loco2021atomistic} and sophisticated enhanced sampling techniques, see section 5.
\end{itemize}
In essence, FeNNix-Bio1 provides the "intelligence" by defining complex molecular interactions, while Tinker-HP serves as the high-performance "engine" necessary to drive these simulations at biological scales. As demonstrated in Figure \ref{fig:perf_Deep-HP}, the integration of FeNNix-Bio1 via the Deep-HP interface yields a performance increase of more than one order of magnitude over state-of-the-art MACE models, bringing foundation-model-driven molecular dynamics into the realm of practical production simulations. Furthermore, due to the strictly short-range nature of the FeNNix-Bio1 architecture, \textbf{these models exhibit superior parallel scalability compared to traditional approaches}, allowing for efficient distributed computing across massive atomic systems. No

\subsection{Chemically-accurate Synthetic Quantum Data\label{sec:ignis} with Classical Methods}
\subsubsection{Ignis: a Large DFT Computations Quantum Database.}Quantum chemistry databases serve as the indispensable "knowledge base" for training neural network potentials, allowing these models to bypass traditional experimental parameterization while maintaining high physical fidelity. Because neural networks are inherently data-driven, their ability to accurately predict molecular behavior depends entirely on the diversity and quality of the underlying quantum mechanical data, such as the energies and forces found in the Ignis or SPICE2 datasets. These databases enable models to learn complex, many-body effects—such as polarization and charge delocalization—that are often simplified or neglected in traditional empirical force fields.
Furthermore, comprehensive databases are critical for ensuring transferability, allowing a foundation model trained on small molecular fragments to successfully extrapolate to large, condensed-phase biological systems like proteins or membranes. By providing a wide range of conformations, including distorted geometries and transition states, these datasets allow neural networks to handle chemical reactivity and bond breaking, tasks that were historically reserved for computationally expensive ab initio methods. Ultimately, as these databases grow through active learning and the inclusion of higher-level quantum data, they push neural networks toward the "chemical accuracy" of $1 \text{ kcal/mol}$, bridging the gap between static structural predictions and dynamic biological reality. \\
In our case, the Ignis database is a comprehensive collection of over 2.2 million DFT synthetic quantum chemistry conformations (energies + forces) specifically curated to ground the FeNNix-Bio1 foundation models in high-level physical reality, see \ref{sec:appendix} for a discussion about the choice of DFT. Its core architecture is built upon the SPICE2(+)-ccECP dataset, which consists of over 2 million SPICE2 configurations and 100,000 ANI-2x samples recomputed using B97M-D3(BJ) Density Functional Theory coupled with correlated consistent effective core potentials (ccECP). To bridge the gap toward complex biological simulations, the database was explicitly augmented with 28,000 configurations of solvated ions—including essential species like $Na^{+}$, $Cl^{-}$, $Mg^{2+}$, and $Zn^{2+}$—alongside 64,080 molecule-water dimers designed to enhance the model's capture of subtle intermolecular interactions. Furthermore, Ignis includes synthetic data for isolated atoms and highly distorted water geometries to enforce correct chemical dissociation limits and improve the overall learning dynamics of intra- and inter-molecular scales. This dataset has already significantly grown, as it is intended to be systematically improvable, evolving through active-learning-driven finetuning processes that expand the foundation model's coverage across relevant areas of chemical space. 
\subsubsection{Beyond DFT: Climbing the Jacob's Ladder up to sCI/DMC.}
\begin{figure}[h!]
    \centering
    \includegraphics[scale=0.45]{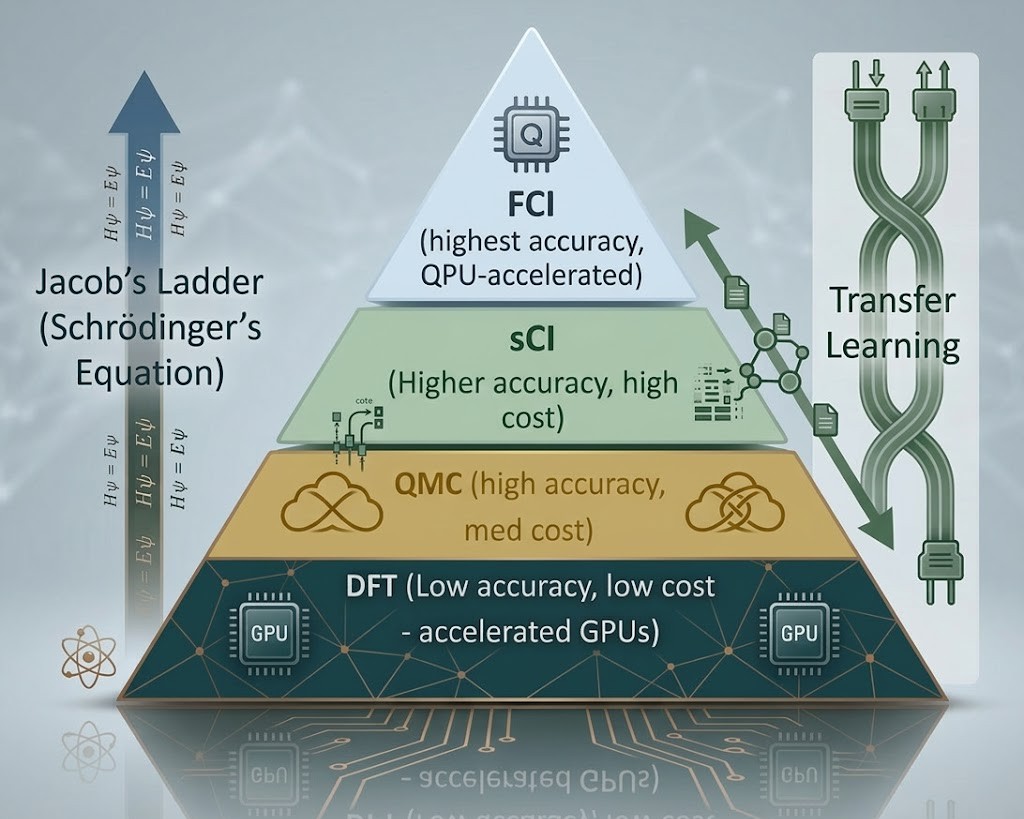}
    \caption{Ignis Database: Jacob's Ladder Strategy for the Quantum Chemistry Computations}
    \label{fig:Jacob-quantum}
\end{figure}

Of course, DFT has limitations in term of accuracy\cite{koch2015chemist}, and towards addressing the discussed Pople diagram (see Fig \ref{fig:pople}), some of us proposed a "Jacob's Ladder" strategy\cite{benali2025}, see Figure \ref{fig:Jacob-quantum}, to generate high-accuracy synthetic datasets (energies and forces) by leveraging multiple layers of increasingly complex quantum chemistry methods. This hierarchical approach begins with the initial Ignis theory level, i.e. DFT, as the foundation layer due to its lower computational cost, which is then refined by higher-accuracy methods. The ladder ascends to Diffusion Monte Carlo, which provides high accuracy at a medium cost, and finally to Selected Configuration Interaction (sCI) combined with DMC for the highest fidelity. By using massively GPU-accelerated software at each rung, we aim to approach "chemical accuracy" (errors below 1 kcal/mol) in large number of cases. It is important to note that obtaining Diffusion Monte Carlo forces is a crucial yet historically difficult component of quantum chemistry, traditionally hindered by the fixed-node approximation and the stochastic nature of the algorithm. To overcome these hurdles at the exascale, we implemented a highly optimized multi-level batching approach that groups walkers and atomic centers to maximize GPU utilization on a supercomputer \cite{benali2025}. This strategy, utilizing a Zero-Variance Zero-Bias force estimator \cite{assaraf2003zero}, achieved sustained performances enabling to perform a large number of computations. Because DMC operates directly in real space, these forces are free from basis set incompleteness errors, effectively reaching the complete basis-set limit.  In that context, sCI-DMC runs are possible as sCI can provide improved initial nodal structures to further improve DMC energies and forces. Despite being approximately 100 times more computationally intensive than energy evaluations, these accurate forces allows for the refinement of the FeNNix-Biol foundation model through transfer/multi-level learning, correcting DFT-level inaccuracies to approach further near-exact quantum precision for large-scale biological simulations.

\begin{table}[h!]
    \centering
    \caption{Summary of the 2025 status for the Ignis Database}
    \label{tab:fennix_database_summary}
    \begin{tabularx}{\textwidth}{l >{\raggedright\arraybackslash}X >{\raggedright\arraybackslash}X}
        \toprule
        \textbf{Dataset Component} & \textbf{Description \& Scope} & \textbf{Quantum Level of Theory} \\
        \midrule
        \textbf{Core Base Set} & $\sim$2.1 million molecular conformations including SPICE2 and 100K ANI-2X configurations selected via active learning. & \textbf{DFT:} $\omega$B97M-D3BJ with ccECP and aug-cc-pVTZ basis sets. \\
        \addlinespace
        \textbf{Solvated Ions} & 28,000 conformations of biologically relevant ions (e.g., Na, Cl, Zn) in varying water cluster sizes (2--10 molecules). & \textbf{DFT:} $\omega$B97M-D3BJ-ccECP. Finetuned with AMOEBA-03 force predictions for monovalent ions. \\
        \addlinespace
        \textbf{Intermolecular Data} & 64,080 molecule-water dimers sampled from the Freesolv database to capture subtle solvation energies. & \textbf{DFT:} Interaction energies computed as the difference between dimer and monomer sum. \\
        \addlinespace
        \textbf{Physical Limits} & Synthetic data for isolated atoms and highly distorted water molecules to ensure correct dissociation behavior. & \textbf{DFT:} Adjusted to match the formation energy of the chosen functional. \\
        \addlinespace
        \textbf{High-Accuracy Subset} & $\sim$40,000 configurations (20,251 with full forces) extracted for ``beyond DFT'' refinement. & \textbf{QMC:} Diffusion Monte Carlo (DMC) operating in the infinite-basis set limit. \\
        \addlinespace
        \textbf{Strong Correlation} & 2,000 configurations specifically targeting significant electronic correlation effects. & \textbf{Multideterminant QMC:} DMC with Selected-CI wavefunctions (1.4M determinants). \\
        \textbf{Near-exact Quantum Reference} & Hundreds of data points converged at the CBS limit to complement sCI/DMC. & \textbf{Quantum Computing:} ADAPT-VQE/DBBSC computations obtained via GPU-accelerated quantum Hyperion emulators. \\
        \bottomrule
          \end{tabularx}
\end{table}

\subsubsection{Software Base: Quantum Chemistry Packages}
To perform these computations, we leverage inhouse GPU-accelerated versions of open-source codes such as PySCF\cite{pyscf} for DFT computations and QMCPack \cite{QMCPACK}. Note that for DMC, our version includes fast generation of DMC forces \cite{benali2025}. We developed our own, GPU-accelerated, massively parallel sCI package\cite{gasperich26}. The essential sCI procedure is described below (see \ref{par:sCI}), and our code builds upon this by incorporating the latest in improved sCI methods and parallel scaling strategies. This set of tools allows us to leverage state-of-the-art algorithms and HPC hardware to push the limit of what is feasible using classical methods.
\subsection{Near-Schrödinger quality Quantum Chemistry Databases with Quantum Computing\label{sec:database}}
\subsubsection{Incorporation High-resolution Quantum Algorithms Results into the Ignis Database}
While the discussed DFT/DMC/sCI strategy is pushing the limits of classical hardware \cite{benali2025}, reaching accuracy beyond these methods is currently hindered by the "exponential wall" of classical computation. However, the advent of FTQC and logical qubits offer a path to overcome this barrier, potentially providing the polynomial speedups necessary for exact, Schrödinger-like Full-CI/CBS computations. In anticipation of this shift, the Ignis dataset strategy already incorporates these quantum algorithms; currently, our Hyperion GPU-accelerated quantum emulator is being utilized to generate hundreds of datapoints converged at the CBS chemical accuracy \cite{traore2024shortcut}. We particularly leverage a specific hybrid quantum-classical framework present in Hyperion and designed to \textbf{achieve high-precision chemical results while drastically reducing the number of qubits required for calculations}\cite{traore2024shortcut}. By integrating 1) Density-Based Basis-Set Correction (DBBSC), a density-based correlation potential to approximate the basis set error of the electron-electron interaction and 2) System Adapted Basis Set (SABS), a  molecular orbital selection strategy to adapt the size of the basis to the qubit budget, with quantum algorithms like ADAPT-VQE, we provide a \textbf{"shortcut" to the complete-basis-set limit}, addressing the "qubit budget bottleneck" that typically limits quantum chemistry to small systems and minimal basis sets. 

\subsubsection{Shortcut to Chemical Accurate Quantum Computing; Up to 10-fold reduction in
qubits requirements with DBBSC} \label{subsec:DBBSC}
\begin{figure}[h!]
    \centering
    \includegraphics[scale=0.45]{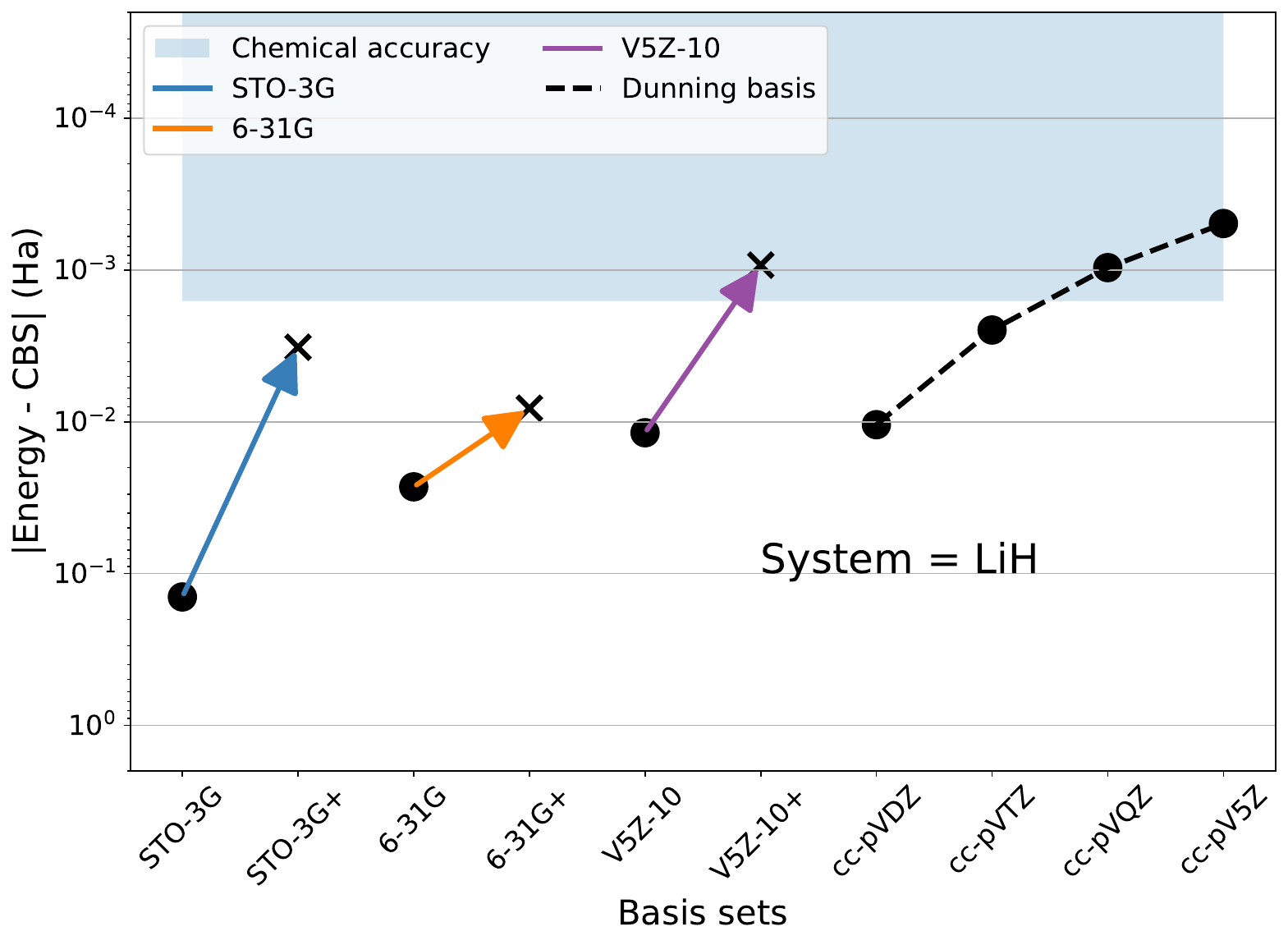}
    \caption{Shortcut to chemically accurate quantum computing via density-based basis-set correction: ground-state energy errors with respect to the extrapolated CBS limit for the LiH molecule. Dot markers correspond to energies without basis-set corrections. Cross markers correspond to the corrected energies. On the x-axis, the labels with a $+$ sign symbolize basis-set corrected values. The blue box corresponds to a range of 1.6 mHa around the extrapolated CBS limit. Full convergence is obtained with only 28 qubits providing a 10-fold reduction in qubits requirements. Data from ref. \cite{traore2024shortcut}.}
    \label{fig:DBBSC}
\end{figure}
The density-based basis set correction comes from a reformulation of the Kohn-Sham formulation of DFT for multi-determinant wave-function. In this picture, the variational problem is written as follow:
\begin{equation}\label{DBBSC}
    E_{0}^{\mathcal{B}} = \min_{\Psi^\mathcal{B}} \left\{ \bra{\Psi^{\mathcal{B}}} \hat{T} + \hat{V}_{\text{ne}} + \hat{W}_{\text{ee}}\ket{\Psi^{\mathcal{B}}} + \bar{E}^{\mathcal{B}} [n_{\Psi_{\mathcal{B}}}]\right\},
\end{equation}
which can be compared to Eq. \eqref{DFT}. In DBBSC, the electron-electron interaction in the basis-set $\mathcal{B}$ is computed in the first term and the density functional approximation is build to recover the basis-set error in the electron-electron interaction. One can find some of this approximations in Ref. \cite{GinTraPraTou2021} and references therein. In order to solve \eqref{DBBSC}, one can formulate the minimization problem as the following eigenvalue equations:
\begin{equation}\label{EigDBBSC}
    \hat{T}^{\mathcal{B}} + \hat{V}_{\text{ne}}^{\mathcal{B}} + \hat{W}_{\text{ee}}^{\mathcal{B}} + \hat{\bar{V}}^{\mathcal{B}}[n_{\Psi^\mathcal{B}}] \ket{\Psi^{\mathcal{B}}} = \mathcal{E}^{B}_{0} \ket{\Psi^{\mathcal{B}}},
\end{equation}
where $\hat{\bar{V}}^{\mathcal{B}}[n^{\mathcal{B}}]$ is a one-electron effective potential computed from the derivative of $\bar{E}^{\mathcal{B}}[n_{\Psi_{\mathcal{B}}}]$ (details in Ref. \cite{GinTraPraTou2021}, also the link between the eigenvalue $\mathcal{E}^\mathcal{B}_0$, and the physical ground-state energy $E_{0}^{\mathcal{B}}$ for which we do not need to add details here). Finally, Eq. \eqref{EigDBBSC} is a self-consistent equation. From the solutions, one can notice the small addition in the value of the ground state energy with respect to what could be obtain with a standard sCI because we are still working on the correlation energy. From this, one can picture the effective potential $\hat{\bar{V}^{\mathcal{B}}}[n_{\Psi_{\mathcal{B}}}]$ as a pertubation potential. This remark justify the second way to implement the DBBSC, using an a posteriori correction that adds classical density-functional and Hartree-Fock (HF) corrections to quantum results:
\begin{equation}
    E_{0}^{\mathcal{B}} \simeq E^{\mathcal{B}} + \bar{E}^{\mathcal{B}} [n_{\Phi^{\mathcal{B}}}],
\end{equation}
where $\Phi^{\mathcal{B}}$ is the solution of the Hartree-Fock equation.
The methodology utilizes two main strategies: an a posteriori correction that adds classical density-functional and Hartree-Fock (HF) corrections to quantum results, and a self-consistent scheme that iteratively updates the Hamiltonian based on electronic density for which we summarize some results in Tab. \ref{tab:dbbsc_summary2}. 

\textbf{SABS}: In the SABS method, one define a strategy to select a subset of $M$ atomic orbitals from a basis set $\mathcal{B} = \left\{  \chi_\mu \right\}_{\mu=1, N_{\text{bas}}}$ such that the overlap between the density in the basis $\mathcal{B}$ and the new reduced basis $\tilde{\mathcal{B}}$ is maximized\cite{traore2024shortcut}.
Through the use of System-Adapted Basis Sets (SABS)\cite{traore2024shortcut} and GPU-accelerated emulation, the study demonstrated that "chemical accuracy" (1.6 mHa) for molecules like $H_2$, LiH, $H_2O$, and $N_2$ can be reached using as few as 10 to 32 qubits—results that would otherwise necessitate hundreds of logical qubits. Figure \ref{fig:DBBSC} highlights the effects of the correction for the LiH molecule that can be fully converged at the absolute chemical accuracy using only 28 qubits, avoiding therefore the brute force needs of more than 290 qubits while providing a ten-fold reduction in qubits requirements. In average, \textbf{qubit reduction range from 4 to 10 depending on the electronic correlation}. Table \ref{tab:dbbsc_summary2} provides other examples including a comparison for N2 with IBM recent computations. 
\paragraph{Comparison with a quantum centric supercomputing approach} Indeed, in their initial state-of-the art hybrid quantum centric supercomputing work\cite{robledo2025chemistry}, IBM also simulated the dissociation of the $N_2$ molecule using Sample-based Quantum Diagonalization (SQD). utilizing a cc-pVDZ (double-zeta) basis set, their approach required a 'brute force' mapping of 58 qubits and over 10,500 gates to produce a qualitatively correct potential energy surface. However, achieving results consistent with reference methods required massive classical computational overhead, involving hundreds of Fugaku supercomputer nodes and yielding an error significantly larger than the chemical accuracy threshold, see Table \ref{tab:dbbsc_summary2}. In contrast, our hybrid DBBSC methodology achieves a higher cc-pVTZ (triple-zeta) quality for $N_2$ using only 32 qubits on 16 H100 nodes at the time using Hyperion-1. This demonstrates the relevance of emulation strategy that allows to explore new algorithms. By drastically reducing qubit requirements while maintaining high-fidelity accuracy, our approach demonstrates a more scalable path toward chemically accurate quantum simulations on near-term hardware. For these systems, Table \ref {tab:cnot} also highlights the capabilities of Hyperion to deal with large circuit encompassing thousands of CNOTS gates.

This hybrid methodology is designed for implementation on Quantum Processor Units (QPUs) and will grow in significance as logical qubit counts increase. By effectively bridging the gap between current hardware limitations and the stringent accuracy requirements for predictive drug design, it enables high-fidelity simulations of complex biological systems. Step by step, we are replacing the Ignis database classical data by their quantum computing counterpart, progressively increasing its quality while improving the chemical space description of our foundation models. Leveraging various forms of learning, \textquote{beyond} DFT FeNNix-Bio potentials are currently under development, see \cite{benali2025} for discussion. 

\begin{table}[h!]
    \centering
    \caption{Comparison of Standard ADAPT-VQE vs ADAPT-VQE + DBBSC}
    \label{tab:dbbsc_comparison}
    \begin{tabularx}{\textwidth}{l >{\raggedright\arraybackslash}X >{\raggedright\arraybackslash}X}
        \toprule
        \textbf{Feature} & \textbf{Standard ADAPT-VQE} & \textbf{ADAPT-VQE + DBBSC} \\
        \midrule
        \textbf{Primary Goal} & Minimize energy within a fixed, finite basis set. & Approach the Complete-Basis-Set (CBS) limit using small basis sets. \\
        \addlinespace
        \textbf{Basis Set Requirements} & Requires large basis sets (e.g., cc-pVTZ) for high-accuracy predictions. & Achieves high accuracy using minimal basis sets (e.g., STO-3G, 6-31G). \\
        \addlinespace
        \textbf{Qubit Resources} & Qubit count scales linearly with the large basis size, often requiring hundreds of qubits. & Drastically reduced; enables accurate simulations with as few as 10--30 qubits. \\
        \addlinespace
        \textbf{Correction Strategy} & None; results are limited by basis-set truncation errors. & Hybrid: Combines quantum solver with classical density-functional and HF corrections. \\
        \addlinespace
        \textbf{Implementation} & Purely variational quantum algorithm. & Can be applied \textit{a posteriori} (Strategy 1) or self-consistently (Strategy 2). \\
        \addlinespace
        \textbf{Convergence Rate} & Slow convergence with respect to basis-set size. & Accelerated convergence to the CBS limit through self-consistent improvement. \\
        \bottomrule
    \end{tabularx}
\end{table}

\begin{table*}[h!] \label{Tab:dbbsc_summary}
\centering
\caption{Comparison of ground-state energy errors (in mHa) and qubit requirements for noiseless ADAPT-VQE. For H$_2$O, LiH, H$_2$ and N$_2$ molecules, we compare standard ADAPT-VQE using the largest basis set that can be fitted in maximum 30 qubit with the use of our best adaptive basis sets and DBBSC. Last column presents the error with the chemical accuracy threshold ($\sim 1.6$ mHa) as the target reference. For complete results, see Ref. \cite{traore2024shortcut}}
\label{tab:dbbsc_summary2}
\begin{tabular}{llccr}
\hline
\textbf{System} & \textbf{Method / Basis} & \textbf{N$_{qubits}$} & \textbf{Energy (Ha)} & \textbf{$\Delta$ vs CBS (mHa)} \\ \hline
\textbf{H$_2$O} & 6-31G (ADAPT-VQE) & 24 & -76.11989 &  258 \\
& V5Z-11 with SC(ADAPT+PBE)+$\Delta$HF** & 30 & -76.33570 & 42.42 \\
& cc-pVTZ (FCI) & 114 & -76.33250 & 46 \\
& FCI (CBS) & 400\textsuperscript{†} & -76.37812 & \\ \hline
\textbf{LiH} & 6-31G (ADAPT-VQE) & 20  & -7.99800  & 27 \\
& V5Z-10 with SC(ADAPT+PBE)+$\Delta$HF* & 28 & -8.02540 & 0.58 \\
& FCI (CBS) & 290\textsuperscript{†} & -8.02482 &  \\ \hline
\textbf{H$_2$} & cc-pVDZ (ADAPT-VQE) & 20 & -1.16275 & 10 \\
& V5Z-8 with SC(ADAPT+PBE)+$\Delta$HF* & 24 & -1.17320 & 0.55 \\
& FCI (CBS) & 220\textsuperscript{†} & -1.17265 &  \\ \hline
\textbf{N$_2$} & STO-3G (ADAPT-VQE) & 16 & -107.65251 & 1772 \\
& V5Z-11 with SC(ADAPT+PBE)+ $\Delta$HF** & 32 & -109.37281 & 52 \\
& cc-pVTZ (FCI) & 116 & -109.37527 & 50 \\
 & FCI(CBS) & 360\textsuperscript{†} & -109.42498 & \\
\end{tabular}
\begin{flushleft}
\small
* Reaches CBS chemical accuracy. \\
** Reaches cc-pVTZ quality.\\
***Sample-based quantum diagonalization (SQD) from IBM on the Fugaku supercomputer coupled to a Heron QPU \cite{robledo2025chemistry}\\
\textsuperscript{†} Number of qubits for the cc-pV5Z basis required to approach the CBS limit in a standard classical FCI scheme.
\end{flushleft}
\end{table*}

\begin{table}[h!]
\centering
\caption{CNOT-gate counts for qubit and qubit-excitation-based (QEB) operator pools for ADAPT-VQE/DBBSC from ref \cite{traore2024shortcut}}
\begin{tabular}{lccc}
\hline
\textbf{Molecule} & \textbf{$N_{qubits}$} & \textbf{$N_{CNOT}$ (QEB)} & \textbf{$N_{CNOT}$ (qubit)}  \\ \hline
\textbf{$H_2O$} & & &  \\
6-31G & 24 & 12657 & 5862   \\
V5Z-11 & 30 & 12647 & 5858   \\ \hline
\textbf{LiH} & & &  \\
6-31G & 20 & 511 & 242   \\
V5Z-10 & 28 & 1083 & 506  \\ \hline
\textbf{$H_2$} & & &  \\
cc-pVDZ & 20 & 233 & 110  \\
V5Z-8 & 24 & 395 & 186   \\ \hline
\textbf{$N_2$} & & &  \\
STO-3G & 16 & 4868 & 2256   \\
V5Z-11 & 32 & 11718 & 5420   \\ \hline
\end{tabular}
\end{table}\label{tab:cnot}

\section{Modern Enhanced Sampling Techniques: From Structural Dynamics to Quantum-Accurate Properties}

\subsection{Deriving Macroscopic Observables from Molecular Simulations}

In the industrial world, molecular simulation acts as a computational microscope. The vast majority of information we derive comes from properties computed as statistical averages~\cite{frenkel2023understanding}. These are based on the Boltzmann and Gibbs distributions, providing information about how atoms behave at specific temperatures and pressures. Such averages can predict the conductivity of battery electrolytes, the structural integrity of new materials or the critical distances between a drug ligand and its protein target~\cite{horton2025accelerated,bedrov2019molecular,de2016role}.
As will be discussed in Subsection~\ref{subsec:qMCMC}, quantum computers are particularly well-suited to accelerate such averages under Gibbs distributions. In fields like drug discovery, we also leverage more involved quantities, calculating complex metrics like absolute free energies of binding~\cite{aldeghi2017predictions} of a drug candidate to a disease-related target, and solubility~\cite{savjani2012drug} of a compound. Computing these properties through quantum-accurate models ensures that our findings aren't mere estimates, but are instead directly inherited from the first principles of physics. 

\subsubsection{The Timescale Challenge: Balancing Accuracy and Dynamics}

Yet, a fundamental conflict remains between accuracy and time; the more accurate models are often constrained by the vast timescale gap of molecular processes. Machine Learning Potentials (MLPs) trained on massive quantum datasets have dramatically reduced the cost of computing the energy of a single configuration. But capturing these static snapshots is not the same as capturing the dynamics of moving between states. To converge these properties, we need simulation timescales that remain out of reach for today’s infrastructure~\cite{bernardi2015enhanced}. For example, simulating a complete ligand binding process to a protein requires microseconds or milliseconds of molecular dynamics~\cite{pan2017quantitative}. Without a method to accelerate the dynamics, the timescale required to connect these static snapshots becomes prohibitive, leaving no choice but to use classical models that fail to capture the quantum essence of the system.

Some researchers try to skip the simulation entirely by using purely data driven techniques leveraging experimental macroscopic values~\cite{Blotz1,passaro2025boltz}, but these \textquote{black box} methods often fail to generalize~\cite{bret2025assessing,adury2026early,vskrinjar2025have} and lack explainability. In drug design, physics-based explainability is essential; knowing exactly how a single water molecule moves in a binding pocket can be the difference between a failed trial and a potent drug~\cite{ladbury1996just,andreev2021addressing}.

\subsection{Overcoming Energetic Bottlenecks with Enhanced Sampling}

\begin{figure}[h!]
	\centering
	\includegraphics[width=0.7\textwidth]{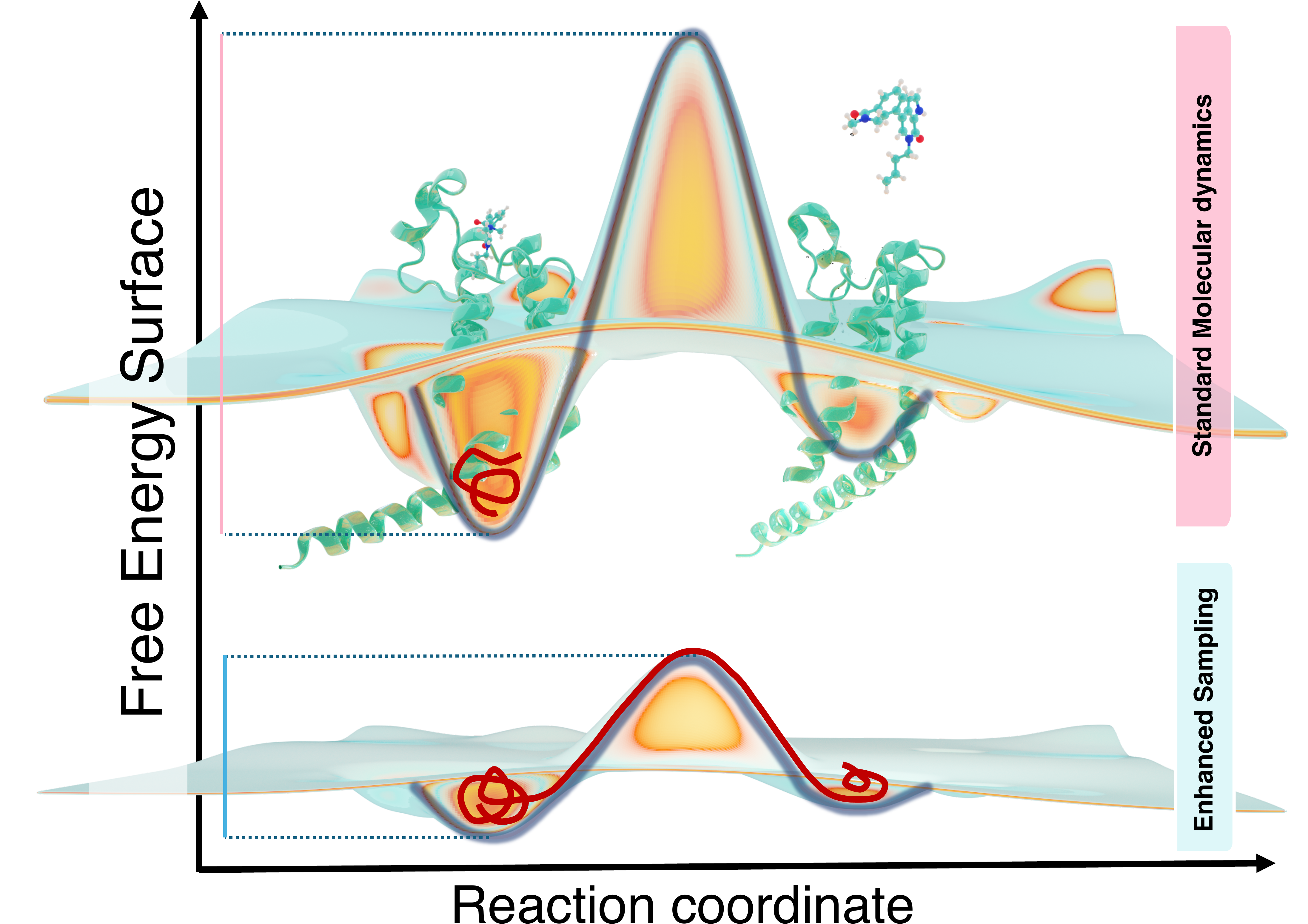
}   
\caption{\textbf{Free energy surface (FES) profiles for a protein-ligand binding event}. The upper panel illustrates the rugged transition state barrier and the high energy required to move between the unbound and bound states. In standard Molecular Dynamics (MD), these high barriers lead to kinetic trapping in local minima. The lower panel demonstrates the application of enhanced sampling techniques, which effectively flatten the free energy landscape. This reduction in barrier height allows the simulation to overcome energetic bottlenecks and explore both states within a feasible computational timescale.
}
	\label{fig:Enhanced_sampling}
\end{figure}

This is where \textbf{Enhanced Sampling becomes indispensable}. By applying a strategic bias to the simulation~\cite{henin2022enhanced}, allowing the overcoming of energetic barriers (see Fig.~\ref{fig:Enhanced_sampling}), we bypass the need to wait for rare events to occur spontaneously, dramatically compressing the time required to sample the relevant physics . Our team developed two specific \textquote{accelerators}: \textbf{Lambda-ABF-OPES}~\cite{ansari2025lambda,lambdaabf} to calculate the \textbf{absolute binding free energy (ABFE)} of a ligand to a macromolecular target and \textbf{Dual-LAO}~\cite{ansari2025fast} to calculate \textbf{the relative binding free energy (RBFE)} between molecules for the same target, a key aspect of the hit optimisation stage of drug discovery. These methods use an "alchemical path"—essentially turning the ligand into a \textquote{ghost} to decouple it from its target. By combining a dynamical alchemical variable ($\lambda$), Adaptive Biasing Force (ABF)~\cite{darve2001calculating,darve2008adaptive,comer2015adaptive}, On-the-fly Probability Enhanced Sampling (OPES)~\cite{Invernizzi2020,Invernizzi2022}, and Thermodynamic Integration (TI)~\cite{straatsma1991multiconfiguration}.

\subsubsection{Lambda-ABF-OPES: Maximizing Convergence and Computational Efficiency}

ABFE calculations are traditionally difficult to converge due to the high dimensionality of the phase space and the significant energy barriers involved. To evaluate the efficiency of our proposed method, we applied it to calculate the ABFE of 11 ligand-protein complexes of BRD4~\cite{ansari2025lambda}, which serves as a realistic and widely recognized benchmark for such alchemical methods. 

\paragraph{Accuracy and speedup} As demonstrated in Ref.~\citenum{ansari2025lambda}, our Lambda-ABF-OPES (LAO) protocol achieves up to a \textbf{9-fold increase in sampling efficiency over the standard L-ABF implementation}~\cite{lambdaabf}. We further compared the total simulation cost of LAO against traditional fixed-$\lambda$ (FL) alchemical methods and modern physics-based methods such as OneOPES (OO)~\cite{rizzi2023oneopes}. For the BRD4 system, our results indicate that LAO is approximately 4.4 times faster than the FL method~\cite{aldeghi2016accurate} and \textbf{between 20 to 58 times faster than OO}~\cite{karrenbrock2024absolute} (see Table~\ref{tab:comprehensive_comparison}). Crucially, this significant gain in speed is achieved while maintaining high accuracy. This confirms that the accelerated convergence of LAO does not compromise the precision required for reliable drug discovery applications.

\paragraph{Computational time saving} As shown in Table~\ref{tab:total_efficiency}, the accelerated convergence of the LAO method leads to substantial savings in computational resources. For this 11 ligand-complex system alone, LAO results in a total saving of approximately 21,384 ns of simulation time compared to the FL method and over 51,264 ns compared to the OO protocol. This reduction in the total computational footprint allows for significantly faster lead optimization cycles in drug discovery without sacrificing the accuracy of the binding affinity predictions.

\begin{table}[h!]
    \centering
           \caption{\textbf{Binding free energy ($\Delta G$, kcal/mol) and computational cost (ns) comparison} across LAO, FL, and OO methods for BRD4 ligands.}
            \label{tab:comprehensive_comparison}
    \scriptsize
    \renewcommand{\arraystretch}{1.2}
    \setlength{\tabcolsep}{2pt} 
    \begin{tabular}{l ccc cccc cc}
        \hline
        \textbf{Cpd.} & $\Delta G_{EXP}$ & $\Delta G_{LAO}$ & $\Delta G_{FL}$ & $\Delta G_{OO}$ & \multicolumn{3}{c}{\textbf{Total Time (ns)}} & \multicolumn{2}{c}{\textbf{LAO Speedup}} \\
        \cline{6-10}
        & & (Ours) & & & LAO\textsuperscript{a} & FL\textsuperscript{b} & OO\textsuperscript{c} & vs. FL & vs. OO \\
        \hline
        L1  & $-9.8 \pm 0.1$ & $-10.56 \pm 0.46$ & $-10.4 \pm 0.6$ & $-10.5 \pm 0.3$ & 192 & 840 & 4000   & \textbf{4.4x} & \textbf{20.8x} \\ 
        L2  & $-9.6 \pm 0.1$ & $-8.26 \pm 0.22$  & $-9.5 \pm 0.4$  & $-7.8 \pm 0.2$  & 192 & 840 & 5600 & \textbf{4.4x} & \textbf{29.2x} \\
        L3  & $-9.0 \pm 0.1$ & $-9.19 \pm 0.30$  & $-9.2 \pm 0.5$  & $-8.8 \pm 0.4$  & 192 & 840 & 4000  & \textbf{4.4x} & \textbf{20.8x} \\ 
        L4  & $-8.9 \pm 0.1$ & $-10.13 \pm 0.15$ & $-9.4 \pm 0.8$  & $-9.5 \pm 0.5$  & 192 & 840 & 4800 & \textbf{4.4x} & \textbf{25.0x} \\ 
        L5  & $-8.8 \pm 0.1$ & $-8.27 \pm 0.52$  & $-8.6 \pm 0.3$  & $-8.4 \pm 0.5$  & 192 & 840 & 4000   & \textbf{4.4x} & \textbf{20.8x} \\ 
        L6  & $-8.2 \pm 0.1$ & $-6.75 \pm 0.30$  & $-9.9 \pm 0.8$  & $-10.0 \pm 0.3$ & 192 & 840 & 11200 & \textbf{4.4x} & \textbf{58.3x} \\ 
        L7  & $-7.8 \pm 0.1$ & $-6.42 \pm 0.65$  & $-5.9 \pm 0.5$  & $-7.9 \pm 0.2$  & 192 & 840 & 4800 & \textbf{4.4x} & \textbf{25.0x} \\
        L8  & $-7.4 \pm 0.1$ & $-5.46 \pm 0.53$  & $-7.8 \pm 0.3$  & $-6.9 \pm 0.3$  & 192 & 840 & 4000  & \textbf{4.4x} & \textbf{20.8x} \\ 
        L9  & $-7.3 \pm 0.1$ & $-7.57 \pm 0.12$  & $-7.7 \pm 0.4$  & $-8.3 \pm 0.3$  & 192 & 840 & 4000  & \textbf{4.4x} & \textbf{20.8x} \\ 
        L10 & $-6.3 \pm 0.0$ & $-7.05 \pm 0.33$  & $-5.9 \pm 0.2$  & $-5.9 \pm 0.2$  & 192 & 840 & 5600 & \textbf{4.4x} & \textbf{29.2x} \\ 
        L11 & $-5.6$         & $-4.98 \pm 0.58$  & $-5.4 \pm 0.2$  & $-6.7 \pm 0.5$  & 192 & 840 & 5600 & \textbf{4.4x} & \textbf{29.2x} \\ 
        \hline
    \end{tabular}

        \vspace{4pt}
    \begin{flushleft}
        \footnotesize 
        \textsuperscript{a} Simulation time per ligand: 36~ns $\times$ 4 walkers for the complex phase $+$ 12~ns $times$ 4 walkers for the solvent phase \\
        \textsuperscript{b} Simulation time per ligand calculated as 73 fixed-$\lambda$ windows (42 for complex, 31 for solvent phase) of 11.5~ns each (0.5~ns NVT and 1.0~ns NPT equilibration, 10~ns production)  \\
        \textsuperscript{c} Simulations using 8 walkers, with total simulation time varying ranging between $4000$ and $11200$~ns \\
    \end{flushleft}
\end{table}

\begin{table}[h!]
    \centering
    \caption{Total computational cost comparison for the 11-ligand BRD4 set. Total time for LAO and FL includes 3 independent repeats per ligand to account for statistical convergence. OO represents the sum of the individual production runs required for the 11-ligand set.}
    \label{tab:total_efficiency}
    \footnotesize
    \renewcommand{\arraystretch}{1.1}
    \setlength{\tabcolsep}{8pt}
    \begin{tabular}{lccc}
        \hline
        \textbf{Method} & \textbf{Calculations} & \textbf{Total Sim. (ns)} & \textbf{Net Savings} \\
        \hline
        LAO (Ours) & 11 Ligands $\times$ 3 Repeats & 6,336 & -- \\
        Fixed-$\lambda$ (FL) & 11 Ligands $\times$ 3 Repeats & 27,720 & \textbf{21,384 ns saved} \\
        oneOPES (OO) & 11 Ligands (Single Runs) & 57,600 & \textbf{51,264 ns saved} \\
        \hline
    \end{tabular}
\end{table}

\subsubsection{Dual-LAO: Navigating Complex Topological Transformations for High-Throughput Relative Binding Free Energies}

RBFE calculations play a critical role in the hit-to-lead optimization phase of structure-based drug discovery; however, their practical utility is frequently constrained by the substantial computational overhead required to achieve converged results, particularly for chemically diverse transformations. To demonstrate the versatility of our approach, we applied the Dual-LAO method to a set of four highly challenging benchmark systems: PWWP1 (contains charge-changing ionic transitions), BRD4 (buried water displacement), P38 (fragment-like large shifts), and CHK1 (scaffold-hopping via ring-breaking). We compared our performance against the state-of-the-art FEP+ protocol commercialized by Schrödinger\cite{ross2023maximal}, which typically requires extensive sampling across numerous $\lambda$-windows to ensure accuracy. 

\paragraph{Accuracy and speedup} As summarized in Table~\ref{tab:RBFE_Comparison_Final}, \textbf{Dual-LAO consistently outperforms the FEP+ baseline in both precision and efficiency}. Our method achieves significantly lower RMSE values—most notably in the PWWP1 and BRD4 systems—while requiring a fraction of the sampling time. Specifically, Dual-LAO provides a speedup of up to 30-fold per edge (averaging 24 ns per transformation) compared to the 480--960 ns required by standard FEP+ protocols. These results indicate that Dual-LAO successfully navigates complex topological changes and solvent rearrangements without the need for the dense $\lambda$-staging traditionally required for convergence. The aggregate impact on computational resources is even more pronounced when considering the entire benchmark set. 

\paragraph{Computational time saving} As detailed in Table~\ref{tab:total_efficiency_dual_top}, across the 56 studied transformations (including three independent repeats to ensure statistical robustness), the Dual-LAO method requires a total of only 4,500 ns of simulation time. In contrast, the standard FEP+ protocol necessitates over 87,000 ns to complete the same set. This results in a net saving of approximately 82,7 $\mu$s of simulation time. Such a massive reduction in the computational footprint facilitates significantly higher throughput in lead optimization, enabling the rapid exploration of vast chemical spaces that were previously cost-prohibitive for high-precision alchemical methods.

\begin{table}[!h]
    \centering
    \caption{\textbf{Performance and Efficiency Benchmark: Dual-LAO vs. FEP+ Method.} Summary of accuracy (RMSE in kcal/mol) and computational cost. Dual-LAO consistently achieves higher accuracy with significantly lower total sampling times compared to standard FEP+ protocol.}
    \label{tab:RBFE_Comparison_Final}
     \footnotesize
    \renewcommand{\arraystretch}{1.1}
    \setlength{\tabcolsep}{8pt}
        \begin{tabular}{llccccc}
            \hline
            \textbf{System} & \textbf{Challenge Type} & \textbf{Method} & \textbf{Sampling/Edge} & \textbf{RMSE} & $R^2$ & Speedup/Edge\\ \hline
            
            \textbf{PWWP1} & Charge-Changing  & \textbf{Dual-LAO} & \textbf{24 ns\textsuperscript{a}} & \textbf{0.46} & \textbf{0.62} & \textbf{30x}\\
             & (Ionic Change) & FEP+ \cite{alibay2022evaluating} & 960 ns\textsuperscript{b}~\cite{ross2023maximal} & 1.14 & 0.79 & -- \\ \hline
            
            \textbf{BRD4} & Buried Water & \textbf{Dual-LAO} & \textbf{24 ns\textsuperscript{a}} & \textbf{0.64} & \textbf{0.76} &  \textbf{15x}\\
            ~ & (Displacement) & FEP+ ~\cite{ross2023maximal} & 480 ns\textsuperscript{c}  & 2.05 & $\approx$ 0 & -- \\ \hline
            
            \textbf{P38} & Fragment-like & \textbf{Dual-LAO} & \textbf{24 ns\textsuperscript{a}} & \textbf{0.82} & \textbf{0.77} & \textbf{15x}\\
             & (Large Shift) & FEP+~\cite{ross2023maximal} & 480 ns\textsuperscript{c} & 0.90 & 0.56 & -- \\ \hline
            
            \textbf{CHK1} & Scaffold-Hopping & \textbf{Dual-LAO} & \textbf{24 ns\textsuperscript{a}} & \textbf{0.58} & \textbf{0.48} &  \textbf{~27x}\\
             & (Ring-Breaking) & FEP+~\cite{ross2023maximal} & 640 ns\textsuperscript{d} & 0.18 & 0.76& --\\ \hline
        \end{tabular}
        \vspace{4pt}
    \begin{flushleft}
        \footnotesize 
        \textsuperscript{a} 4 ns $\times$ 4 walkers complex) + (2 ns $\times$ 4 walkers solvent. \\
        \textsuperscript{b} 24 $\lambda$ windows simulated for 20 ns per window per phase. \\
        \textsuperscript{c} 12 $\lambda$ windows simulated for 20 ns per window per phase. \\
        \textsuperscript{d} 16 $\lambda$ windows simulated for 20 ns per window per phase. \\
        \textbf{Note:} All RMSE values are in kcal/mol. Dual-LAO consistently outperforms benchmarks in both efficiency and precision.
    \end{flushleft}
\end{table}

\begin{table}[h!]
    \centering
    \caption{\textbf{System-Specific Computational Savings with Dual-LAO.} Comparison of cumulative sampling time (ns) required across all 56 transformations. By reducing the sampling per edge to a unified 24 ns, Dual-LAO achieves a total reduction of 82,860 ns (approx. 82,7 $\mu$s).}
    \label{tab:total_efficiency_dual_top}
    \footnotesize
    \renewcommand{\arraystretch}{1.1}
    \setlength{\tabcolsep}{2pt}
    \begin{tabular}{lcccc}
        \hline
        \textbf{System} &\textbf{No. Edges} & \textbf{FEP+ Total (ns)} & \textbf{Dual-LAO Total (ns)} & \textbf{Net Savings (ns)} \\
        \hline
        PWWP1 & 30 $\times$ 3 Rep. & 46,080 & 2160 & \textbf{93,920}  \\
        BRD4 & 11 $\times$ 3 Rep. & 15,840 & 990 & \textbf{14,850}  \\
        P38 & 7 $\times$ 3 Rep. & 10,080 & 630 & \textbf{9,450}  \\
        CHK1 & 8 $\times$ 3 Rep. & 15,360 & 720 & \textbf{14,640} \\
        \hline
       \textbf{Total} & \textbf{56  $\times$ 3 Rep.} & \textbf{86,360} & \textbf{4,500} & \textbf{82,860} \\
        \hline
    \end{tabular}
\end{table}

Ultimately, combining modern enhanced sampling techniques with the latest machine learning potentials makes the computation of quantum-accurate properties a practical reality. This powerful synergy holds the promise of significantly improving the utility of molecular simulations in industrial applications like drug discovery. Moreover, the next section highlights further possible enhancements leveraging Quantum Computing.
    
\subsection{Quantum Speedups for Statistical Mechanics and Molecular Simulation}\label{subsec:qMCMC}

As we have discussed, molecular dynamics ultimately reduce to the computation of statistical averages. More precisely, one is often interested in estimating expectation values with respect to the Gibbs distribution associated with a physical system. In practice, these quantities are typically computed using Markov Chain Monte Carlo methods, where a Markov chain is used to generate samples from the target distribution. Quantum computers are particularly well suited to accelerate such tasks.

\subsubsection{End-to-end quadratic speedups for equilibrium sampling and observable estimation}

In practice, Gibbs measures evaluations as described in the above sections can be accessed through Markov Chain Monte Carlo (MCMC), whose efficiency is governed by the spectral gap and mixing properties of the underlying Markov process.
In complex energy landscapes, mixing times can be prohibitively long, and achieving small statistical error requires a large number of samples. 

Quantum algorithms offer two distinct avenues for acceleration. First, given access to samples, amplitude estimation provides a quadratic improvement in precision over classical Monte Carlo sampling~\cite{Brassard_2002}. This replaces the classical $\mathcal O(1/\epsilon^2)$ sampling complexity for error $\epsilon$ with $\mathcal O(1/\epsilon)$. Second, quantum walks associated with reversible Markov chains exhibit a quadratic improvement in spectral gap dependence~\cite{Szegedy}, suggesting a quadratic reduction of the asymptotic mixing rate from $\mathcal O(1/\delta)$ to $\mathcal O\left(1/\sqrt{\delta}\right)$, where $\delta$ is the classical gap. Together, these ingredients suggest \textbf{the possibility of end-to-end quadratic speedups for quantum equilibrium sampling and observable estimation}.

However, asymptotic improvements in mixing or precision do not automatically translate into practical speedups. Any realistic advantage requires a fully end-to-end implementation that includes state preparation, controlled reflections, amplitude amplification, and ultimately fault-tolerant error correction. In particular, quantum walk speedups presuppose nontrivial initial state preparation. Circumventing this requirement is known as the \textbf{quantum walk mixing problem}~\cite{Richter_2007}. 

A complementary approach is provided by continuous-time quantum Gibbs samplers based on physically motivated Lindbladians, as developed in~\cite{Chen2025EfficientQT, single_ancilla_gsprep}. These algorithms simulate open quantum system dynamics whose fixed point is the Gibbs state. Crucially, because the evolution contracts arbitrary initial states toward equilibrium, this framework requires far less a priori knowledge of the target thermal state and alleviates the initial state bottleneck that limits previously mentioned quantum walk-based methods. However, it does not provide a clear spectral gap amplification mechanism.

Thus, quantum algorithms provide compelling asymptotic evidence for accelerated Gibbs sampling, via both quadratic precision improvements and gap amplification. Achieving genuine practical speedups hinges on overcoming the initial state preparation barrier and integrating these primitives into resource-efficient, fault-tolerant architectures~\cite{Claudon_2025, Claudon_2024, claudon2026quantumcircuitsmetropolishastingsalgorithm, claudon2025simplealgorithmreflecteigenspaces}. Resolving these challenges would bring quantum advantage in statistical mechanics and its applications significantly closer to reality. In other words, quantum algorithms promise to accelerate drug discovery by facilitating binding free energy estimations and the study of protein-ligand complexes, which Section~\ref{sec:FM} explains are central to the field. Moreover, some of us have shown that current quantum devices such as \textbf{Quantinuum's H2 and 98-qubit Helios} can already achieve small quantum Markov Chain Monte Carlo simulations~\cite{qp_cqt_quantinuum}, opening a new route for quantum-accelerated atomistic molecular simulations. Let us detailed these recent experiments.
\subsubsection{Experimental Realization of the Markov Chain Monte Carlo Algorithm on a Quantum Computer}

The experiments reported in this section were carried out on the \textbf{H2 and Helios Quantinuum trapped-ion quantum computers}. The objective of the experiments is \textbf{to demonstrate, on present-day Noisy Intermediate-Scale Quantum (NISQ) hardware, a minimal instance of quantum Markov Chain Monte Carlo workflow}. The present experiments combine a quantum encoding of a simple Markov chain with amplitude estimation in order to illustrate this approach on real devices.

Table~\ref{tab:lcu_monte_carlo} reports the experimental results obtained when using the prepared stationary state $\pi$ to estimate the expectation value $E_\pi(f)=0.5$ of a function $f$ through Quantum Amplitude Estimation. In the context of this simple two-state Markov chain, the experiment provides a minimal demonstration of a quantum Markov Chain Monte Carlo procedure. A quantum circuit first prepares the stationary distribution, after which amplitude estimation evaluates the mean. Each successful state preparation produces an estimate for $E_\pi(f)$. The possible outputs are discrete, and correspond to the values of $0$, $0.5$, and $1$. The table therefore reports how often each of these outcomes was observed over repeated runs of the circuit on the two devices. In the absence of noise, the algorithm would return $0.5$ in every successful run, as indicated in the "Expected" row. In practice, the overwhelming majority of measurements indeed concentrate on this value, while a small number of outcomes appear at $0$ and $1$. Despite the imperfect gates of current hardware, \textbf{the implemented circuits reliably capture the target expectation value}. 

\begin{table}[h!]
\centering
\begin{tabular}{lccc}
\toprule
\textbf{Measured $E_\pi(f)$:}   &  \textbf{0}   & \textbf{0.5} & \textbf{1} \\
\midrule
\textbf{H2-1}                   & 12            & 444 & 29 \\
\textbf{Helios}                 & 8             & 464 & 16 \\
\textbf{Expected}               & 0             & 500 & 0 \\
\bottomrule
\end{tabular}
\caption{\centering Measurements of $E_\pi(f)$ with $10^3$ shots via the linear combination of unitaries method. The state preparation was successful $485$ times on H2-1 and $488$ times on Helios. Data from ref. \cite{qp_cqt_quantinuum}.}
\label{tab:lcu_monte_carlo}
\end{table}

We also experimentally test one of our quantum walk constructions designed for the Metropolis-Hastings algorithm, the most widely used Markov Chain Monte Carlo method. The theoretical framework and circuit construction are presented in~\cite{claudon2026quantumcircuitsmetropolishastingsalgorithm}.

\begin{figure}[h!]
    \centering
    \includegraphics[width=\linewidth]{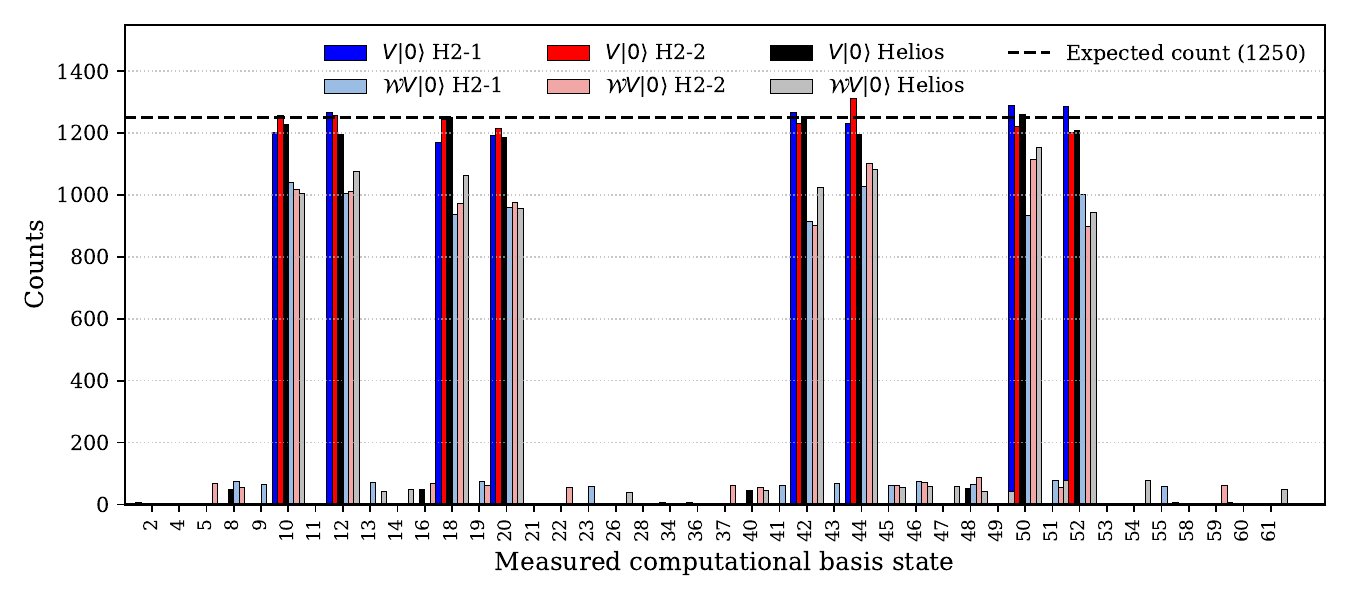}
    \caption{\centering Measurement statistics of $10^4$ measurements of $V\ket0$ and $\mathcal W V\ket0$ on Quantinuum's H2-1, H2-2 and Helios devices. We expect $V\ket0$ to be a $1$ eigenvector of $\mathcal W$, thus should remain unchanged after the application of the walk operator. The eigenstate is $V\ket0=\mathcal WV\ket0=\frac1{\sqrt8}\left(\ket{10}+\ket{12}+\ket{18}+\ket{20}+\ket{42}+\ket{44}+\ket{50}+\ket{52}\right).$ Data from ref. \cite{qp_cqt_quantinuum}.}
    \label{fig:histogram}
\end{figure}

To validate the implementation of the quantum walk, we perform the following experiment. The target fixed point of the walk operator is prepared by applying a unitary $V$ to the initial state $\ket0$. The computational basis statistics of this state are measured. The operator $\mathcal W$ is then applied, and the same statistics are measured again. The statistics are expected to remain unchanged. Figure~\ref{fig:histogram} shows the measurement histograms obtained on the H2-1, H2-2 and Helios devices. In this particular case, the state under study is $V\ket0=\mathcal WV\ket0=\frac1{\sqrt8}\left(\ket{10}+\ket{12}+\ket{18}+\ket{20}+\ket{42}+\ket{44}+\ket{50}+\ket{52}\right)$.

Our results show that, \textbf{at the current noise levels, circuits of depth corresponding to roughly $250$ elementary gates can still yield quantitatively meaningful Monte Carlo estimates, even when applied directly to the physical qubits}. By leveraging the full quadratic speedups of Markov Chain Monte Carlo algorithms, this approach offers \textbf{broad utility in our drug design pipeline as it impacts statistical physics and computational chemistry but also Bayesian inference, and machine learning optimization}.

\section* {Conclusion}
The convergence of these next-generation algorithms has been already implemented within a drug discovery pipeline \cite{Atlas26}. It enables the large-scale, high-fidelity prediction of novel therapeutics for complex challenges in areas that require new treatments such as oncology, inflammation or respiratory diseases. It marks a \textbf{transformative shift in drug discovery}, moving the field from trial-and-error methodologies \textbf{toward a future of quantitative, quantum-accurate precision}. Indeed, the proposed integrated approach (\ref{fig:summary}) establishes a unique competitive position by transcending the inherent trade-offs between chemical accuracy and computational scalability that currently constrain the industry.
Unlike classical AI drug discovery platforms that often rely on \textquote{black box} data-driven techniques, which can fail to generalize or provide physical explainability, our framework utilizes the \textbf{FeNNix-Bio1} foundation model (and associated \textbf{Ignis proprietary database}) to ensure results are directly inherited from first principles, providing the \textquote{intelligence} for interactions while leveraging the scalable Tinker-HP engine. While pure-play quantum approaches remain constrained by the 'qubit bottleneck' of NISQ-era hardware, our hybrid methodology—leveraging Hyperion emulators and Density-Based Basis-Set Correction (DBBSC)—achieves chemical accuracy today. By using only 10 to 36-40 qubits, \textbf{we deliver up to ten-fold reduction in resource overhead compared to brute-force methods enabling to explore systems far beyond the 300 qubits limits}. As hardware scales, \textbf{this framework, which is QPU-ready, will significantly accelerate the deployment of QPUs for high-fidelity quantum data production}. Furthermore, compared to usual Pharma internal efforts that frequently rely on empirical force fields requiring tedious manual parameterization, this tripartite synergy automates high-precision simulations for dynamic, reactive systems with over millions of atoms using machine learning foundation models. By mapping an exponentially large Hilbert space onto a linear number of qubits to bypass the \textquote{exponential wall} of classical many-body problems and by utilizing quantum-enhanced sampling to provide quantum acceleration for simulations, this approach serves as a superior, complementary frontier for modeling the reactive reality of complex cellular environments. While \textbf{neural network architectures are also expected to evolve alongside emerging quantum capabilities}\cite{alexeev2025artificial}, \textbf{this tripartite synergy is already addressing critical pharmacological challenges}. By leveraging current NISQ technology, these hybrid systems enable complex water placement studies in proteins through quantum optimization and achieve chemical accuracy in electronic structures via state-of-the-art GPU-accelerated quantum emulation. This progress establishes \textbf{a clear trajectory toward near-exact, Schrödinger-level, precision} as hardware and Quantum Phase Estimation (QPE) algorithms mature. As quantum hardware transitions toward fault-tolerance, it will also emerge as the \textbf{\textquote{beyond GPU} frontier - providing quadratic speedups for statistical mechanics}, including Monte Carlo and molecular dynamics simulations. This technological leap bridges the gap between static protein snapshots and the dynamic, reactive reality of complex cellular environments. Finally, while drug discovery represents a rigorous benchmark for simulation accuracy, the HPQC and ML synergy extends far beyond life sciences; these methodologies are equally transformative for modeling electrolytes, next-generation batteries, and advanced materials. 

\section{Technical Appendix} \label{sec:appendix}
\subsection{Classical Quantum Chemistry}
\subsubsection{Hartree-Fock Theory}
At the lowest level of the Pople diagram, Hartree–Fock (HF) theory treats electrons within a mean-field approximation, neglecting their instantaneous correlated motion. The solution of the Hartree-Fock approximation is a compact single determinant $\Phi_{\text{HF}}(\{ c_i\}_{i=1, N_{\text{MO}}}; \textbf{x}_1, \textbf{x}_2, ..., \textbf{x}_{N_{\text{elec}}})$, where $\{ c_i\}_{i=1, N_{\text{MO}}}$ are a set of parameter to optimize in this problem. Starting with an Hartree-Fock calculation is a good starting point to have a crude qualitative description of the molecular system but not quantitatively comparable with experiments. To allow the latter, on must take into account the instantaneous correlated motions. For that there are the two main family of correlated methods: 1) methods based on density functional theory (DFT).; 2)methods based on the wave-function approximation (WFT) that we will introduce in the following paragraphs, 

\subsubsection{Density-functional Theory}
The Hohenberg-Kohn theorem demonstrating the mapping from the potential (in our case electron-nucleus) to the ground-state density. Thus, DFT can claim a considerable reduction of the computational cost of quantities comparable with experiments in comparison with wave-function based approaches. In the Kohn-Sham formulation of DFT, one solves the following minimization problem in terms of single-determinant wave functions $\Phi$:
\begin{equation}\label{DFT}
    E_0 = \min_{\Phi} \left\{ <\Phi| \hat{T} + \hat{V}_{\text{ne}} |\Phi> + E_{\text{Hxc}}[n_\Phi] \right\},
\end{equation}
where $E_{\text{Hxc}}$ is the Hartree-Exchange-correlation density functional and $n_{\Phi}$ is the electronic one-electron density of the single-determinant wave function $\Phi$. We note that DFT allows to accounts for correlation effects through a single-determinant formulation where the correlation is hidden in a density functional.
In practice, DFT suffers from the unknown exact expression of $E_{\text{Hxc}}[n_\Phi]$. Therefore, in practice one must choose among a zoo of approximations which are sufficient to make density-functional approximation a standard in chemistry applications but which lacks 1) systematic improvements of the approximations 2) systematic understanding of successful of unsuccessful results. In practice, DFT is the "workhorse" of quantum chemistry because it shifts the focus from the complex, $3N$-dimensional wavefunction (where $N$ is the number of electrons) to the 3-dimensional electron density. This mathematical shift reduces the computational complexity from exponential to a much more manageable cubic scaling ($O(N^3)$), allowing researchers to model systems with hundreds or even thousands of atoms. By using functional approximations to handle electron correlation and exchange, DFT provides a remarkable balance: it is accurate enough to predict chemical properties like bond lengths and reaction energies, yet efficient enough to be run on standard laboratory computers. In the context of the design of NNPs, this balance allows DFT to be used to compute large scale databases of millions of energies and forces.\newline
\subsubsection{Wave-function Theory} In this family of methods, one focuses on the form and parameterization of the wave-function $\Psi(\{ \theta\}; \textbf{x}_1, \textbf{x}_2, ..., \textbf{x}_{N_{\text{elec}}})$ and the associated methods to optimize the problem parameters $\{\theta\}$ by taking into account the exact electron-electron interaction. In this section, we will mention, among all the available approaches, selected CI and DMRG who are expressed as variational problem but also coupled-cluster and diffusion Monte-Carlo which can be qualified as projector method where the standard implementations leads to non-variational problems. 

\paragraph{Selected Configuration Interaction (sCI) methods}\label{par:sCI}
sCI \cite{huron1973iterative,garniron2019quantum} construct systematically improvable multideterminant wavefunctions by iteratively selecting the most important determinants from the Full-CI expansion. These methods achieve high accuracy while using only a small fraction of the full configurational space. At iteration $k$, the sCI wavefunction is a linear combination of the set of determinants $\{\ket{\Phi_i^{(k)}}\}_{i=1, \tilde{N}_{\text{det}}^{(k)}}$ with coefficients $\{d_i^{(k)}\}$:
\begin{equation}
    \ket{\Psi^{(k)}_{\text{sCI}} (\{ d_i^{(k)}\}_{i=1, \tilde{N}_{\text{det}}^{(k)}}; \textbf{x}_1, \textbf{x}_2, ..., \textbf{x}_{N_{\text{elec}}})} = d_0^{(k)} \ket{\Phi_{\text{HF}}} + \sum_{i=1}^{\tilde{N}_\text{det}^{(k)}} d_i^{(k)} \ket{\Phi_i^{(k)}},
\end{equation}
where $\tilde{N}_\text{det}^{(k)}$ is the number of included determinants at iteration $k$.
To expand the wavefunction and explore more of the full configurational space, new determinants are selected from those connected to the current wavefunction. Our Hamiltonian is at most a two-electron operator, so the number of connections to any determinant in $\{\ket{\Phi_i^{(k)}}\}$ is quartic in the system size ($O(N_\text{elec}^2 N_\text{MO}^2)$). Some selection criterion is evaluated for these connected determinants, and the best $N_\text{sel}^{(k)}$ of these are added to the wavefunction to obtain $\{\ket{\Phi_i^{(k+1)}}\}_{i=1, \tilde{N}_{\text{det}}^{(k+1)}}$, where  $\tilde{N}_{\text{det}}^{(k+1)}=\tilde{N}_{\text{det}}^{(k)} + N_{\text{sel}}^{(k)}$.

This class of methods provides a controllable path toward the exact solution (by iteratively increasing $\tilde{N}_\text{det}$), and it also has enormous flexibility in all parts of the iterative process (e.g., periodically pruning unimportant determinants, using stochastic methods, incorporating different selection criteria, using parameterized filtering/screening techniques)\cite{schriber_aci_2017,garniron2019quantum,sharma_shci_2017,tubman_asci_2020}.
The full configurational space grows exponentially with system size, so sCI methods will still hit the exponential wall: achieving progressively smaller improvements in energy requires an exponentially growing number of determinants as the Full-CI limit is approached; however, these methods traverse the space in a way that can yield incredibly accurate results for problem sizes that are not feasible with other high-accuracy methods.

\paragraph{Coupled Cluster approaches} These methods, particularly CCSD(T)~\cite{ccsdt}, are often considered the “gold standard”, provide high accuracy but exhibit steep computational scaling (typically \(O(N^7)\)) and can fail for strongly correlated, multireference systems.  In this approach, we consider an exponential ansatz for the wave-function:
\begin{equation}
\ket{\Psi_{\text{CCSD}}} = e^{\hat{T}_1 + \hat{T}_2} \ket{\Phi_{\text{HF}}},
\end{equation}
where $\hat{T}_1$ and $\hat{T}_2$ are cluster of excitation operators such as $\hat{T}_1$ groups one-electron excitation operators, and similar with $\hat{T}_2$. Solving the CCSD problem is, roughly, optimizing the parameters of the excitation operators through a set of amplitudes equations based on a projection onto the Hartree-Fock solution.
Alternative high-accuracy approaches include the Density Matrix Renormalization Group (DMRG) and Diffusion Quantum Monte Carlo (DMC).

\paragraph{Density Matrix Renormalization Group} DMRG is ~\cite{White1992,DMRG,Schollwck2011} a popular optimization technique to approximate the ground-state energy of strongly correlated quantum systems, it can be seen as a numerical algorithm for the efficient truncation of the Hilbert space of low-dimensional quantum systems. It was developed to overcome the curse of dimensionality, where the dimension of the quantum many-body Hamiltonian scales exponentially with the number of degrees of freedom, making exact diagonalization impossible for larger systems. 

In the original formulation of White \cite{White1992}, DMRG algorithm uses a truncated iterative diagonalization ( a decimation procedure) that divides a large system into smaller numerical solvable blocks. This formulation uses two main strategies. The infinite-system approach iteratively increases the system size while truncating the basis, retaining only the eigenvectors that correspond to the highest eigenvalues of the reduced density matrices. The finite-system approach refines this initial approximation by keeping the total system size fixed and employing a sweeping protocol.  By systematically growing one block while shrinking the other, this back-and-forth sweeping optimizes the wave function variationally, ensuring the estimated energy  decreases until convergence is reached.
A few years after its invention, DMRG was reformulated using the language of tensor networks, which allows for much more efficient computer coding. The finite-system DMRG is cast as a variational optimization routine where the Hamiltonians are represented as Matrix Product Operators (MPO) and the wavefunction as an MPS.  Rather than explicitly constructing and diagonalizing density matrices to truncate the system, this approach applies a Singular Value Decomposition (SVD) to the local tensors within the MPS. Discarding the lowest singular values during the SVD yields optimal data compression, this is mathematically equivalent to retaining the highest eigenvalues of the density matrix in the original formulation.

While DMRG is an incredibly powerful tool for specific types of physics problems, DMRG is practically only efficient for one-dimensional (1D) quantum models. The method's success relies on truncating the Hilbert space to a maximum number of retained states, denoted by the bond dimension $D$. If a quantum state is highly entangled, such as in two or three-dimensional systems that obey boundary area laws, the required bond dimension $D$ scales exponentially with the system size. In such cases, the algorithm encounters the very exponential wall it was originally designed to bypass. Last but not least, one can consider Diffusion Monte Carlo (DMC).
\paragraph{Diffusion Monte Carlo (DMC)} DMC \cite{DMC} is a stochastic quantum chemistry method that solves the many-body Schrödinger equation by evolving an initial wavefunction in imaginary time. Unlike traditional ab initio methods that rely on finite basis sets, DMC operates directly in real space, making it inherently free from basis-set incompleteness errors and allowing it to reach the infinite-basis (CBS) limit by construction. The algorithm works by ensuring that any trial state not orthogonal to the ground state will converge to that ground state in the long-time limit. To handle the fermion sign problem, DMC typically employs the fixed-node (FN) approximation, which constrains the nodal surface of the many-body wavefunction to match that of a reference trial wavefunction, such as a Slater-Jastrow form. Due to its "embarrassingly parallel" nature, DMC is ideally suited for exascale supercomputing, enabling highly accurate energy and force evaluations for complex molecular systems containing thousands of electrons on exascale-class supercomputers.
\subsubsection{Basis Sets}
\paragraph{General case}In quantum chemistry, basis sets are collections of one-electron orbital functions, typically Gaussian-type orbitals (GTOs) , used to represent the electronic Hamiltonian and describe the molecular wavefunction. Because a finite number of functions can only approximate the true, continuous electronic distribution, the choice of basis set directly dictates the accuracy and computational cost of a simulation. The Complete Basis-Set (CBS) limit represents the theoretical point where the basis set becomes infinitely large, meaning the mathematical representation is no longer constrained by the number of functions used. Reaching the CBS limit is essential for achieving "chemical accuracy"—an error within approximately 1.6 mHa—as exact solutions like Full Configuration Interaction (FCI) are defined only at this limit. Since using an infinite basis is physically impossible, researchers typically estimate the CBS limit through extrapolation schemes or utilize corrective methods, such as Density-Based Basis-Set Correction (DBBSC), to approach it with significantly fewer quantum resources. Alternatively, among the classical quantum chemistry methods, the primary advantage of Diffusion Monte Carlo (DMC) is its ability to operate directly in real space, which makes it inherently free from basis-set incompleteness errors and allows it to reach the CBS limit by construction.
\paragraph{Quantum Computing} In quantum computing, the choice of basis sets is directly related to the number of qubits required for a simulation, as the encoding of the molecular problem determines how many quantum resources are consumed. In the standard second-quantized framework often used for chemistry, the electronic Hamiltonian is mapped such that one spin-orbital corresponds to one qubit. Consequently, using a larger and more accurate basis set (moving toward the CBS limit) linearly increases the number of qubits needed. Therefore, achieving chemically useful results typically requires basis sets much larger than minimal ones. For complex molecules, the qubit count for these large basis sets quickly exceeds the capacity of current NISQ hardware and early FTQC (Fault-Tolerant Quantum Computing) devices. Techniques like DBBSC (Density-Based Basis-Set Correction) act as a "shortcut" by allowing high-accuracy predictions—comparable to large basis sets—while using minimal basis sets that require far fewer qubits. For example, reaching "triple-zeta" quality for an $N_2$ molecule, which might normally require 100 qubits, can be achieved with only 16 qubits using these correction methods.Alternatively, in first-quantized real-space formulations, the relationship changes: the qubit count scales logarithmically with the spatial resolution of the grid rather than linearly with the number of orbitals
\section* {Acknowledgments}
The authors thank Marie-Céline Bezat, Christophe Jurczak and Robert Marino for careful reading of the manuscript. This work has received funding from the European Research Council (ERC) under the European Union's Horizon 2020 research and innovation program (grant agreement No 810367), project EMC2 (JPP). Support from the PEPR EPIQ - Quantum Software (ANR-22-PETQ-0007, JPP) and HQI (JPP) programs is acknowledged.

\bibliographystyle{unsrt}
\bibliography{biblio}

\end{document}